\documentclass[a4paper, fleqn, usenatbib]{mnras}

\usepackage{newtxtext, newtxmath}

\usepackage[T1]{fontenc}
\usepackage{ae, aecompl}

\usepackage[dvipdfmx]{graphicx}	
\usepackage{amsmath}	
\usepackage{amssymb}	
\usepackage[hang, small, bf]{caption}
\usepackage[subrefformat=parens]{subcaption}
\usepackage{color}

\newcommand{\GADGET}{{\small GADGET-3}}
\newcommand{\Msun}{{M_\odot}}

\title[Osaka feedback model]{Osaka Feedback Model: Isolated Disk Galaxy Simulations}

\author[Shimizu, et al.]
{Ikkoh Shimizu$^{1}$\thanks{E-mail: shimizu@astro-osaka.jp}, 
Keita Todoroki$^2$, Hidenobu Yajima$^{3,4,5}$, 
Kentaro Nagamine$^{1,6,7}$, \\
$^{1}$Theoretical Astrophysics, Department of Earth \& Space Science, Graduate School of Science, Osaka University, 1-1 Machikaneyama, Toyonaka, Osaka 560-0043, Japan \\
$^{2}$Department of Physics \& Astronomy, University of Kansas, 1082 Malott, 1251 Wescoe Hall Dr., Lawrence, KS 66045-758, USA \\
$^{3}$Frontier Research Institute for Interdisciplinary Sciences, Tohoku University, Sendai 980-8578, Japan \\
$^{4}$Astronomical Institute, Tohoku University, Sendai 980-8578, Japan \\
$^{5}$Center for Computational Sciences, University of Tsukuba, Ten-nodai, 1-1-1 Tsukuba, Ibaraki 305-8577, Japan \\
$^{6}$Department of Physics \& Astronomy, University of Nevada, Las Vegas, 4505 S. Maryland Pkwy, Las Vegas, NV 89154-4002, USA \\
$^{7}$Kavli IPMU (WPI), The University of Tokyo, 5-1-5 Kashiwanoha, Kashiwa, Chiba, 277-8583, Japan \\
}

\pubyear{2017}

\begin{document}
\label{firstpage}
\pagerange{\pageref{firstpage}--\pageref{lastpage}}
\maketitle

\begin{abstract}
We study various implementations of supernova feedback model and present the results of our `Osaka feedback model' using isolated galaxy simulations performed by the smoothed particle hydrodynamics (SPH) code \GADGET. 
Our model is a modified version of Stinson et al.'s work, and we newly add the momentum kick for SN feedback rather than only thermal feedback.  
We incorporate the physical properties of SN remnants from the results of Chevalier and McKee \& Ostriker, such as the effective radius of SN bubble and the remnant life-time, in the form of Sedov-Taylor (ST)-like solutions with the effect of radiative cooling. 
Our model utilizes the local, physical parameters such as density and temperature of the ISM rather than galactic or halo properties to determine the galactic wind velocity or mass-loading factor. 
The Osaka model succeeds in self-regulating star formation, and naturally produces galactic outflow with variable velocities depending on the local environment and available SN energy as a function of time.
An important addition to our previous work by Aoyama et al. is the implementation of the {\small CELib} chemistry library which allows us to deal with the time-dependent input of energy and metal yields for type Ia \& II supernovae (SNe) and asymptotic giant branch (AGB) stars. 
As initial tests of our model, we apply it to isolated galaxy simulations, and examine various galactic properties and compare with observational data including metal abundances. 
\end{abstract}



\section{Introduction}
\label{sec:intro}

The importance of feedback effects for galaxy formation and evolution has been recognized since the early days of galaxy formation studies \citep[e.g.,][]{Larson74, Blumenthal84, Silk85, Dekel86}.  
In the current paradigm of $\Lambda$ cold dark matter (CDM) model, without the feedback, too many low-mass galaxies form in low-mass dark matter halos, making the faint-end slope of galaxy stellar mass function too steep at low redshifts ($z$). 
Massive galaxies also form too many stars at low-$z$, and cannot form a proper red sequence.  
These are known as the overcooling problem in galaxy formation \citep[e.g.,][]{White1991}. 
Currently, massive stars, supernovae (SNe), and supermassive black holes (i.e., active galactic nuclei; AGN) are considered to be the primary sources for such feedback effects. 

In the early cosmological hydrodynamic simulations, thermal energy from SN explosions was injected in the ambient interstellar medium (ISM) following the star formation (i.e., thermal feedback) \citep[e.g.,][]{Cen92a, Katz1996}.  
However, such energy was quickly radiated away from high-density gas, and did not suppress subsequent star formation as the gas cooled rapidly. 
This was the `numerical overcooling problem' in the CDM simulations of galaxy formation. 
To avoid this problem, some researchers resorted to shutting off the radiative cooling for a certain period of time to artificially enhance the heating of gas by the feedback \citep[e.g.,][]{Thacker2001, Springel2003, Stinson2006}. 
Another strategy is to introduce a multiphase ISM model \citep[e.g., ][]{Springel2003, Keller2014}. 
In this model, they assume that each gas element has cold high-density gas and hot low-density gas in pressure equilibrium. 
The model can avoid the problem by storing feedback energy in the hot phase where the cooling is inefficient. 

Another problem related to feedback is the inability to regulate baryon content in galaxies.
Fixing the former problem of excessive SFR does not necessarily fix the latter problem. 
To regulate baryon content in simulated galaxies, people also developed `kinetic feedback' model, where some fraction of SN energy is given to the ambient gas as kinetic energy of winds. 
In the case of SPH simulations, this energy was given to the ambient gas particles \citep[e.g.,][]{Springel2003, Oppenheimer2006, Vecchia2008, Choi2011, Okamoto2014}, and in the case of Eulerian mesh simulations it was given to the fluid elements in the ambient cells \citep[e.g.,][]{Cen05}. 
Furthermore, in order to reproduce the galactic wind phenomena observed in the local Universe \citep[e.g., M82, as examined by][]{Lehnert99,Strickland09} and in high-$z$ star-forming galaxies such as the Lyman break galaxies \citep{Shapley03, Steidel10}, wind particles are often decoupled from hydrodynamic interactions for a while by hand so that these particles can escape from high-density star-forming regions \citep[e.g.,][]{Springel2003, Oppenheimer2006, Choi2011, Okamoto2014}. 

Different simulations have adopted different combinations of thermal and kinetic feedback with varying efficiencies. 
While we refrain from explaining the details of individual models here (some examples are given in latter Section), in most cases, the wind velocity and the mass-loading factor are taken to be proportional to some power of virial velocity of the halo or velocity dispersion of the system. 
In the simplest form, these parameters are set to constant values \citep[e.g.,][]{Springel2003, Vecchia2008}. 
Different power-indices may represent different underlying physics such as the energy-driven or momentum-driven wind \citep{Murray2005, Dave2006, Oppenheimer2006, Oppenheimer2010, Choi2011, Puchwein2012, Okamoto2014}. 
These models still rely on the properties of dark matter halo or galaxy as a whole, and the parameters described above are not the outcome of self-consistent treatment of physical processes on small scales, mainly due to lack of resolution in large-scale cosmological simulations. 

More recently, \citet{Vecchia2012} proposed a model which deposit thermal energy stochastically to heat up gas particles to a desired temperature, high enough ($T>10^7~{\rm K}$) for the gas to give feedback to circum gas before radiative cooling efficiently works.
They also showed that their model can reproduce strong galactic wind. 
However, as the authors pointed out, the feedback efficiency is sensitive to the numerical resolution. 
These problems can be recognized in almost all feedback models. 
Moreover, the SN explosion site does not necessarily correspond exactly to the location of the star-forming site due to the stochastic model. 
In order to solve these problems, \citet{Keller2014} implemented a superbubble feedback model based on the fact that the feedback from star clusters forms superbubbles and behave quite differently from isolated SN. 
They introduced models for three subgrid physics: thermal conduction, evaporation, and multiphase nature of ISM. 
In their model, the feedback energy is stored in hot, low-density gas in order to mitigate the overcooling problem. 
They successfully suppressed the star formation activity and made strong galactic winds without using many parameters for their simulation. 
\citet{Agertz2013} studied the momentum and energy budget for stellar feedback considering stellar evolution, i.e., stellar radiation, stellar wind, type-II SN (SN-II) and type-Ia SN (SN-Ia). 
Especially, for the SN-II feedback, they used the momentum of the terminal phase of the Sedov-Taylor (ST) phase. 
Similar approaches are also taken by \citet{Hopkins2011,  Kimm2014, Hopkins2018}. 
In their model, they consider that the momentum of gas is boosted in the ST phase by the thermal pressure from a hot bubble. 
These feedback schemes are sometimes called `mechanical feedback' \citep{Hopkins2011, Hopkins2018}. 
Interestingly, these authors argued that the numerical resolution dependency of their model results is very weak. 

So far, many studies have focused on the individual SN explosion rather than multiple explosions without directly resolving the individual explosion except for the \citet{Keller2014}. 
Moreover, almost all studies have not been able to directly resolve the individual explosion due to the limitation of the computer performance. 
It has long been known that the solution for multiple SN events is qualitatively different from a single explosion \citep[e.g., ][]{Castor1975, Weaver1977, MacLow1988, Keller2014}.
This means that using the analytic solution of single SN explosion might lead to wrong results. 
Therefore, we should be cautious in using these methods. 
However, this is still a challenging issue for the treatment of multiple SN explosions in cosmological simulations, and we leave further refinement of the model to our future work. 

In recent years, new sets of large-scale cosmological hydrodynamic simulations have been performed, such as Illustris \citep{Vogelsberger2014}, EAGLE \citep{Schaye2015}, and IllustrisTNG \citep{Pillepich2018}. 
However, these large cosmological simulations still have a difficulty in breaking the barrier of $\sim$kpc spatial resolution while simultaneously solving for the large-scale structure on $\sim$100\,Mpc scales. 

Ideally, we would like to resolve the small-scale physics on sub-kpc scales, while simultaneously considering the cosmological effects such as the large-scale structure traced by dark matter and the cosmic inflow of pristine gas. 
In the recent years, it has become easier to perform zoom-in cosmological hydrodynamic simulations, and a number of simulations are being performed with resolution better than 100\,pc, although the number of sample galaxies is limited compared to those in a cosmological full box \citep[e.g.,][]{Agertz15, Muratov2015, Stewart2017}. 

Given these situations, it is still worthwhile to test the detailed impact of star formation and feedback on small scales employing isolated disk galaxies with higher resolution and finer time-steps \citep[e.g.,][]{Kim14, Keller2014}.   
With such simulations, we can achieve higher resolution with much less computational resources, and attempt to develop a more physically motivated feedback model based on the local physical quantities, without artificially introducing additional parameters. 

Furthermore, recent works using high-resolution zoom-in cosmological hydrodynamic simulations have emphasized the importance of early stellar feedback (ESFB) \citep[e.g.,][]{Ceverino09, Fall2010, Hopkins2011, Agertz2013, Aumer2013, Stinson2013, Agertz15, Rosdahl2015}. 
The stellar wind and radiation might sweep the ISM near the star-forming region prior to SN explosions, thereby making the impact of SN feedback stronger and efficient. 
In these simulations, efficient self-regulation of star formation is achieved by early stellar feedback (ESFB) and SN feedback, making the resulting stellar masses more consistent with the stellar-to-halo-mass ratio (SHMR) obtained from abundance matching techniques \citep[e.g.,][]{Moster2013, Behroozi13}. 
On the other hand, the model used in \citet{Vecchia2012} and \citet{Keller2014, Keller2015, Keller2016} reproduced a set of observational results such as SHMR, without the treatment of ESFB. 
It is still a matter of debate whether the ESFB model, which is often not motivated well physically, is necessary or not. 

Throughout this paper, we adopt {\it Planck} $\Lambda$CDM cosmology with following cosmological parameters: 
$(\Omega_{\rm{M}}, \Omega_{\Lambda}, \Omega_{\rm B}, \sigma_8, h) = (0.3089, 0.6911, 0.0486,  0.8159, 0.6774)$, 
where $h = H_0 / 100 {\rm ~km ~s^{-1} ~Mpc^{-1}}$ 
\citep{Planck2015}. 
We assume the Chabrier initial mass function (IMF) with a mass range of 0.1--120 $\rm M_\odot$ \citep{Chabrier2003} for both observational data and our simulation, and our stellar masses include the remnant mass as discussed in \citet{Shimizu2013}.


\section{Simulation Setup}
\label{sec:sim_set}

We use a modified version of the Tree-PM smoothed particle hydrodynamics (SPH) code \GADGET, which is the successor of  {\small GADGET-2} \citep{Springel2005}. 
Our code includes the time-step limiter \citep{Saitoh2009} and the density-independent formulation of SPH \citep{Hopkins2013, Saitoh2013}. 
We adopt the quintic spline kernel \citep{Morris1996}, and the number of neighbour particles for each SPH particle is set to $128 \pm 8$. 
The radiative cooling is calculated using the {\small Grackle-3} chemistry and cooling library \footnote{https://grackle.readthedocs.org/}  \citep {Smith2017}, which solves the primordial chemistry network for atomic H, D, He, as well as molecular H$_2$ and HD. 
The library also includes photo-heating and photo-ionization under the UV background (UVB), 
and we employ the UVB value at $z=0$ for a isolated disk galaxy. 
In order to avoid artificial numerical fragmentation when the Jeans mass at low temperatures is not resolved, 
we introduce a Jeans pressure floor following \citet{Hopkins2011, Kim2016}: 
\begin{equation}
P_{\rm Jeans} = \frac{1}{\gamma \pi} N_{\rm Jeans}^2 G \rho_{\rm gas} r_{\rm sys}^2,
\end{equation}
where $\gamma = 5/3$, $N_{\rm Jeans} = 4.0$ and $r_{\rm sys}$ is chosen from the larger one of either the smoothing length or the gravitational softening of an SPH particle. 
In this prescription, we ensures that the Jeans length is always resolved with $N_{\rm Jeans}$ system lengths. 

We use three different initial conditions for our isolated disk galaxy simulations as summarized in Table~\ref{tbl1}: `M12' and `M12hi' are taken from the AGORA project \citep[M12:][]{Kim2016}, and `M10' is a dwarf galaxy used by \citet{Vecchia2008, Vecchia2012}. 
The total mass of each galaxy is $10^{12}~(10^{10})~{\rm M_{\odot}}$ for the M12 (M10) galaxy. 
We employ $10^5$ dark matter particles, $10^5$ gas (SPH) particles,
$10^5$ and  $1.25\times 10^4$ collisionless particles that represent the stars in the disk and bulge, respectively. 
We adopt a fixed gravitational softening length of $\epsilon_{\rm grav} = 80~ (20)$\,pc for the M12 (M10) galaxy, 
but allow the minimum gas smoothing length to reach $10$ per cent of $\epsilon_{\rm grav}$. 
The final gas smoothing length reaches $\sim 30~(10)$\,pc for the M12 (M10) galaxy as the gas becomes denser owing to radiative cooling. 

For a convergence check, we use the higher resolution galaxy `M12hi' with mass $10^{12}~{\rm M_{\odot}}$ which is exactly the same as M12 galaxy except for the mass resolution. 
In M12hi simulation, we set $\epsilon_{\rm grav} = 40~{\rm pc}$.  

\begin{figure*}
\includegraphics[width = 165mm]{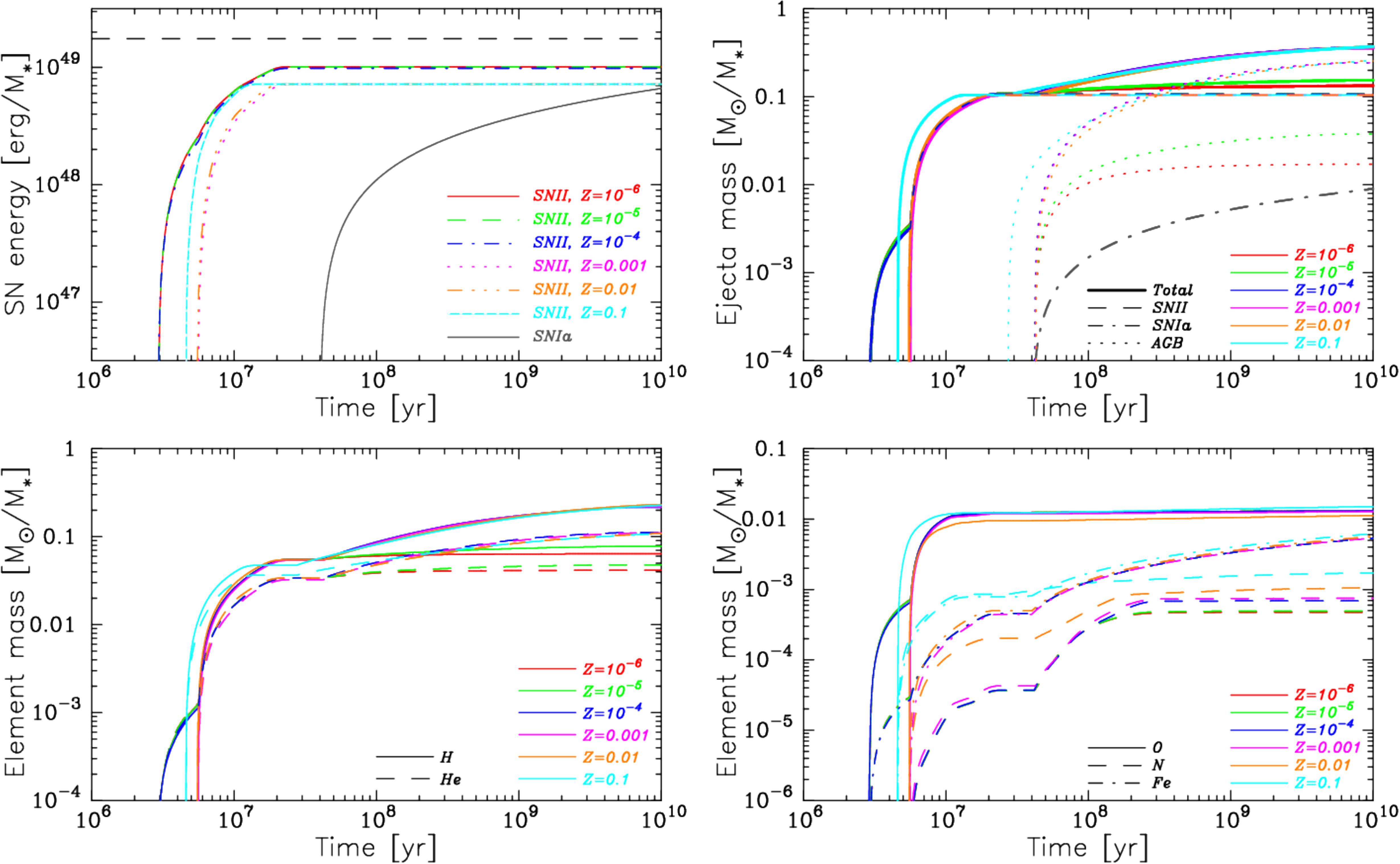}
\caption{{\small CELib} outputs for various metallicities. 
Each panel shows SN (SN-II and SN-Ia) energy output ({\it top left}), the ejecta (SN-II, SN-Ia) mass ({\it top right}), yield pattern of primordial elements (H, He; {\it bottom left}) and some metal elements (O, N, Fe; {\it bottom right}). 
The corresponding lines are shown in the legend. }
\label{fig:celib}
\end{figure*}

One of new features of the present work is the usage of the {\small CELib} library \citep{Saitoh2016, Saitoh2017}, which allows a separate treatment of Type II supernovae (SN-II), Type Ia SNe (SN-Ia), and asymptotic giant (AGB) stars.  
We consider the stellar lifetime and metallicity-dependent metal yield and mass loss from SN-II, SN-Ia and AGB stars based on {\small CELib}.  
We also calculate the time-dependent SN rate with this library. 
We adopt the delay-time distribution function of SN-Ia with a power law of $t^{-1}$ for the SN-Ia event rate \citep[e.g.,][]{Totani2008, Maoz2012}. 
Single SN explosion energy is set to $1.0 \times 10^{51}$\,erg. 
{\small CELib} library can treat the evolution of the 13 important elements with H, He, C, N, O, Ne, Mg, Si, S, Ca, Fe, Ni, and Eu. 
These elements are major coolants in the inter stellar medium (ISM). 
Eu is mainly a product and tracer of $r$-process in the neutron star mergers. 
Figure \ref{fig:celib} represents the {\small CELib} output for various metallicities adopted in this work. 
As shown in this figure, the metal abundance strongly depends on the time evolution of the star cluster. 
In order to explore the evolution, we need to reasonably resolve the SN and AGB events rather than to integrate these events. 
In the following, we describe how we treat resolved SN and AGB events in our simulations. 

Some part of this work was already employed in \citet{Aoyama2017}, which focused on the implementation of dust model without the {\small CELib} library. 
This paper does not deal with the dust model, but focus more on the details of the feedback model implementation with the {\small CELib} library.

\begin{table*}
\begin{center}
\begin{tabular}{lcccc}\hline
Galaxy name &  & M12 & M10& M12hi \\ \hline
Parameter & Symbol & Value & Value & Value \\ \hline
Gas mass  & $M_{\rm gas}$ & $8.59 \times 10^{9} ~{\rm M_{\odot}}$ & $1.21 \times 10^{8} ~{\rm M_{\odot}}$ & $8.59 \times 10^{9} ~{\rm M_{\odot}}$ \\
Dark matter mass & $M_{\rm DM}$ & $1.25 \times 10^{12} ~{\rm M_{\odot}}$ & $9.51 \times 10^{10} ~{\rm M_{\odot}}$ & $1.25 \times 10^{12} ~{\rm M_{\odot}}$ \\
Disc mass & $M_{\rm disk}$ & $4.30 \times 10^{9} ~{\rm M_{\odot}}$ & $2.81 \times 10^{8} ~{\rm M_{\odot}}$ & $4.30 \times 10^{9} ~{\rm M_{\odot}}$ \\
Bulge mass & $M_{\rm bulge}$ & $3.44 \times 10^{10} ~{\rm M_{\odot}}$ & $1.41 \times 10^{8} ~{\rm M_{\odot}}$ & $3.44 \times 10^{10} ~{\rm M_{\odot}}$ \\
Total mass & $M_{\rm tot}$ & $1.3 \times 10^{12} ~{\rm M_{\odot}}$ & $1.00 \times 10^{10} ~{\rm M_{\odot}}$ & $1.3 \times 10^{12} ~{\rm M_{\odot}}$ \\ \hline
Virial radius & $R_{\rm vir}$ & $205 ~{\rm kpc}$ & $35.1 ~{\rm kpc}$ & $205 ~{\rm kpc}$ \\
Virial velocity & $V_{\rm vir}$ & $200 ~{\rm km/s}$ & $35.1 ~{\rm km/s}$ & $200 ~{\rm km/s}$ \\
Scale length & $r_{\rm disk}$ & $3.43 ~{\rm kpc}$ & $0.533 ~{\rm kpc}$ & $3.43 ~{\rm kpc}$ \\
Scale height & $h_{\rm disk}$ & $0.343 ~{\rm kpc}$ & $0.0533 ~{\rm kpc}$ & $0.343 ~{\rm kpc}$ \\ \hline
Number of gas particle & $N_{\rm gas}$ & $1.00\times 10^{5}$ & $1.00\times 10^{5}$ & $1.00\times 10^{6}$ \\
Number of dark matter & $N_{\rm DM}$ & $1.00\times 10^{5}$ & $1.00\times 10^{5}$ & $1.00\times 10^{6}$ \\
Number of disc particle & $N_{\rm disk}$ & $1.00\times 10^{5}$ & $1.00\times 10^{5}$ & $1.00\times 10^{6}$ \\
Number of bulge particle & $N_{\rm bulge}$ & $1.25\times 10^{4}$ & $1.25\times 10^{4}$ & $1.25\times 10^{5}$ \\ \hline 
Gas particle mass & $m_{\rm gas}$ & $8.59 \times 10^{4} ~{\rm M_{\odot}}$ & $1.21 \times 10^{3} ~{\rm M_{\odot}}$ & $8.59 \times 10^{3} ~{\rm M_{\odot}}$ \\
Dark matter particle mass & $m_{\rm DM}$ & $1.25 \times 10^{7} ~{\rm M_{\odot}}$ & $9.51 \times 10^{4} ~{\rm M_{\odot}}$ & $1.25 \times 10^{6} ~{\rm M_{\odot}}$ \\
Disc particle mass & $m_{\rm disk}$ & $3.44 \times 10^{5} ~{\rm M_{\odot}}$ & $2.81 \times 10^{3} ~{\rm M_{\odot}}$ & $3.44 \times 10^{4} ~{\rm M_{\odot}}$ \\
Bulge particle mass & $m_{\rm bulge}$ & $3.44 \times 10^{5} ~{\rm M_{\odot}}$ & $1.13 \times 10^{4} ~{\rm M_{\odot}}$ & $3.44 \times 10^{4} ~{\rm M_{\odot}}$ \\ \hline
Grav. softening length & $\epsilon_{\rm grav}$ & $80$ pc & $20$ pc & $40$ pc \\ 
Final gas smoothing length & $h_{\rm sml}$ & $\sim$30 pc & $\sim$10 pc & $\sim$10 pc \\ 
\hline
\end{tabular}
\caption{Physical parameters of the initial conditions for the idealized isolated disk galaxies. 
M12 is same as the AGORA project \citep{Kim2016}, and M12hi is the same as M12 galaxy but with 10 times higher mass resolution. 
M10 is $10^{10}\,{\rm M_{\odot}}$ galaxy which is used in \citet{Vecchia2008, Vecchia2012}. 
The character $M$ is used for the galactic masses, while $m$ is used for that of a single particle. }
\label{tbl1}
\end{center}
\end{table*}


\section{Star Formation and Feedback Models}

\subsection{Star Formation}
\label{sec:SF}

We assume that the star formation (SF) takes place when gas density exceeds a threshold density ($n_{\rm H} > 10~{\rm cm}^{-3}$ for the isolated disk galaxy simulation and $n_{\rm H} > 0.1~{\rm cm}^{-3}$ for the cosmological simulation), and the gas temperature is sufficiently low ($T < 10^4$\,K). 
The gas particles that satisfy above conditions spawn star particles stochastically with Chabrier IMF with mass range $0.1$ to $120~ {\rm M_{\rm \odot}}$ \citep{Chabrier2003}. 
As in ordinary galaxy formation simulations, a star particle represents a star cluster with a range of stellar masses of IMF. 
The number of star particles that can be spawned from a single gas particle is set to $2$, with a mass of about half of the initial gas mass. 
Note that the initial star particle mass is not exactly one half of the initial gas particle mass, because the gas particles can accrete recycled ISM from past SNe or AGB stars. 
 
Our SF prescription is similar to previous works \citep{Katz1992, Nagamine2001, Springel2003, Stinson2006, Kim14, Kim2016}.  
The star formation rate (SFR) density $\dot{\rho_*}$ is given by 
\begin{equation}
\dot{\rho_*} = c_* \frac{\rho_{\rm gas}}{t_{\rm ff}}, 
\end{equation}
where $c_*$ is the SF efficiency, $\rho_{\rm gas}$ is the gas density and $t_{\rm ff}$ is the local free-fall time. 
We adopt $c_* = 0.05$ as the fiducial value, which roughly reproduces the observed relation between the surface gas density and the surface SFR density, i.e., the Kennicutt-Schmidt law  \citep{Kennicutt1998}. 
The star particles are stochastically spawned from gas particles with the following probability \citep{Katz1992, Springel2003}:
\begin{equation}
{\cal P_*} = \frac{m_{\rm gas}}{m_*} \left[ 1 - \exp{\left( \frac{dt}{t_{\rm SF}} \right)} \right], 
\end{equation}
where $m_{\rm gas}$, $m_*$, $dt$ and $t_{\rm SF}$ are the gas particle mass, initial star mass, computational time-step and the SF time-scale $(t_{\rm SF} \equiv t_{\rm ff} / c_*)$, respectively. 
The initial star particle mass is defined by $m_{*} = m_{\rm gas}^{\rm init} / n_{\rm spawn}$, where $m_{\rm gas}^{\rm init}$ is the initial gas particle mass, and $n_{\rm spawn}$ is the number of star particles that can be spawned from a gas particle, respectively. 
We adopt $n_{\rm spawn} = 2$ throughout this paper. 
We note that the value of $n_{\rm spawn}$ strongly affects the ESFB and SN feedback, because the input feedback energy from a star particle depends on its mass. 
We discuss this issue in later sections. 


\subsection{Supernova Feedback}
\label{sec:SNFB}

The Sedov-Taylor (ST) solution provides analytic solutions for the size of a single SN bubble and the duration of the hot phase \citep{Sedov1959, Taylor1950}. 
According to this solution, the hot SN bubble size, $R_{\rm SN}$, is 
\begin{equation}
R_{\rm SN} = 23.4 ~E_{51}^{0.29} n_{0}^{-0.42} ~ {\rm [pc]},
\end{equation}
where $E_{51}$ is the SN energy normalised by $10^{51}$\,erg, and $n_0$ is the ambient hydrogen density. 
In this equation, we assume that the ST (adiabatic) phase lasts until $33\%$ of the input SN energy is lost by radiative cooling \citep{Draine2011}.  
The hot (adiabatic) phase duration time, $t_{\rm ST}$, is 
\begin{equation}
t_{\rm ST} = 4.93 \times 10^4 ~E_{51}^{0.22} n_{0}^{-0.55} ~ {\rm [yr]}. 
\end{equation}

Unfortunately, these small scales in both space and time are unresolved in current cosmological hydrodynamic simulations, therefore we need to implement a subgrid model for SN feedback. 
When the above feedback scales are unresolved, just dumping thermal energy into the ambient ISM is insufficient in cosmological simulations, because the energy is quickly radiated away via radiative cooling (i.e., the over-cooling problem), and as a result, SN feedback becomes very inefficient. 
In order to solve this problem, people have often turned off the radiative cooling and/or hydrodynamic interaction for wind particles for a certain period of time, however, this treatment is unphysical as we described in Section~\ref{sec:intro}.

\citet{Chevalier1974} and \citet{McKee1977} performed numerical simulations of SN bubble evolution under a more realistic situation than the simple ST solution.  
They included not only the cooling via infra-red, UV and X-ray radiation, but also the effects of magnetic field on the remnant, and found that the hot phase continues longer than the ST phase. 
In their result, the duration time of hot phase ($t_{\rm hot}$) and the hot SN bubble radius ($R_{\rm bub}$) are obtained as
\begin{eqnarray}
t_{\rm hot} &=& 8.3 \times 10^5 ~E_{51}^{0.31} n_{0}^{0.27} \tilde{P}_{04}^{-0.64} ~ {\rm [yr]}, \label{timehot} \\
R_{\rm bub} &=& 54.95 ~E_{51}^{0.32} n_{0}^{-0.16} \tilde{P}_{04}^{-0.20} ~ {\rm [pc]}, \label{rbub}
\end{eqnarray}
where $\tilde{P}_{04} \equiv 10^{-4} P_{0} k_{\rm B}^{-1}$, and $P_{0}$ is the ambient gas pressure, and $k_{\rm B}$ is the Boltzmann constant. 
They also estimate the surviving time of the low-density cavity as follows: 
\begin{equation}
t_{\rm sur} = 7.1 \times 10^6 E_{51}^{0.32} n_{0}^{0.34} \tilde{P}_{04}^{-0.70} ~ {\rm [yr]}, 
\label{timesur}
\end{equation}
which is roughly ten times longer than $t_{hot}$. 
In \citet{Stinson2006, Stinson2013}, they adopt Eqn.\,(\ref{timesur}) as the cooling shut-off time. 

In our Osaka feedback model, motivated by \citet{Stinson2006}, we adopt Eqns.\,(\ref{timehot}) \& (\ref{rbub}), and use the density and pressure of SPH particles in the vicinity of star particles for $n_0$ and $P_0$. 
We regard $R_{\rm bub}$ as the SN bubble radius, and only the gas particles in this radius are affected by the SN feedback and receive the energy and ejecta from SNe. 
We turn off the cooling only for $\Delta t < t_{\rm hot}$ rather than $t_{\rm sur}$ which is adopted by \citet{Stinson2006}, and always keep the hydrodynamic interaction on even during this phase. 
We note that in the worst case of this model, if the duration time of the phase of one SN explosion is larger than the interval time between SN events from the same star particle, the adiabatic phase continues at least 40 Myrs which corresponds to the lifetime of the minimum SN-II progenitor star. 
As described in \citet{Agertz2013}, such cooling shut-off model might maximize the effect of SN feedback.

In our previous work \citep{Todoroki2014, Kim2016, Aoyama2017}, 
we assumed that star particles instantaneously explode as SN-II after a delay time of $4$\,Myr . 
However, in the present study, we adopt a model with multiple SN explosions accounting for stellar lifetimes as we described above. 
Using the {\small CELib} library, we calculate the SN rate of a star particle using its age and metallicity. 
The duration of SN explosions, $t_{\rm SN}$, is given by 
\begin{equation}
t_{\rm SN} = t_{\rm max}^{\rm SN} - t_{\rm min}^{\rm SN},
\end{equation}
where $t_{\rm max}^{\rm SN}$ and $t_{\rm min}^{\rm SN}$ are the beginning and the end of SN explosions for a stellar population represented by a star particle. 
We take following time-scales from Fig.\,\ref{fig:celib} and Fig.\,3 of \citet{Saitoh2017}: 
$(t_{\rm min}^{\rm SNII}, t_{\rm max}^{\rm SNII}) = (2.2 \times 10^6 - 4.6 \times 10^6, 1.4 \times 10^7 - 4.6 \times 10^7)$~[yr] for SN-II, 
and $(t_{\rm min}^{\rm SNIa}, t_{\rm max}^{\rm SNIa})  = (4 \times 10^7, 1.4 \times 10^{10})$~[yr] for SN-Ia which depend on stellar metallicity. 

We consider the stellar lifetime and metallicity-dependent metal yield and mass loss from SN-II, SN-Ia and AGB stars based on {\small CELib}.  
We also calculate the time-dependent SN rate with this library. 
We adopt the delay-time distribution function of SN-Ia with a power law of $t^{-1}$ for the SN-Ia event rate \citep[e.g.,][]{Totani2008, Maoz2012}. 
Single SN explosion energy is set to $1.0 \times 10^{51}$\,erg. 

In order to deposit SN feedback energy and metal yield gradually rather than instantaneously, we allow the SN feedback to take place with following logarithmic time interval, so that the deposited SN energy of each event approximately equals 
\begin{equation}
\log_{10} dt = \frac{\log_{10}{(t_{\rm max}^{\rm SN})} - \log_{10}{(t_{\rm min}^{\rm SN})}}{n_{\rm fb}}, 
\label{eq:snfbdt}
\end{equation}
where $n_{\rm fb}$ is the number of times that the SN energy is injected during $t_{\rm SN}$. 
The energy and metals deposited during this time interval is also calculated by the CELib library. 
In order to discuss the metal evolution in galaxies in more detail, a larger value of $n_{\rm fb}$ is more preferable.  
We adopt $n_{\rm fb} = 8$ as our fiducial value. 
According to \citet{Kim2015}, if the integrated SN energy is the same, the final momentum integrated from multiple SNe is slightly smaller than that of a single SN, although the value strongly depends on the gas density around SNe sites and the time interval between SNe. 
They also argued that multiple SNe from massive star cluster form a superbubble, which strongly affects the properties of the host galaxy in the case of more realistic situation. 

We discuss the dependency on $n_{\rm fb}$ in later section in more detail, but there are two limits on the value of $n_{\rm fb}$:
1) the released energy for one SN event should be at least $10^{51}$\,erg, and 2) the number of particles in the hot bubble should be more than two. 
If $n_{\rm fb}$ becomes very large, then the number of gas particles in the hot bubble decreases to one or zero eventually, in which case the SN energy is assigned to the nearest gas particle. 
As a result, the wind velocity and heat-up temperature directly correlate with $n_{\rm fb}$ and are uniquely determined.  
Therefore, there is a maximum value for a possible $n_{\rm fb}$.  

We assume that a fraction $\epsilon_{\rm K}^{\rm SN}$ of the SN energy is converted to the kinetic energy of the wind, and the remaining fraction, $\epsilon_{\rm T}^{\rm SN} = 1 - \epsilon_{\rm K}^{\rm SN}$, is deposited as thermal energy. 
This is one of the differences from \citet{Stinson2006} and this explicit momentum kick helps clearing gas out from star-forming regions than in the case of thermal only feedback. 
According to \citet{Durier2012} and \citet{Keller2014}, the difference between thermal form and kinetic form or combination of both form is very small or nothing for the high-resolution simulations in which the SN bubble can be fully resolved. 
However, it is unclear whether the difference between them is still small (or nothing) in the case of the lower resolution. 
We think that it might be still worth studying the selection effects. 
Here we adopt $\epsilon_{\rm K}^{\rm SN} = 0.3$ \citep[e.g.,][]{Chevalier1974, Durier2012}. 
Both energy components are distributed to the gas particles within the SN bubble radius, $R_{\rm bub}$ (see Eqn.~\ref{rbub}). 
Hence, the energy that an $i$-th gas particle receives from nearby SN explosions is computed as
\begin{equation}
\Delta E_{\rm i} = \frac{m_{\rm i} W(r_{\rm i}, R_{\rm bub})}{\sum_{\rm j = 1}^{\rm N} m_{\rm j} W(r_{\rm j}, R_{\rm bub})} E_{\rm SN},  
\end{equation}
where $E_{\rm SN}$ is the SN energy from the star particle of concern, $W$ is the SPH kernel function, 
$r_{\rm i}$ is the distance between $i$-th gas particle and the star particle. 
We use $R_{\rm bub}$ as the smoothing length for SN feedback.  
We also distribute the SN ejecta (gas and metals) in the same manner as the SN energy. 
The effects of SN-II and SN-Ia are computed similarly, except for the time-delay of SN-Ia. 

In \citet{Hopkins2011} and \citet{Kimm2014}, they considered the terminal momentum in the snowplow phase (the momentum-conserving phase). 
On the other hand, here we focus on the outflow velocity in the Sedov-Taylor phase (adiabatic phase) as follows. 
The wind velocity in the ST solution is given by 
\begin{equation}
V_{\rm ST} = 181\, E_{51}^{0.07} n_{0}^{0.14} ~ [{\rm km/s}]. 
\end{equation}
We find that blindly using this formula for the wind velocity leads to the violation of energy conservation. 
This is partly due to lack of numerical resolution, because when the mass of a star particle is large, the associated total SN energy ($E_{51}$) also becomes large proportionally, making the value of $V_{\rm ST}$ unphysical for a collection of SNe. 
To avoid this problem, some researchers use the terminal momentum for the estimation of the wind velocity \citep{Hopkins2011, Agertz2013, Kimm2014}. 
On the other hand, we opt to use a simpler formula based on the energy conservation law for the wind velocity, 

\begin{equation}
V_{\rm wind} = \sqrt{\frac{2 E_{\rm K}^{\rm SN}}{m_{\rm gas}}}, 
\end{equation}
where $E_{\rm K}^{SN}$ is the kinetic SN feedback energy received by a gas particle, and $m_{\rm gas}$ is the mass of a gas particle. 
In the following, we call the kinetic component of our model as `kinetic feedback'. 
The direction of wind particles is randomly chosen which is called as `isotropic winds'. 

When the resolution is not sufficiently high, we note that sometimes there are no gas particles within the SN bubble radius $R_{\rm bub}$. 
In that case, we assign the feedback energy and ejecta to the nearest gas particle from the star particle of concern. 
With this treatment, there is a possibility that a single gas particle receives the entire SN energy and ejecta, which could cause an unphysical, abrupt increase of metallicity for a particular gas particle. 
In order to avoid this problem, we introduce smoothing when estimating the metallicity, instead of just taking particle's metallicity \citep{Okamoto2005}, and this smoothed metallicity is used to estimate the photo-heating and radiative cooling rates. 
We also note that our model exactly conserves the SN energy but the momentum conservation might break down in the case of a few gas particle in $R_{\rm bub}$. 


\subsection{Early Stellar Feedback}
\label{sec:ESFB}

The strong UV radiation and stellar winds emitted by young, massive stars ionize and heat up the ambient gas. 
This mechanism not only suppresses star formation, but it may also smooth out the clumpy gas distribution due to increased thermal pressure. 
This effect is often referred to as the `early stellar feedback' (ESFB). 
Subsequent SN feedback may work more effectively with ESFB, and it implies that ESFB is very important for early evolutionary stages of galaxies \citep[e.g.,][]{Fall2010}. 

However, many previous cosmological simulations have ignored ESFB, and in such simulations, star-forming regions tend to remain dense, cool too much, and form new stars before SN explosions take place. 
As a result, such simulations overproduce stars in the early phase, and often failed to reproduce the results on stellar-to-halo mass ratio (SHMR) from both observations and the abundance matching technique \citep[e.g.,][]{Behroozi13}. 
Therefore, ESFB might offer one of the solutions for the overcooling problem. 

Several recent studies argued that ESFB is necessary to reproduce the observed SHMR \citep[e.g.,][]{Agertz2013, Stinson2013, Okamoto2014}, and there are mainly two ways to implement it in simulations. 
One is to focus on the radiation pressure feedback from young massive stars. 
In this model, the stellar radiation imparts its momentum to the ambient gas and dust, and pushes them out of star-forming regions \citep[e.g.,][]{Hopkins2012b, Wise2012, Agertz2013, Okamoto2014}. 
Another approach is a simple thermal feedback, where the ambient gas receives thermal energy as a result of ESFB. 
In this approach, the ambient gas around young stars is ionized and heated to a few $\times~ 10^4 $\,K  \citep[e.g.,][]{Hasegawa2013, Pawlik2017}. 
\citet{Stinson2013} argued that 10 per cent of total stellar radiation from massive young stars is necessary for ESFB in order to reproduce the observed SHMR. 
At the same time, we also note that some researchers argued that additional energy such as ESFB is not necessary to reproduce the observations \citep[e.g.,][]{Vecchia2012,Keller2014,Keller2015,Keller2016}. 
It is still much debated whether only SNII feedback is enough to reproduce the observations or not. 

In the present work, we implement a model for ESFB similarly to \citet{Stinson2013}, and examine its effectiveness in enhancing the impact of SN feedback. 
In this implementation, we do not artificially turn off the radiative cooling for a certain period of time as in  \citet{Stinson2013}. 
Since the duration time of ESFB is very short and unresolved in low-resolution simulations, almost all ESFB energy could be deposited all at once, in which case gas heating by ESFB would work effectively. 
If the simulation time-step is too large, ESFB events may never occur. 
In order to avoid these situations, we set the minimum time-step to be small enough to resolve the ESFB time-scale as well as SN events. 
We deposit the thermal energy at a constant rate in each time-step, $dt_{\rm ESFB}$, until $t_{\rm min}^{\rm SNII}$ is reached. 
Then, the fractional thermal energy, $\Delta E_{\rm ESFB}$, deposited by a star particle in $dt_{\rm ESFB}$ is 
\begin{equation}
\Delta E_{\rm ESFB} = \epsilon_{\rm ESFB} \frac{E_{\rm bol}}{n_{\rm esfb}} m_\ast, 
\label{eq:esfbfbdt}
\end{equation}
where $\epsilon_{\rm ESFB}$ is the ESFB efficiency, $E_{\rm bol}$ is the bolometric luminosity emitted by massive stars normalized by stellar mass, $n_{\rm esfb}$ is the number of ESFB energy deposition of a star particle, and $m_\ast$ is the star particle mass.  
Here, $dt_{\rm ESFB} = t_{\rm min}^{\rm SNII} / n_{\rm fb}$, and we adopt $\epsilon_{\rm ESFB} = 0.1$, $E_{\rm bol} = 2.0 \times 10^{50}~ {\rm erg / \Msun}$ following \citet{Stinson2013}, and $n_{\rm esfb} = 8$ as our fiducial model. 
Note that it is possible to heat the gas particles to above $\sim 10^5$\,K with this treatment. 
This might lead to unphysical results, because the photoheating that we consider as our ESFB model should not heat the gas to above a few $\times 10^4~{\rm K}$ \citep[e.g., ][]{Hasegawa2013, Pawlik2017}. 
In order to prevent this overheating by the ESFB model, we put a cap on the maximum temperature $T_{\rm cap}=20,000$\,K. 
This treatment for the ESFB is different from the original Stinson's model \citep{Stinson2013} in which there is no upper limit for the heated temperature.


\section{Results \& Discussions}
\label{sec:results}
To explore the effects of each feedback process, at first, we compare our isolated disk galaxy simulations with available observational data such as the Kennicutt-Schmidt (KS) law and radial profile of some physical quantities. 
Then, we explore our model properties using the cosmological simulations. 
Based on the feedback models that we described above, we run a series of simulations as summarized in Table~\ref{tbl2}. 

Our fiducial run (`K30T70') is the one that includes all feedback processes, which are the early stellar feedback (ESFB), type-II supernova (SN-II), and type-Ia supernova (SN-Ia). 
K0T100 and K100T0 runs are same as the fiducial run but only thermal feedback ($\epsilon_{\rm K}^{\rm SN} = 0, \epsilon_{\rm K}^{\rm SN} = 1$) which is very similar to \citet{Stinson2006} and only kinetic feedback ($\epsilon_{\rm K}^{\rm SN} = 1, \epsilon_{\rm K}^{\rm SN} = 0$), respectively. 
The Cool-on run is the same as our fiducial run but has always cooling on.  
Additional five runs are those with ESFB and SN-II (`ESFB-SNII'), ESFB-only, SNII-only, SNIa-only, and without feedback at all (`No-FB'). 

In all of these runs (including the No-FB run), we always consider the metal yield from SN-II, SN-Ia and asymptotic giant branch (AGB) stars calculated by the {\small CELib} library. 
The ESFB run includes only thermal feedback by construction, and all other runs include the kinetic feedback.
For comparison, we also simulate the stochastic thermal feedback model based on \citet[][`Sto-TH' model]{Vecchia2012} with heat-up temperature $T=10^{7.5}~{\rm K}$ and the constant velocity wind model based on \citet[][`Sto-CW' model]{Vecchia2008} with the wind velocity $V_{\rm wind}=600$\,km\,s$^{-1}$ and mass loading factor $\eta = 2$ which is similar to the SH03 model. 
In these stochastic models, the SN energy is released in  multiple time-steps similarly to our Osaka model and the total available energy for the SN feedback is exactly the same as our fiducial model. 
Note that the integrated SN energy from a star particle in our model is about three times lower than that of the model used in \citet{Vecchia2012}. 
Therefore our implementations of stochastic models are slightly different from the original model by \citet{Vecchia2008, Vecchia2012}.  

\begin{table}
\begin{center}
\begin{tabular}{lc}\hline
Run name & Notes \\ \hline
K30T70  & fiducial run with all feedback models \\
No-FB & without any feedback \\
K100T0  & fiducial run but 100\% kinetic feedback \\
K0T100  & fiducial run but 100\% thermal feedback \\
Cool-on  & fiducial run but always cooling on \\
ESFB-only & only ESFB model \\
SNII-only & only SN-II feedback \\
SNIa-only & only SN-Ia feedback \\
ESFB-SNII & with ESFB \& SN-II feedback\\
Sto-TH & stochastic thermal feedback model $^a$ \\
Sto-CW & constant wind model $^b$ \\
\hline
\end{tabular}
\caption{List of simulations with different feedback models compared in this paper. 
($a$): The Sto-TH run is based on the stochastic thermal feedback model of \citet{Vecchia2012} with heat-up temperature $T=10^{7.5}~{\rm K}$.
($b$): The Sto-CW run is based on the constant velocity wind model of \citet{Vecchia2008} with $V_{\rm wind}=600$\,km\,s$^{-1}$ and $\eta = 2$. }
\label{tbl2}
\end{center}
\end{table}

\subsection{Distribution of gas density and temperature}
\begin{figure*}
\centering
\begin{minipage}[l]{1.0\textwidth}
\centering
\includegraphics[width = 170mm]{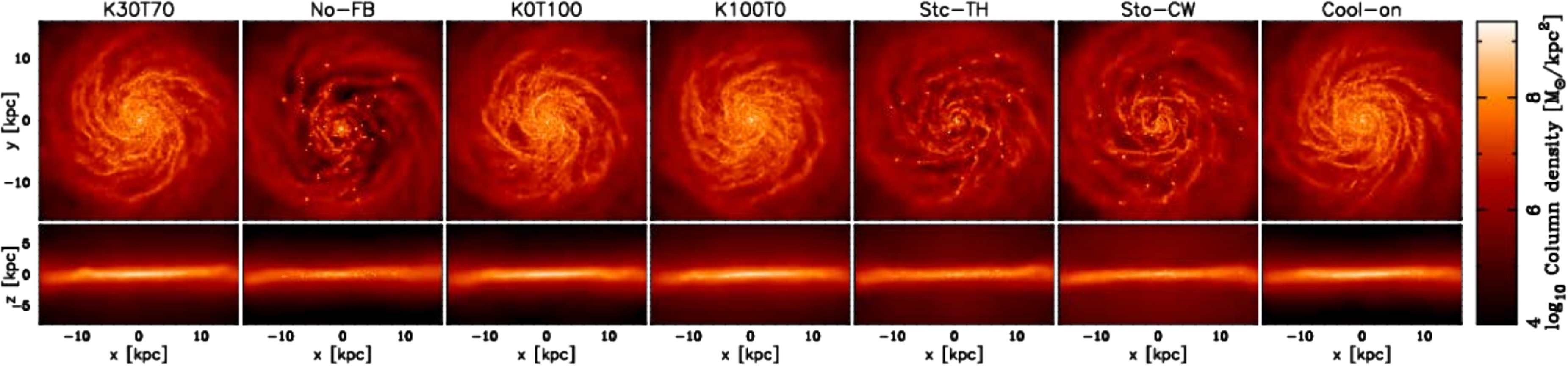}
\caption{Projected gas density (i.e., column density) of simulated galaxies at $t = 1$\,Gyr for different runs (from left to right:  T30K70, No-FB, K0T100, K100T0, Sto-TH, Sto-CW, Cool-on runs). 
Top and bottom panels show the face-on and edge-on views of the simulated galaxy. } 
\label{fig:gpmm12}
\end{minipage}
\vfill{}
\begin{minipage}[c]{1.0\textwidth}
\centering
\includegraphics[width = 170mm]{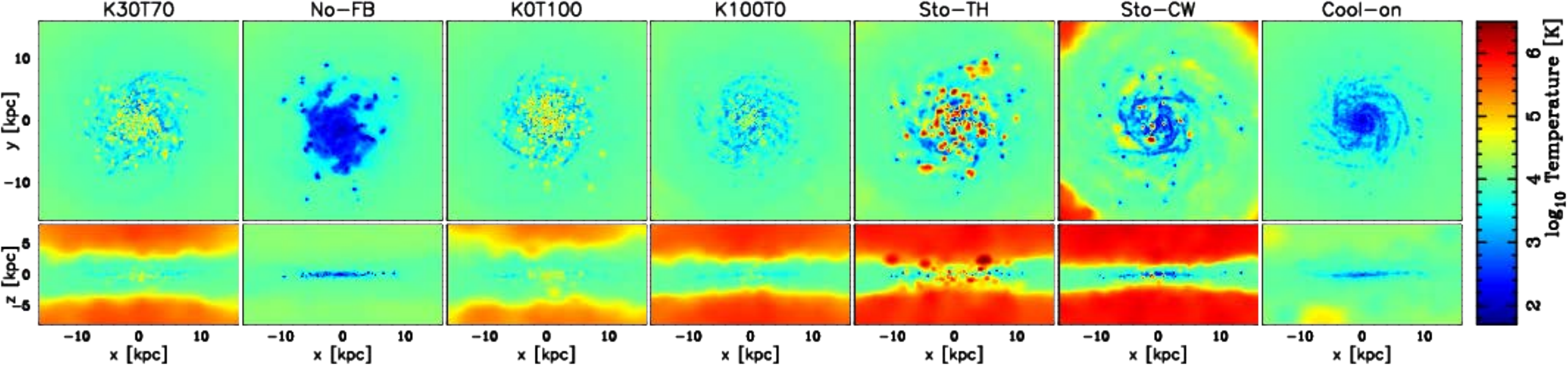}
\caption{Same as Fig.~\ref{fig:gpmm12} but for temperature, weighted by density squared. }
\label{fig:tpmm12}
\end{minipage}
\vfill{}
\begin{minipage}[c]{1.0\textwidth}
\centering
\includegraphics[width = 170mm]{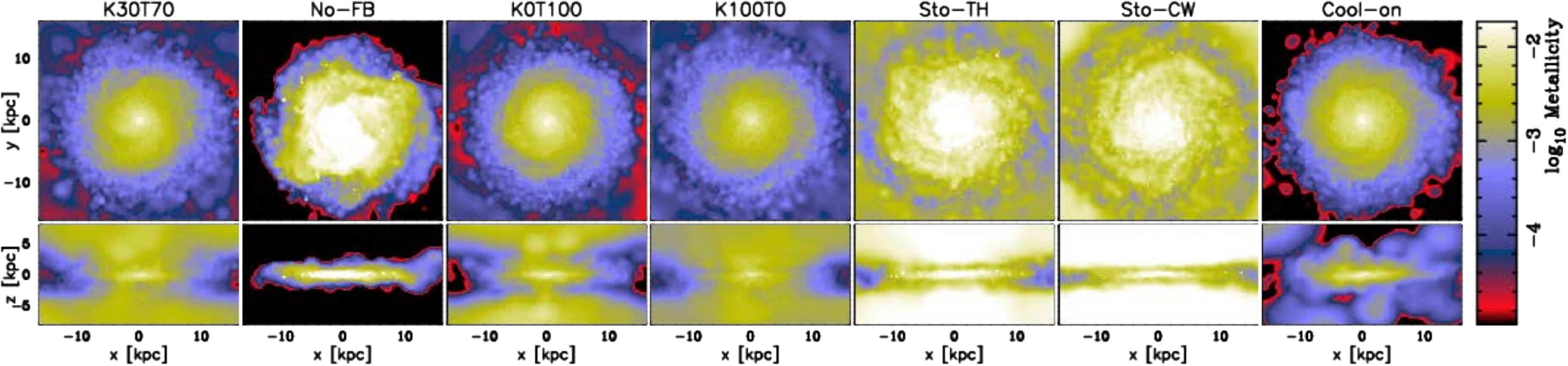}
\caption{Same as Fig.~\ref{fig:gpmm12} but for metallicity. }
\label{fig:mpmm12}
\end{minipage}
\vfill
\begin{minipage}[c]{1.0\textwidth}
\centering
\includegraphics[width = 170mm]{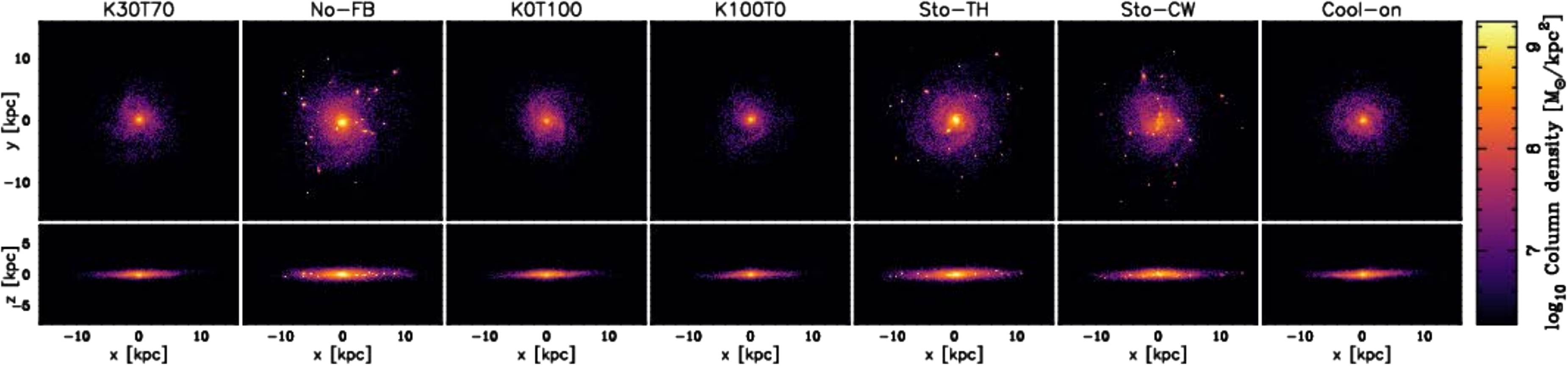}
\caption{Same as Fig.~\ref{fig:gpmm12} but for projected stellar mass density. }
\label{fig:spmm12}
\end{minipage}
\end{figure*}

\begin{figure*}
\centering
\begin{minipage}[l]{1.0\textwidth}
\centering
\includegraphics[width = 170mm]{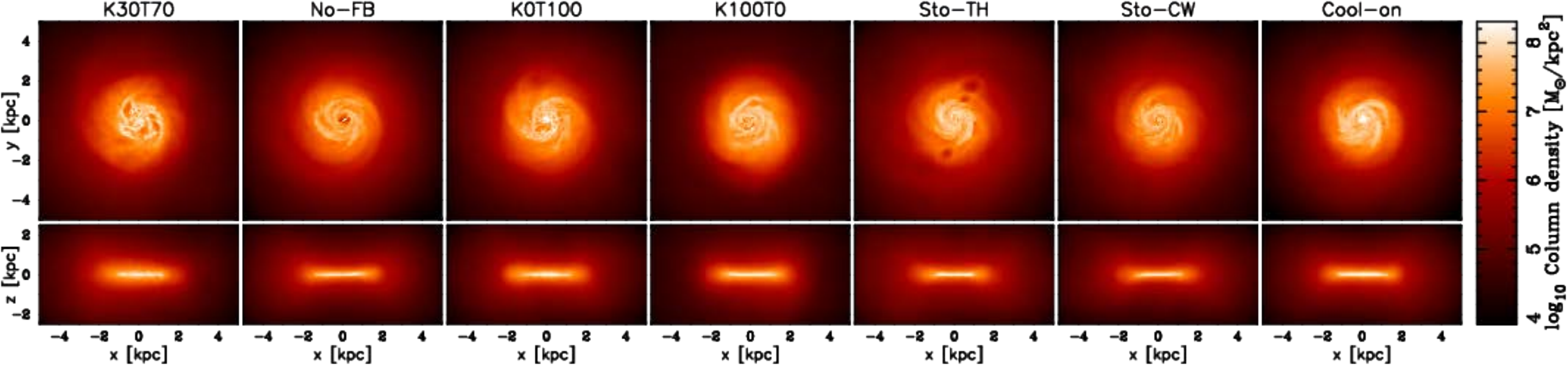}
\caption{Same as Fig.~\ref{fig:gpmm12} but for M10 galaxy. } 
\label{fig:gpmm10}
\end{minipage}
\vfill{}
\begin{minipage}[c]{1.0\textwidth}
\centering
\includegraphics[width = 170mm]{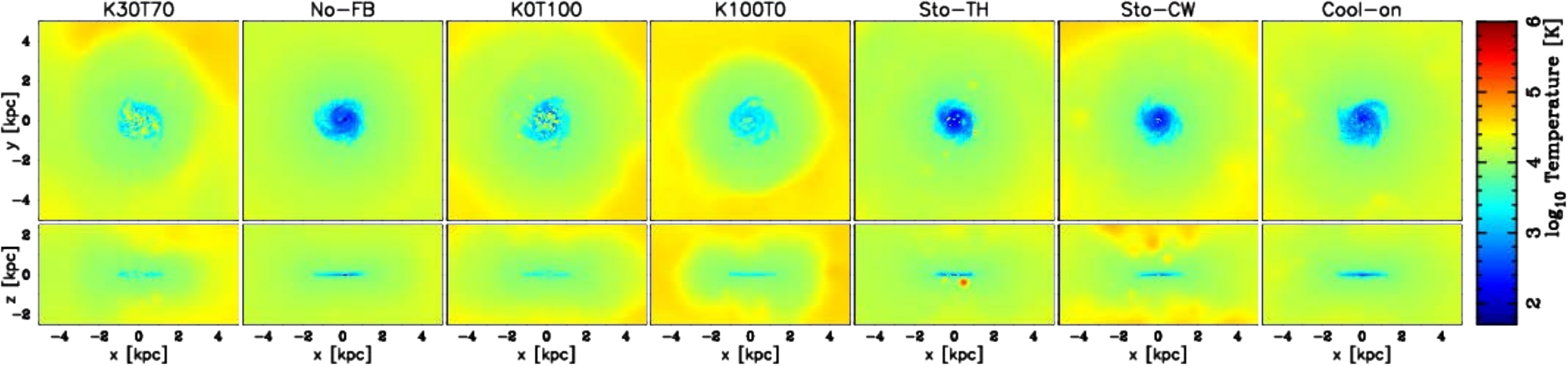}
\caption{Same as Fig.~\ref{fig:tpmm12} but for M10 galaxy. }
\label{fig:tpmm10}
\end{minipage}
\vfill{}
\begin{minipage}[c]{1.0\textwidth}
\centering
\includegraphics[width = 170mm]{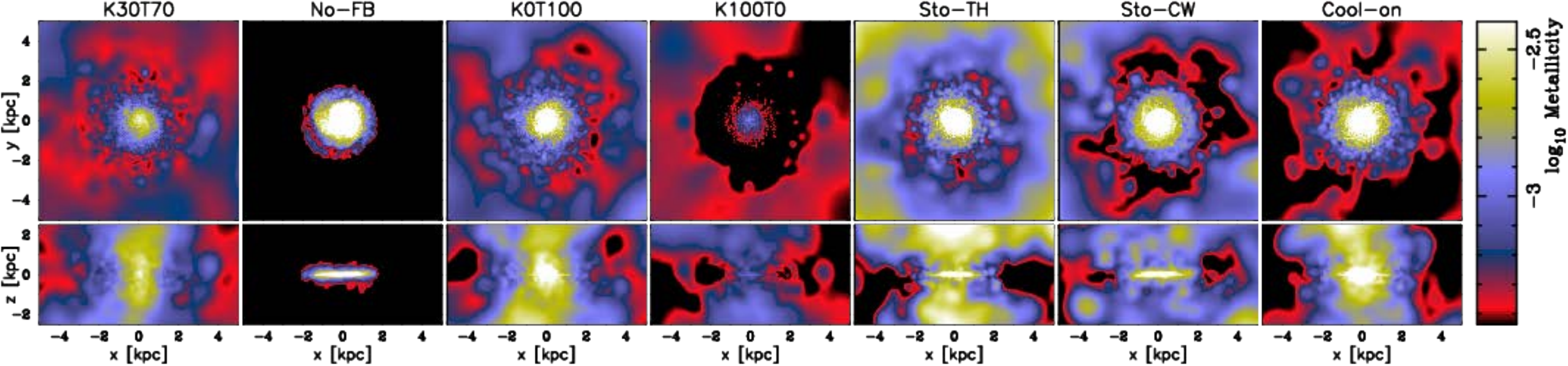}
\caption{Same as Fig.~\ref{fig:mpmm12} but for M10 galaxy. }
\label{fig:mpmm10}
\end{minipage}
\vfill
\begin{minipage}[c]{1.0\textwidth}
\centering
\includegraphics[width = 170mm]{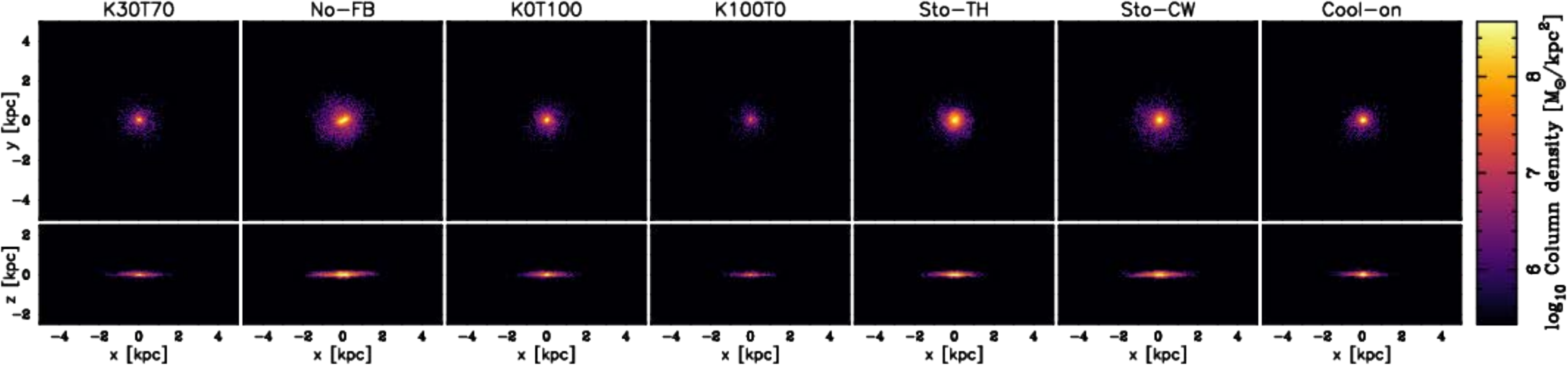}
\caption{Same as Fig.~\ref{fig:spmm12} but for M10 galaxy. }
\label{fig:spmm10}
\end{minipage}
\end{figure*}

First, we explore the effects of feedback on the distribution of gas, stars and metals in this section. 
Figures~\ref{fig:gpmm12} and \ref{fig:gpmm10} show the projected gas density of each run at $t = 1$\,Gyr for the M12 and M10 runs. 
For the M12 galaxy, our fiducial run (K30T70), K0T100 and K100T0 with all feedback processes (K30T70) have well-developed, nicely organized, gaseous spiral arms with little gas clumps even though the energy fraction between kinetic and thermal are different. 
Interestingly, the gas structure in our model without cooling shut-off is very similar to the models with cooling shut-off. 
This implies that the cooling shut-off might not be so important for the gas structure for massive galaxy. 

In the No-FB run, the gas is able to cool rapidly without being heated by the feedback, and therefore form these dense gaseous clumps. 
The regions in-between the spiral arms appear as dark, low column-density regions, and the gas is rapidly consumed into stars in high-density gaseous clumps. 
Similar clumpy structures can be seen in Sto-CW and Sto-TH models as well as No-FB run. 
This is because all of star formation sites are not always suppressed by the SN feedback in the stochastic models and there is a time-delay before the SN feedback causes any significant effects to suppress star formation.

On the the hand, for the M10 galaxy, we find that the situation dramatically changes. 
Unlike the massive galaxy cases, the gas structure of the Osaka models (K30T70, K0T100, K100T0) shows very irregular shapes and spiral arms are destroyed by the feedback. 
Similar structures can be seen in the Sto-CW and Sto-TH runs. 
This is because the feedback effectively works and disturbs the gas structure due to shallower potential. 
However, in the Cool-on model, smooth spiral arms can be seen and the structure seems to be not affected by the SN feedback. 
This may imply that for the small galaxies, the thermal feedback is more efficient than massive galaxies case. 
We find the irregular bubble structures in K30T70 and K0T100 runs which consists of many SN bubbles. 
We note that these larger bubbles might be a bit unphysical because our model is based on single SN explosion. 
As pointed out by \citet{Keller2014}, such model cannot treat multiple SN explosion accurately.   
We see large spherical hot bubbles due to the strong thermal feedback, and these are due to single SN explosion rather than multiple explosions. 

Figures~\ref{fig:tpmm12} and \ref{fig:tpmm10} show the gas temperature projection weighted by density--squared. 
From the comparison of gas and temperature distribution, the spatial distribution of cool gas apparently reflect the high density gas. 
We see that the No-FB and Cool-on runs do not have the red, hot components, and most of the gas in the disc has temperature $T< 10^4$\,K. 
The models including the thermal feedback (K30T70, K0T100 and Sto-TH) show large hot bubbles in orange and red colour. 
On the other hand, many hot bubbles can also be recognized in kinetic-only feedback models. 
This is because we always turn on the hydrodynamic interaction between the wind particles and the ISM in both models and this causes the shock heating and hot bubbles.  

The Sto-TH and Sto-CW runs show the highest ISM temperatures, and the red, large hot bubbles can be seen at the surface of the galactic disc. 
The circumgalactic (CGM) and intergalactic medium (IGM) heating for these two models also work well even though the suppression for the star formation activity might be weaker than our Osaka model. 
On the other hand, the CGM gas is also heated up to $T\gtrsim 10^5$\,K, which might be a little excessive, as was pointed out earlier \citep[e.g.,][]{Oppenheimer2006, Choi2011}. 

We find that the bubble sizes for the M10 galaxy runs become smaller than that of M12 runs, because M10 runs can resolve denser gas and the bubble size automatically shrinks in accordance with Eqn.~(\ref{rbub}). 
Moreover, the temperature around the galaxy for M10 runs looks lower than that for M12 runs. 
This is because almost all wind particles cannot stay near the galaxy and they escape into the IGM due to shallower potential of the M10 galaxy. 
These gas particles can cool adiabatically (see also Fig. \ref{fig:pdm10}), which is quite different from the M12 runs. 

\subsection{Metal enrichment}
The distribution of metals can reveal the impact of feedback more clearly. 
Galactic winds from star-forming galaxies carry metals into the CGM and IGM, and the enrichment pattern as a function of distance from galaxies could tell us about mass-loading rate and kinetic energy of SN feedback. 
Figures~\ref{fig:mpmm12} and \ref{fig:mpmm10} show the projected gas metallicity for M12 and M10 runs. 
The runs with momentum kick can enrich not only the ISM but also the CGM and possibly the IGM. 
From the comparison between K0T100 and K100T0 runs, we find that the momentum kick model can easily enrich the CGM and IGM than the case of thermal-only feedback.
The thermal-only feedback models can also do metal enrichment with the help of thermal pressure. 
Despite the fact that the difference between K30T70 and Cool-on models is only whether cooling is enabled or not for the SN feedback, this makes a significant difference in the metal pollution. 
This suggests that though the kinetic feedback is more efficient to enrich the CGM and IGM than the thermal feedback, the metal enrichment process by the thermal pressure is still non-negligible. 
We find that the metallicity around the galaxy for M10 runs is smaller than that for M12 runs. 
This might be the same reason as to why the temperature around the galaxy is relatively low. 

\subsection{Stellar distribution}
Figures~\ref{fig:spmm12} and \ref{fig:spmm10} show the projected stellar density for each run. 
Similarly to the projected gas distribution for all Osaka model, even the Cool-on model indicates a smooth distribution of stars with very few stellar clumps. 
In the AGORA code comparison project \citep{Kim2016}, almost all runs showed some clumpy structures similar to that of our No-FB, Sto-CW and Sto-TH runs. 
This suggests that our SN feedback model is able to disperse dense gas and suppress star formation in dense clumps. 
Many previous numerical simulations suppressed the overcooling and star formation by removing the gas from star-forming regions by strong winds without the hydrodynamic interaction. 
However, in our Osaka feedback model, we do not shut off the hydrodynamic interaction, and it can still suppress the star formation in dense gaseous clumps, although the mass outflow is weaker than that of the Sto-CW and Sto-TH run. 
Furthermore, in our Osaka model except for the Cool-on model, the compactness of the stellar distribution becomes prominent as increasing the fraction of kinetic energy. 
This implies that the kinetic feedback suppresses the star formation activities more efficiently and makes compact stellar distribution than the thermal feedback. 

\subsection{Star Formation History}
\begin{figure}
\includegraphics[width = 83mm]{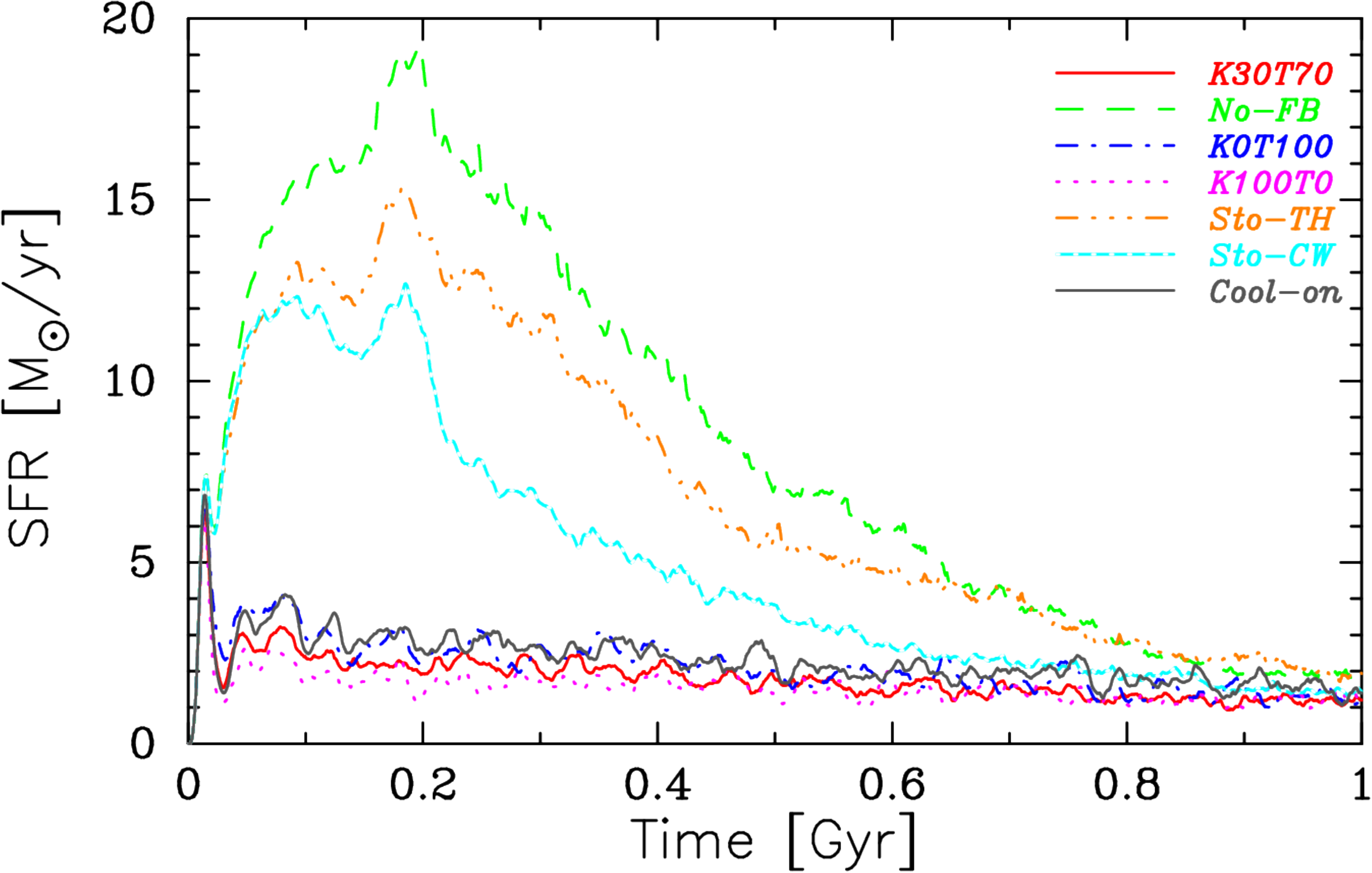}
\caption{Star formation histories of different runs for M12 runs as shown in the legend. }
\label{fig:sfhm12}
\end{figure}

\begin{figure}
\includegraphics[width = 83mm]{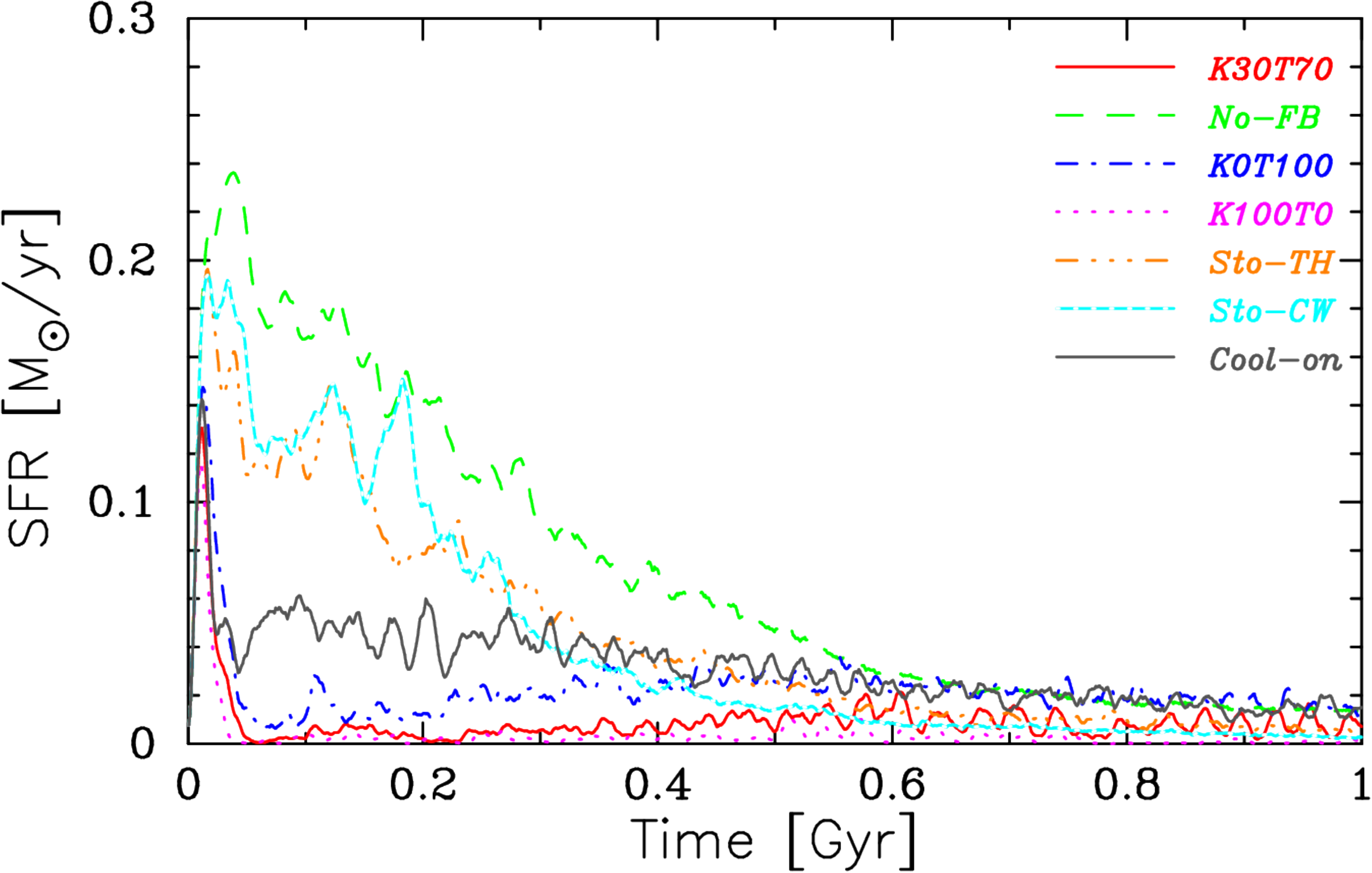}
\caption{Same as Fig.~\ref{fig:sfhm12} but for M10 galaxy. }
\label{fig:sfhm10}
\end{figure}

SFR is one of the key quantities in understanding galaxy evolution, and SF history (SFH) serves as a quantitative measure of the impact of star formation and feedback. 
Figures \ref{fig:sfhm12} and \ref{fig:sfhm10} show the star formation history for M12 and M10 runs. 
Interestingly, all Osaka models strongly suppress the star formation even when the cooling during the SN feedback is always turned on. 
On the other hand, the suppression for the stochastic models is weaker than the Osaka model. 
There might be two reasons. 
The first one is the star formation threshold density is higher than that of the original papers \citep{Vecchia2008, Vecchia2012} and this results in a weaker effect for the SN feedback. 
The second is the total available feedback energy from one star particle is a few times smaller than the original papers \citep{Vecchia2008, Vecchia2012}. 
The horizontal line in the left panel of Fig.~\ref{fig:celib} shows their available energy, which is higher than that of Osaka model. 
Actually, if we increase available SN energy to the same level of what is used in \citet{Vecchia2012}, the stochastic thermal feedback model becomes more efficient. 


\subsection{Kennicutt-Schmidt (KS) Relation}
\begin{figure}
\includegraphics[width = 83mm]{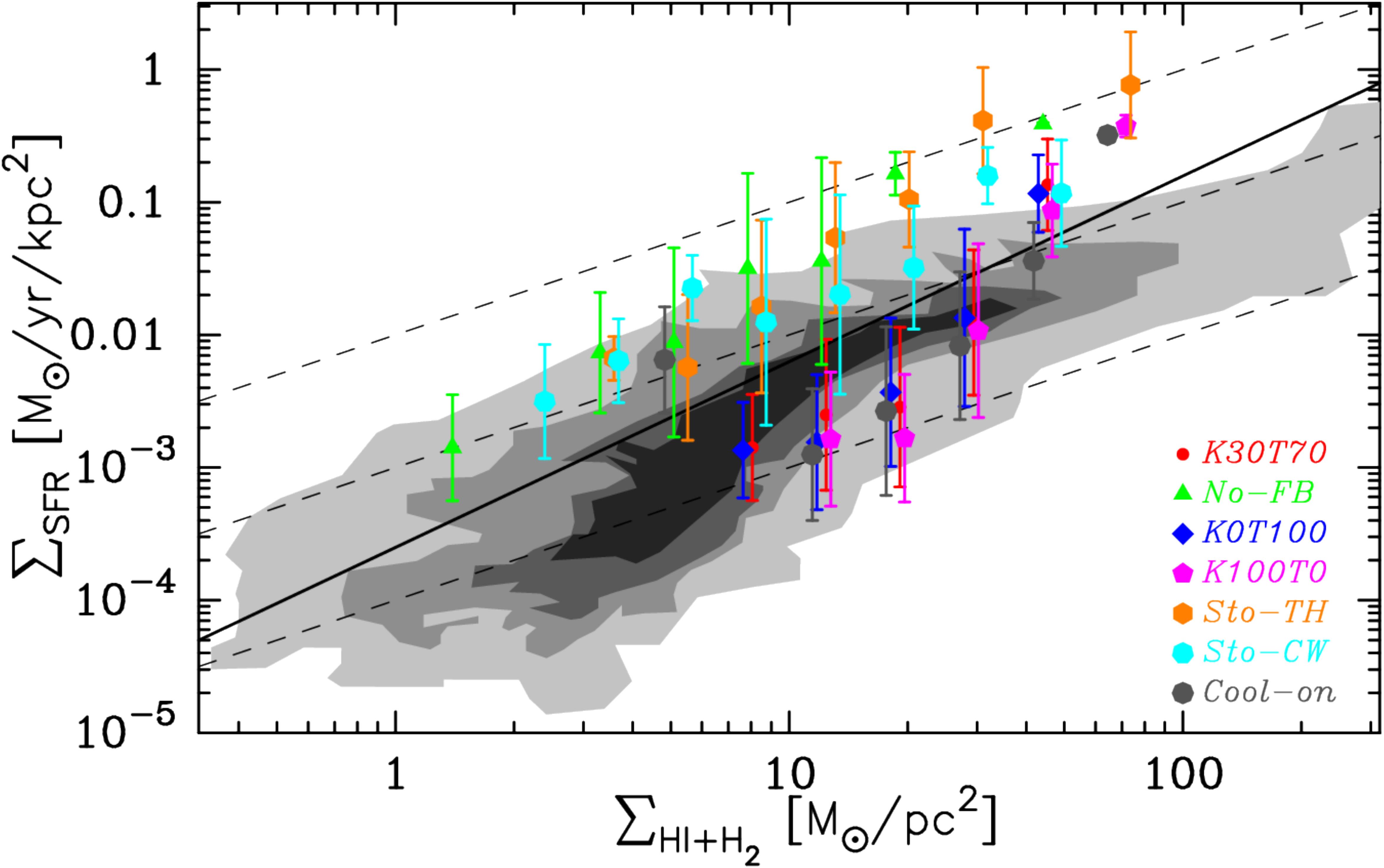}
\caption{The Kennicutt-Schmidt relation of each model averaged in $750 \times 750$\,pc patches at $t = 1$ Gyr for M12 galaxy.}
\label{fig:kslawm12}
\end{figure}

\begin{figure}
\includegraphics[width = 83mm]{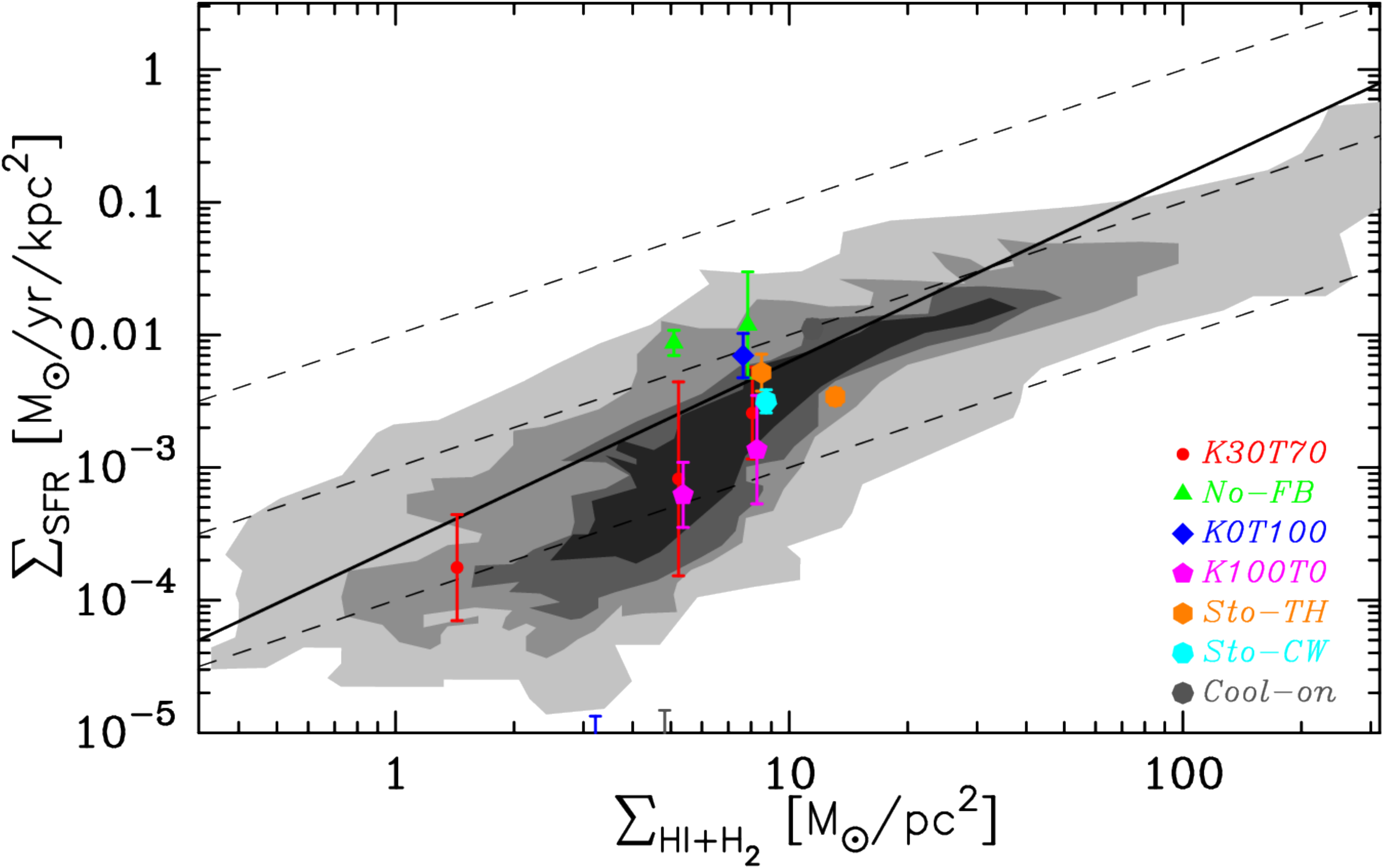}
\caption{Same as Fig.~\ref{fig:kslawm12} but for M10 galaxy.}
\label{fig:kslawm10}
\end{figure}

Observationally, it is well known that the surface density of SFR ($\sum_{\rm SFR}$) strongly correlates with that of the gas \citep[e.g.,][]{Kennicutt1998, Wong2002, Bigiel2008, Leroy2008, Kennicutt2012}, which is often referred to as the Kennicutt-Schmidt (KS) relation. 
The KS relation can be used to calibrate the feedback models in simulations or check their validity. 
Figures~\ref{fig:kslawm12} and \ref{fig:kslawm10} are the KS relation for M12 and M10 runs. 
For M12 runs, our Osaka models reproduce the observational KS relation within error-bars even though the values of our models are somewhat lower than observations at $\Sigma_{\rm HI+H_2} \sim 10-20$\,M$_\odot$ pc$^{-2}$. 
On the other hand, the value of $\Sigma_{\rm SFR}$ in the stochastic models and the No-FB tend to be higher than observations. 

Note that we adjust the value of the star formation efficiency $c_*$ to match the observation, but we never do such adjustment for the stochastic models and we use the same treatment as in the Osaka model. 
In \citet{Vecchia2008} and \citet{Vecchia2012}, the KS law is automatically reproduced without any parameter tuning because their implementation for the star formation is based on the KS law itself. 
Thus, we note that it might be an unfair comparison. 
However, it is clear that the stochastic SN feedback is weaker for the massive galaxy. 
We find, in any case, the stochastic models sensitively depend on the IMF, the star formation threshold density and the feedback yield models. 
In the case of M10 runs, all models are consistent with the observation and we cannot see apparent difference in all models. 


\begin{figure*}
\includegraphics[width = 170mm]{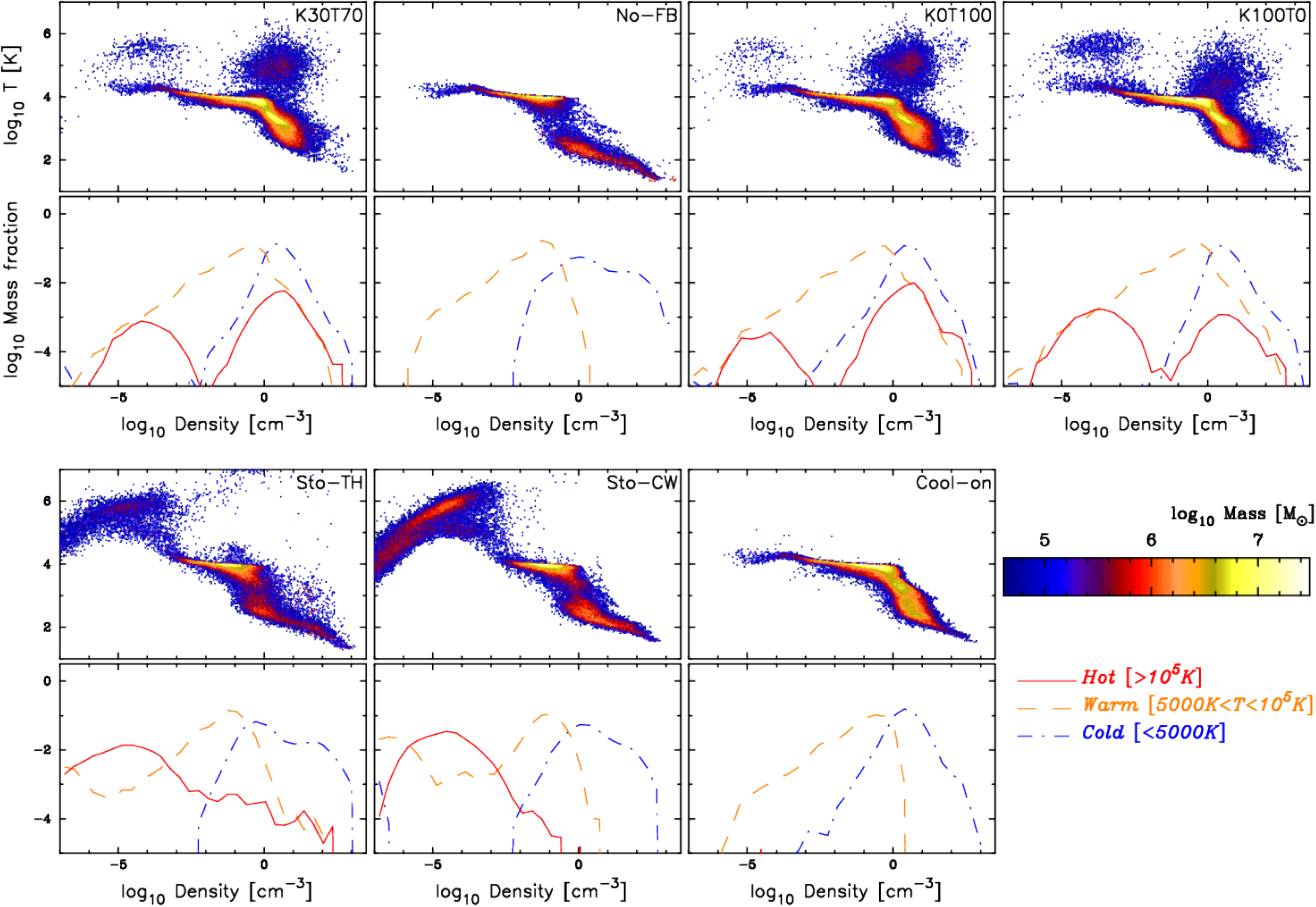}
\caption{For each model, we show the density--temperature phase diagram in the top panel, and the mass-weighted gas density distribution function in the bottom panel. 
The gas is divided into three phases of temperature: hot ($T> 10^{5}~{\rm K}$; red solid line), warm ($5000~{\rm K} < T < 10^5~{\rm K}$; blue dot-dashed) and cold $(< 5000~{\rm K}$; green dashed).}
\label{fig:pdm12}
\end{figure*}
\begin{figure*}
\includegraphics[width = 170mm]{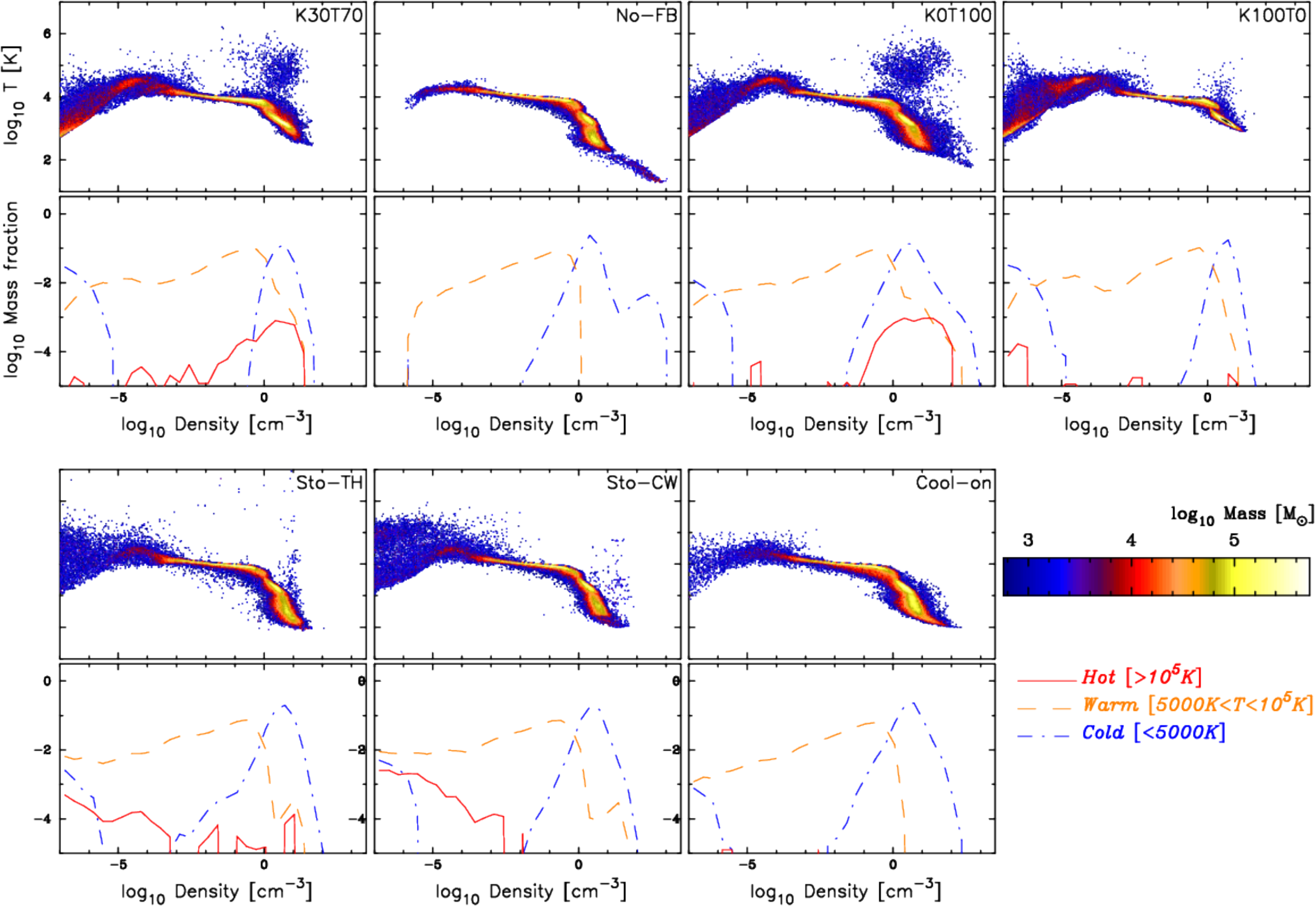}
\caption{Same as Fig.~\ref{fig:pdm12} but for M10 galaxy.}
\label{fig:pdm10}
\end{figure*}

\subsection{Density--Temperature Phase Diagrams}
In Figs.~\ref{fig:pdm12} and \ref{fig:pdm10}, we present the density--temperature phase diagram (upper panels) and the mass-weighted distribution functions of gas density (bottom panel) at t = 1 Gyr for M12 and M10 runs. 
For the M12 runs, except for the Cool-on model, all Osaka models have high-density ($n_{\rm H}= 1-10~{\rm cc^{-1}}$) and warm gas ($< 10^6~{\rm K}$) that also can be seen in \citet{Stinson2006}. 
Interestingly, the Cool-on model has a similar distribution to No-FB model. 
This means the cooling has a strong impact on the physical state of gas particles. 
Actually, in our feedback implementation, the typical heat-up temperature $\Delta T _{\rm i}$ for a gas particle is given by the following equation: 
\begin{eqnarray}
\Delta T _{\rm i} &=& \frac{2}{3} \frac{\mu m_{\rm H}}{k_{\rm B}} \frac{n_{\rm SNII} E_{\rm SNII}^{51} m_{\rm *}}{\sum_{< r_{\rm bub}} m_{\rm gas}} \nonumber \\
                  &=& 1.33 \times 10^5 \left( \frac{\mu}{0.6} \right) \left( \frac{n_{\rm SNII}}{7.0 \times 10^{-3} {\rm M_{\odot}^{-1}}} \right) \nonumber \\
                  && \times \left( \frac{n_{\rm ngb}}{8} \right)^{-1} \left( \frac{n_{\rm fb}}{8} \right)^{-1} \left( \frac{n_{\rm spawn}}{2} \right)^{-1}~[{\rm K}], 
\label{eq:deltat}
\end{eqnarray} 
where $\mu$, $m_{\rm H}$, $k_{\rm B}$, $n_{\rm SNII}$ and $E_{\rm SNII}^{51}$ are the mean molecular weight, the proton mass, and the Boltzmann constant, the number of SN-II explosion per unit stellar mass and SN-II energy from a single explosion normalised by $10^{51}~{\rm erg}$, respectively.  
As long as we use this feedback scheme, almost all gas particles cannot be heated to above $10^6~{\rm K}$. 
The fundamental problem is that one gas particle spawns one or more star particles.
This means that the mass of a star particle is equal to the gas particle mass or smaller. 
This causes the large gas mass to be heated to unexpectedly low temperature as pointed out by \citet{Vecchia2012, Keller2014}. 

Some solutions for this problem are considered.
One of simplest solutions is changing the mass ratio between a gas particle and a star particle. 
If the mass of a star particle is set to be heavier than a gas particle (i.e., a star particle is born out of many gas particles), the input SN energy increases with the mass of star. 
This procedure is equivalent to making the denominator of Eq.\,(\ref{eq:deltat}) larger. 
This may result in the heating of gas to a desired temperature. 
On the other hand, we can get similar effect if the mass of a gas particle affected by feedback is set to be small using the particle splitting method \citep[e.g., ][]{Chiaki2015}. 
However, these solutions assume that the number of gas particles (or the total mass) in bubble radius does not change with increasing input energy. 
According to Eq.~(\ref{rbub}), the total mass in the radius is proportional to the input SN energy, and the above assumption is not realistic.  

As another solution, we consider accumulating thermal energy from many stars. 
If many star particles form around a gas particle, the gas particle gets large SN energy from these star particles. 
This might lead to a high temperature due to the accumulation of thermal energy from each star. 
However, this situation will almost never occur in our simulation as we describe below. 
We conclude that as long as we adopt Stinson-type feedback scheme, it is difficult to overcome this problem in which the temperature of almost all gas particles affected by the SN feedback is $< 10^{6}~{\rm K}$.

We find that the temperature of a few gas particles exceed $\sim 10^{6}~\rm K$ by gaining energy from some SN-II explosion events, though such events are very rare and the temperature of almost all gas particles affected by the SN feedback is below $10^6~{\rm K}$. 
\citet{Keller2014} and \citet{Vecchia2012} pointed out that such (high-density and warm $< 10^6~{\rm K}$) gas is unphysical because the cooling time in such a region is very short ($t_{\rm cool} = 10^4 - 10^6$ yrs). 
The reason why our model have such gas is mainly because of the typical cooling shut-off time ($10^6 - 10^7$ yrs, see also bottom-left panel of Fig.~\ref{fig:nfb}), which is longer than the cooling time of the gas in the region. 
However, sometimes we recognize the existence of such gas in the stochastic models just after the SN feedback. 
With our current resolution, we can resolve such short time-scale, and the existence of such gas might be acceptable if we have sufficient resolution. 
Nevertheless, if too many gas particles stay in such a phase for a long time, it will be unphysical, and we have to treat this high-density, warm gas carefully when we study the ISM and CGM. 
Fortunately, the fraction of such gas is not the dominant component in high-density region (most dominant components are cool gas $T < 5000~{\rm K}$). 
Therefore, these gas does not strongly affect the ISM study. 

Interestingly, the phase diagram for Cool-on runs is very similar to the No-FB run even though the Cool-on model can suppress the star formation as well as other Osaka models. 
This suggests that the gas particles affected by the SN feedback for Cool-on model satisfy the temperature threshold but do not satisfy the SF density threshold because these gas escape the star-forming region by the outflow. 
We note that the Cool-on model can suppress star formation as long as the density threshold is high enough, but if we adopt the lower threshold, then this model might be unable to suppress star formation. 

Finally, another possible way to interpret this warm gas at $\sim 10^5~{\rm K}$ is that it is reminiscent of the hot component of the multiphase ISM model by \citet{Springel2003}. 
In their model, the gas above the SF threshold density had an effective equation of state with temperature higher than $10^4 ~{\rm K}$ for the SN-heated gas \citep[e.g.,][]{Robertson2004}, but its density was still described by that of the cold component of the subgrid model. 
In our model, once the gas gets heated by SN feedback, it might stay at densities above SF threshold, but at the temperatures above the star-forming regions and do not participate in star formation. 
This is one of the reasons for adding the kinetic wind explicitly as part of the SN feedback model in \citet{Springel2003}.  

\begin{figure*}
\includegraphics[width = 170mm]{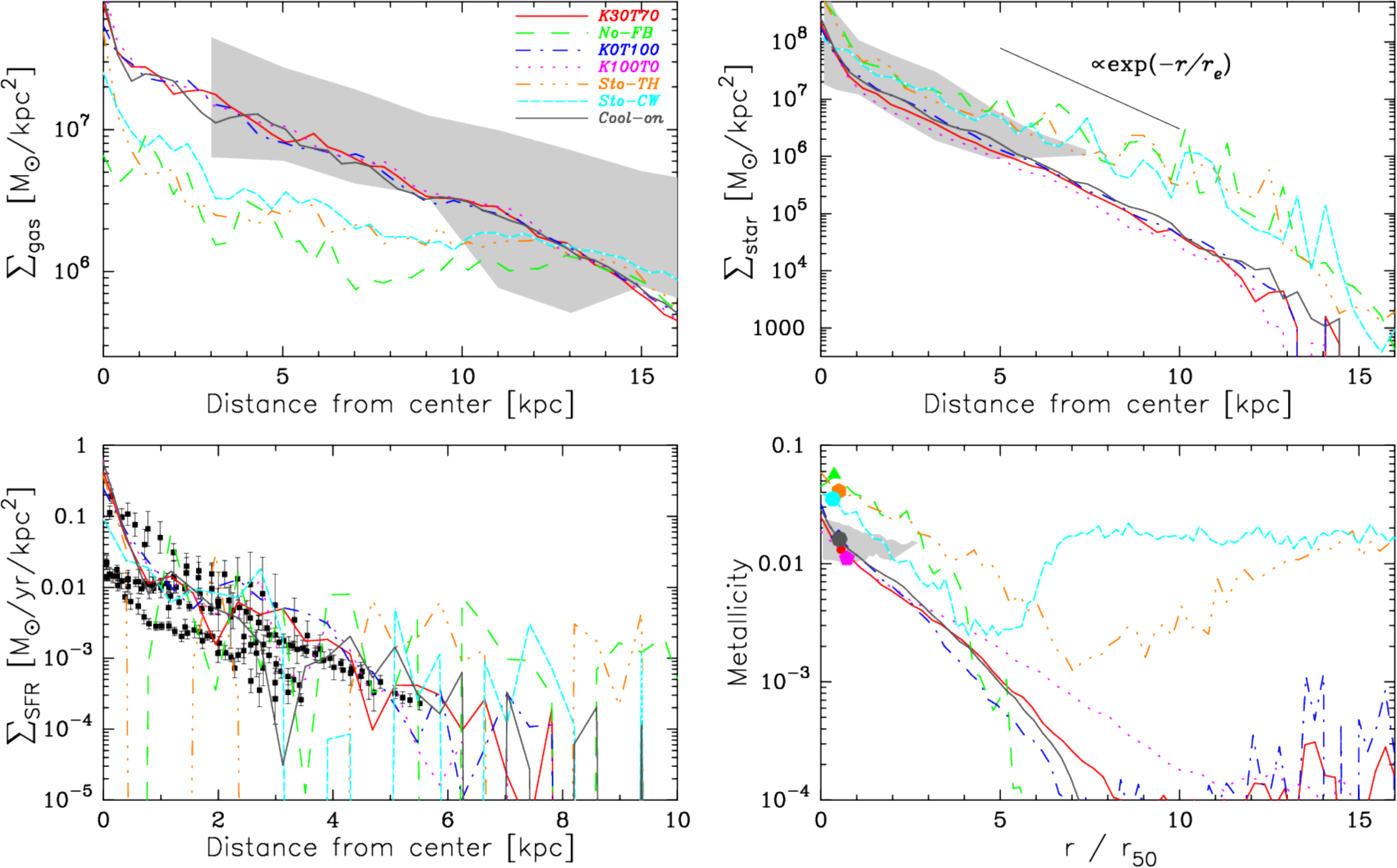}
\caption{Radial profiles of physical quantities at $t = 1$ Gyr for each run, as indicated in the legend of top-left panel. 
{\it Top left panel}: Total gas surface density profiles. The shaded area indicates the range of observational data from 33 nearby spiral galaxies \citep{Bigiel2012}. 
{\it Top right}: Stellar mass surface density profiles.  An exponential profile is also shown to guide the eye. 
{\it Bottom left}: SFR surface density profiles. The horizontal axis is the distance from galactic centre. 
The points with error-bars represent the observational data of nearby galaxies with similar stellar masses \citep{Leroy2008}. 
{\it Bottom right}:  Projected gas metallicity profile.  Here, the metallicity is bare value and is not divided by the solar value.  The horizontal axis is the distance from galactic centre normalized by $r_{50}$. 
The points show $r=1\,{\rm kpc}$ in real space. 
The shaded area shows the observational data from the SDSS-IV MaNGA survey \citep{Belfiore2017}.}
\label{fig:rpm12}
\end{figure*}
\begin{figure*}
\includegraphics[width = 170mm]{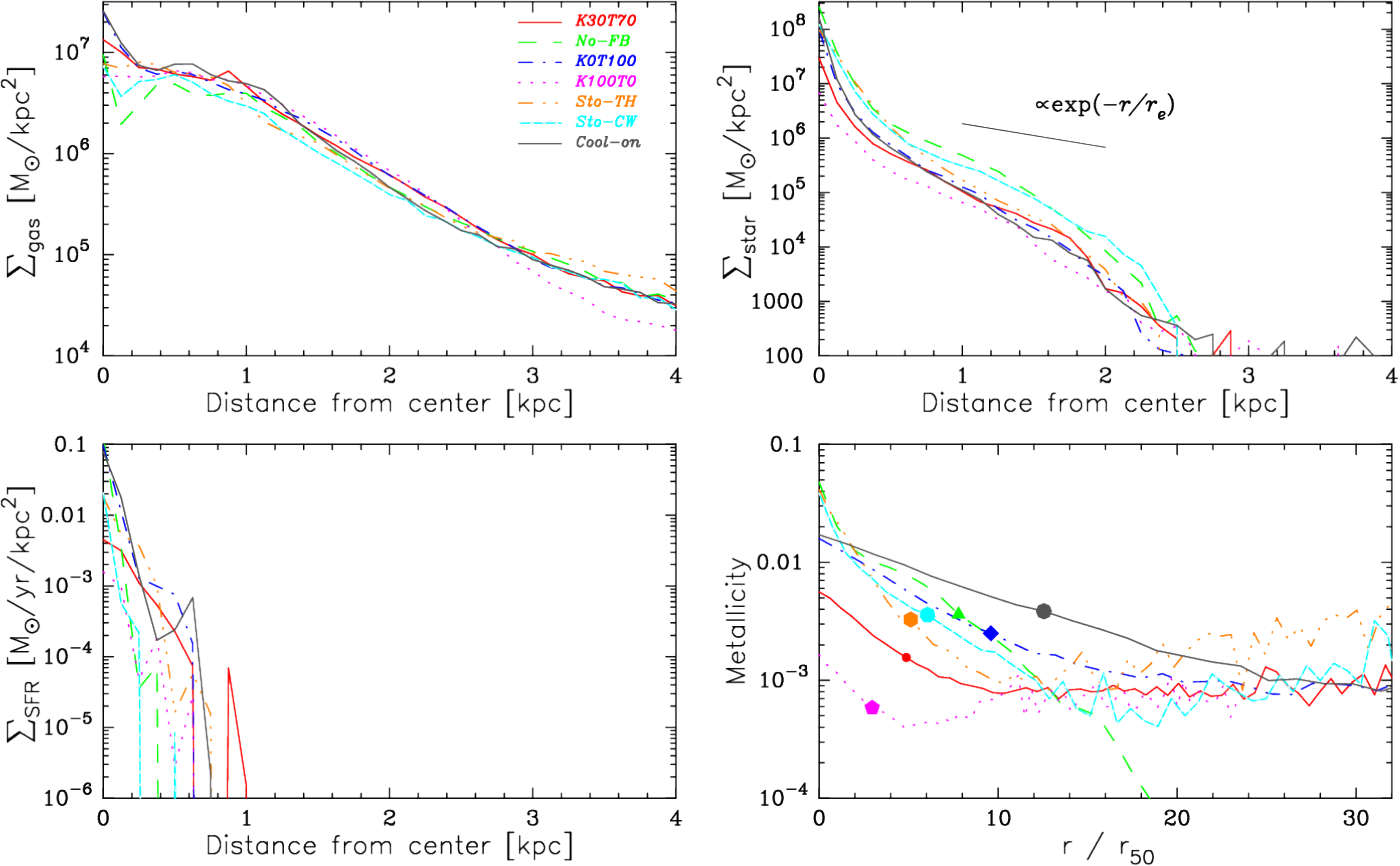}
\caption{Same as Fig.~\ref{fig:rpm12} but for M10 galaxy.}
\label{fig:rpm10}
\end{figure*}

\subsection{Radial Profiles of Physical Quantities}
In Figs.~\ref{fig:rpm12} and \ref{fig:rpm10}, we discuss the radial profiles of various quantities of our simulated galaxies at $t = 1$\,Gyr for M12 and M10 runs. 
Since we are dealing with isolated galaxies here, examining $t=1$\,Gyr output is by no means a truly realistic comparison to the real Milky Way or nearby spiral galaxies, but it does provide a baseline for the future cosmological simulations and comparison with observations. The AGORA code comparison project was also performed with a similar spirit \citep{Kim14, Kim2016}.   

The top-left panel of Fig.~\ref{fig:rpm12} shows the total gas surface density. 
The Osaka models (K30T70, K0T100, K100T0) are all consistent with the shaded observational data, and show very similar profiles with each other. 
We find that the Osaka model with cooling-on (Cool-on) also matches the observation.  
The No-FB run shows much lower gas surface density than other runs and also underpredicts the observational data, because too much gas is converted into stars due to lack of SN feedback. 
The Sto-TH and Sto-CW runs show similar profiles to the No-FB, but this is not just due to gas consumption. 
As shown in Figs.\,\ref{fig:sfhm12} and \ref{fig:sfhm10}, the SFR in Sto-TH and Sto-CW are lower than that of No-FB, which 
suggests that much of the gas is removed by the strong galactic wind as well as the gas consumption by star formation. 

The top-right panel of Fig.~\ref{fig:rpm12} shows the stellar mass surface density profiles, which give a consistent picture as the gas surface density. 
The No-FB run shows the highest stellar mass density, and the Sto-TH and Sto-CW runs are same or somewhat below the No-FB run. 
The other runs (all Osaka models) show lower stellar density profiles, with the K100T0 run being the lowest in the intermediate radial range of $r=5-12$\,kpc.  
The Osaka models agree with an exponential profile of $\exp(-r/r_{\rm e})$ at $r=2-12$\,kpc, with an effective radius of $r_{\rm e}=1.77$\,kpc. 

In the bottom-left panel of Fig.~\ref{fig:rpm12}, we show the SFR surface density profiles. 
The observational data for galaxies with similar stellar masses to our simulation is shown with black data points \citep{Leroy2008}, and all of our runs are within the range of observed data. 
For the No-FB, Sto-TH and Sto-CW runs, the gas and SF is concentrated in dense clumps, which causes discontinuous jumps in the SFR surface density profile.  
This feature can be seen as clumpy stellar structure in Fig.\,\ref{fig:spmm12} for No-FB, Sto-TH and Sto-CW runs. 

Finally, we present the gas metallicity profile in the bottom-right panel of Fig.~\ref{fig:rpm12}, together with the observational data from the SDSS-IV MaNGA survey, which used 550 nearby galaxies \citep[shaded region,][]{Belfiore2017}.
The metallicity profile can elucidate how the metals are transported from star-forming regions to the ISM and CGM. 
The observational data is often presented with galactic radius normalized by the half-light radius of each galaxy. 
Here, we take the radius within which 50\% of total stellar mass is contained as $r_{50}$.
For our simulated galaxy, $r_{50} \approx 2.0 \sim 4.2$\,kpc at $t=1$\,Gyr. 
The metallicity profiles of the Osaka model runs (K30T70, K0T100 and K100T0) are consistent with the observation at $r/r_{50} < 2$, although the overall slopes are somewhat steeper than the observed metallicity gradient. 
This could be because the current observations cannot probe the metallicity profile at very large radii of the galactic discs, as is evident from our comparison. 
The metal enrichment in the outer region of Cool-on run is more efficient than that of K0T100. 
This means that the momentum kick has stronger impact on this mechanism than thermal pressure in our model. 
The Osaka model runs cause the metal enrichment of the ISM at large galactic radii of $r/r_{50} > 6$. 
The No-FB run produces too much stars and metals, overpredicting the metallicity at all radii, but cannot transport the metals to $r/r_{50} > 6$. 
The Sto-TH and Sto-CW runs overpredict the observation in the shorter radius ($r/r_{50} < 6$). 
These runs can also distribute metals to larger radii ($r/r_{50} > 6$) unlike the No-FB run.
Moreover, the Sto-CW run shows an interesting feature of increasing metallicity profile at $r/r_{50} > 3$, which is caused by the strong galactic wind model. 
This suggests that Sto-TH and Sto-CW models generate strong outflows even though star formation is not suppressed significantly. 

From all the panels of Fig. \ref{fig:rpm12}, we conclude that the Osaka models produce reasonable radial surface density profiles of gas, stars, SFR and metallicity. 
On the other hand, the profiles for M10 runs are more sensitive to the details of SN feedback than in the M12 runs, because M10 galaxy has a shallower potential. 
The trend for the gas distribution is similar to the M12 runs, but the difference between the models is smaller than the M12 runs. The star formation in M10 is not active except for the very early phase. 
Therefore, SNe do not occur frequently, resulting in the small difference among different runs.
However, we find that the metallicity gradient for each run clearly shows different features even in the Osaka models. 
Our fiducial run (K30T70) and K100T0 run obviously show very low metallicity from inner region to outer region. 
This reflects that both models have low star formation activities (see also Fig.\,\ref{fig:sfhm10}). 
Furthermore, the metallicity in the K100T0 run is much lower than that of our fiducial run, which is due to the effect of strong wind velocity and strong outflow.

\subsection{Mass Outflow Rate \& Mass Loading Factor}
Galactic outflow plays an important role in suppressing star formation in galaxies. 
As the supernova bubbles overlap and percolate, they eventually erupt out of the galactic disc, forming a galactic fountain and supergalactic winds as we see in the M82 galaxy. 
Therefore it would be interesting to see whether the feedback model that is based only on the local ISM properties can reproduce reasonable mass outflow rates and mass-loading factors of the wind escaping from the galactic disc. 
At a given time $t$, we measure the mass outflow rate at a height $h$ from the galactic plane as 
\begin{equation}
\dot{M}_{\rm out}(h, t) = \sum_i^{N} m_i W(|z_i - h|)\, v_{i, z} \, \pi \sqrt{h_{i, {\rm sml}}^2 - (z_i - h)^2}, 
\end{equation}
where $m_i$, $W$, $z_i$, $v_{i, z}$ and $h_{i, {\rm sml}}$ are the mass, kernel function, $z$ coordinate, $z$-direction outflow velocity, and smoothing length of the $i$-th SPH particle, respectively. 
Here we take $h = 1$\,kpc and $4$\,kpc as the two representative heights for measuring the outflow rate. 
Once we obtain $\dot{M}_{\rm out}(h, t)$ for the galaxy, the mass loading factor 
can be computed as 
\begin{equation}
\eta(h, t) = \frac{\dot{M}_{\rm out}(h, t)}{\dot{M}_{*}(t)}, 
\end{equation}
where $\dot{M}_{*}(t)$ is the total instantaneous SFR of the galaxy at time $t$. 
Since the Osaka feedback model is based on local physical quantities, {\em $\dot{M}_{\rm out}(h, t)$ and $\eta(h, t)$ are the outcomes of feedback in the simulation, varying spatially and temporarily, rather than predetermined constant factors.} 
This modelling will allow us to compute $\eta$ as functions of halo mass and redshift when we apply our model to a zoom-in cosmological hydrodynamic simulation \citep[e.g.,][]{Muratov2015}. 
In contrast, in the Sto-CW run, the wind velocity and $\eta$ are manually set to constant values of $v_{\rm wind}=600$\,km\,s$^{-1}$ and $\eta = 2$. 
For the Sto-TH run, we do not explicitly set the wind velocity but the thermal pressure by high temperature gas ($T=10^{7.5}~{\rm K}$) can drive strong outflows. 
 
\begin{figure*}
\includegraphics[width = 170mm]{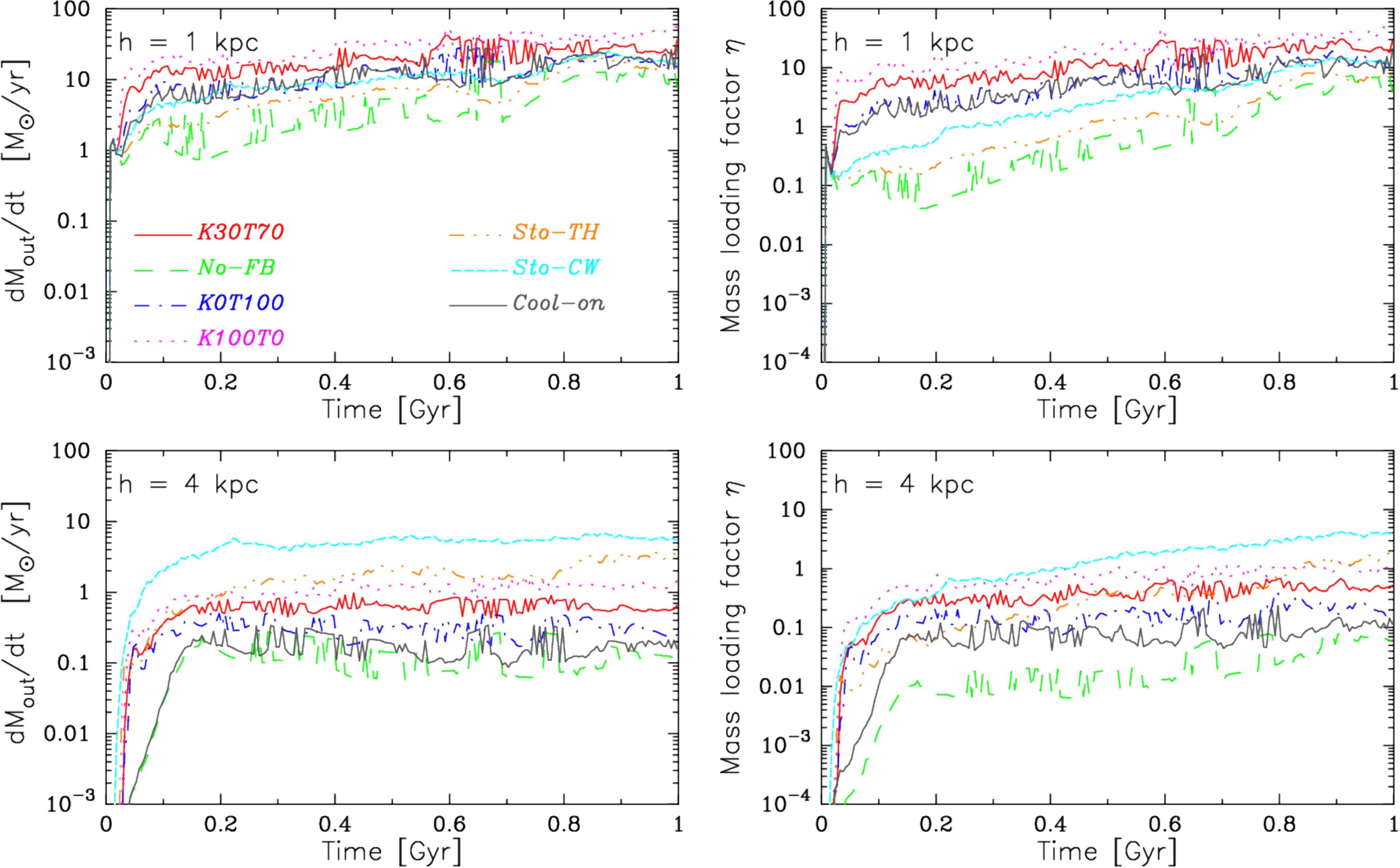}
\caption{Mass outflow rate ({\it left panels}) and mass loading factor ({\it right}) as a function of time. 
Top and bottom panels show these quantities measured at the height of $h=1~ {\rm kpc}$ and $4~ {\rm kpc}$ from simulated galactic plane, respectively. 
The line types are noted in the panel. } 
\label{fig:mlfm12}
\end{figure*}

\begin{figure*}
\includegraphics[width = 170mm]{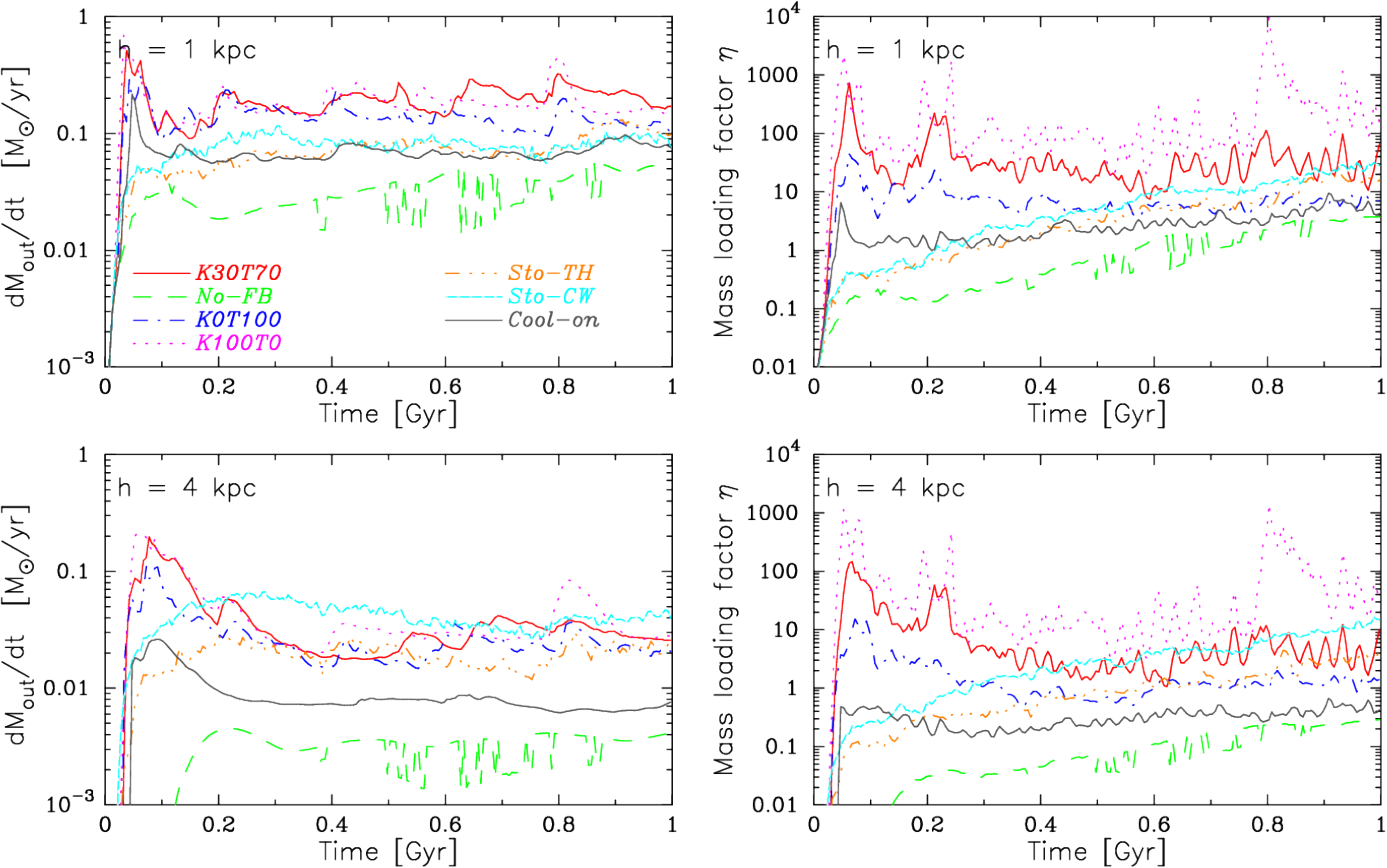}
\caption{Same as Fig.~\ref{fig:mlfm12} but for M10 galaxy.} 
\label{fig:mlfm10}
\end{figure*}

Figures~\ref{fig:mlfm12} and \ref{fig:mlfm10} present the time evolution of mass outflow rate ($\dot{M}_{\rm out}$, left panels) and mass loading factor ($\eta$, right panels) at $h = 1$\,kpc (top row) and 4\,kpc (bottom row). 
By definition, if SFR is roughly constant, we expect $\dot{M}_{\rm out}$ and $\eta$ to be also roughly constant in time, once the galaxy reaches a steady state. 
This is true for the Osaka model, and $\dot{M}_{\rm out}$ is roughly constant with $\dot{M}_{\rm out} \sim 5-20~{\rm M_{\odot} yr^{-1}}$ for $h=1$\,kpc. 
However, $\eta$ gradually increases over time in most of the runs, because the SFR is slowly declining in our isolated galaxy (see Fig.~\ref{fig:sfhm12}). 
At $h=1$\,kpc, $\eta$ increases by more than a decade, as the SFR declines by a similar factor. 
From the comparison between K30T70 and Cool-on runs, we find that the thermal pressure is important to drive the outflows similarly to the momentum kick. 
In contrast to the Osaka models, $\dot{M}_{\rm out}$ and $\eta$ of the Sto-TH and Sto-CW runs are lower than the Osaka models despite of large outflow velocities. 
This is because the outflow velocities of both models are too large, thus almost all gas particles would not stay near the galactic plane, and not captured at $h=1-4$\,kpc. 
Conversely, the velocities of our Osaka models is moderate and gas particles can remain in the disc.

On the other hand, $\dot{M}_{\rm out}$ of M10 runs shows a similar trend with M12 runs but the value is smaller than M12 runs. 
Interestingly, $\eta$ in M10 runs show higher values than that in  M12 runs. 
This means that the SN feedback works more effectively due to shallower potential. 

It is also interesting to examine the dependence of $\dot{M}_{\rm out}$ on the height from the galactic plane. 
For the Osaka runs, $\dot{M}_{\rm out}$ rapidly decreases with increasing height, from $10-20~{\rm M_{\odot}yr^{-1}}$ at $h = 1$\,kpc to $\dot{M}_{\rm out} < 1~{\rm M_{\odot}yr^{-1}}$ at $h = 4$\,kpc. 
On the other hand in the Sto-TH and Sto-CW runs,  $\dot{M}_{\rm out}$ changes little between $h = 1$\,kpc and $4$\,kpc. 
Significantly stronger $\dot{M}_{\rm out}$ and higher $\eta$ for the Sto-CW and Sto-TH runs are the consequences of high wind velocity. 
The strong outflow in both runs can escape from the galaxy without falling back onto the disc. 
On the other hand, the velocities of almost all wind particles in the M12 simulations are not high enough to escape from the galaxy, and such gas particles fall back onto the disc and form a galactic fountain. 
Yet a small number of gas particles can reach the outer region of the galaxy. 
In the cases of M10 runs, the situation dramatically changes. 
Although the Osaka models can launch gas outflow escaping from galaxies, $\dot{M}_{\rm out}$ is not large.


\subsection{Distribution of Cooling Shut-off Time}
\begin{figure}
\includegraphics[width = 80mm]{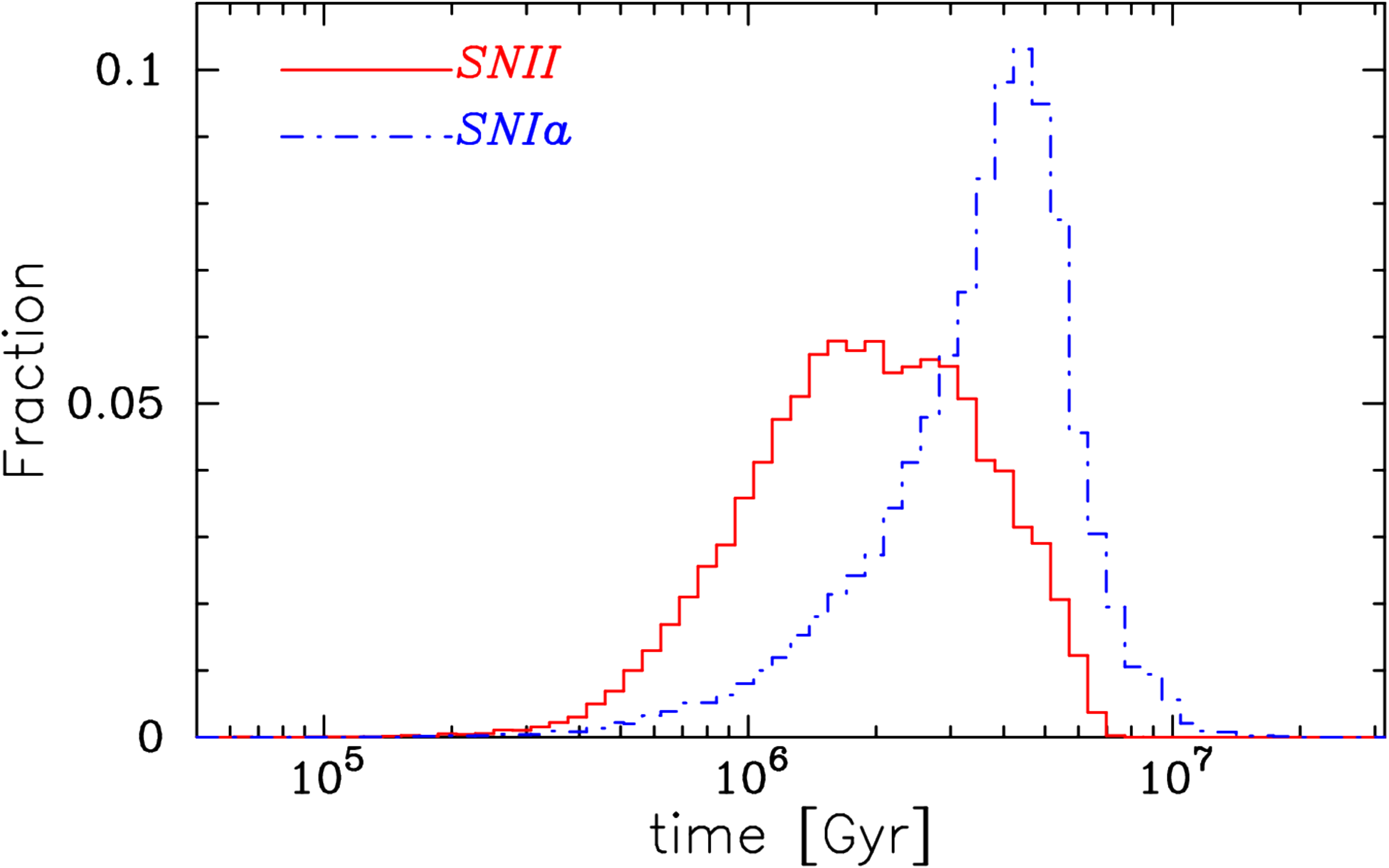}
\caption{Distribution of the cooling shut-off time in the fiducial run. 
The two histograms correspond to SN-Ia and SN-II feedback as shown in the legend.
As we described in Section~\ref{sec:SNFB}, we turn off cooling only for $\Delta t < t_{\rm hot}$. } 
\label{fig:tdm12}
\end{figure}

As described in the previous subsection, we have seen the importance of cooling shut-off. 
The effect helps gas transportation into the CGM and IGM by the thermal pressure. 
Here, we study the probability distribution function of the duration time. 
Figure~\ref{fig:tdm12} shows the distribution of cooling shut-off time during the hot phase for the fiducial run. 
As we described in Section~\ref{sec:SNFB}, we turn off cooling only for $\Delta t < t_{\rm hot}$.  
We find that the typical duration time is several times of $10^6~{\rm yr}$, which is much shorter than those in \citet{Stinson2006} due to our higher resolution. 
Interestingly, our SN feedback model is efficient enough to suppress star formation even though the cooling shut-off time is shorter than the time adopted by \citet{Stinson2006}, which was $\sim 30$ Myrs. 

\subsection{Distribution of Wind Velocities}
As described above, the outflow rate and mass loading factor sensitively depend on the SN affected gas mass and wind velocity. 
In our Osaka model, the SN hot bubble radius $r_{\rm bub}$ (Eqn. \ref{rbub}) controls the feedback efficiency, and this parameter is uniquely derived from the physical conditions around stellar particles. 
This is one of the major advantages of our Osaka model. 
However, $r_{\rm bub}$ has the dependencies on the gas density and input SN energy. 
Actually, $r_{\rm bub}$ increases with decreasing (increasing) ambient gas density (input SN energy). 
While, if $r_{\rm bub}$ become larger, the gas particles in this radius also increases and the received energy of each gas particle decreases. 
This suggests that there is a possibility of self-regulation of the energy. 
In other words, the wind velocity might not have strong dependency on these physical quantities.  
Therefore, in this subsection, we focus on the wind velocity. 
After this subsection, we explore how deposited SN energy which relates to the $n_{\rm fb}$ has an impact on the galaxy evolution. 
Moreover, we also discuss the resolution dependency after subsection. 

\begin{figure}
\includegraphics[width = 80mm]{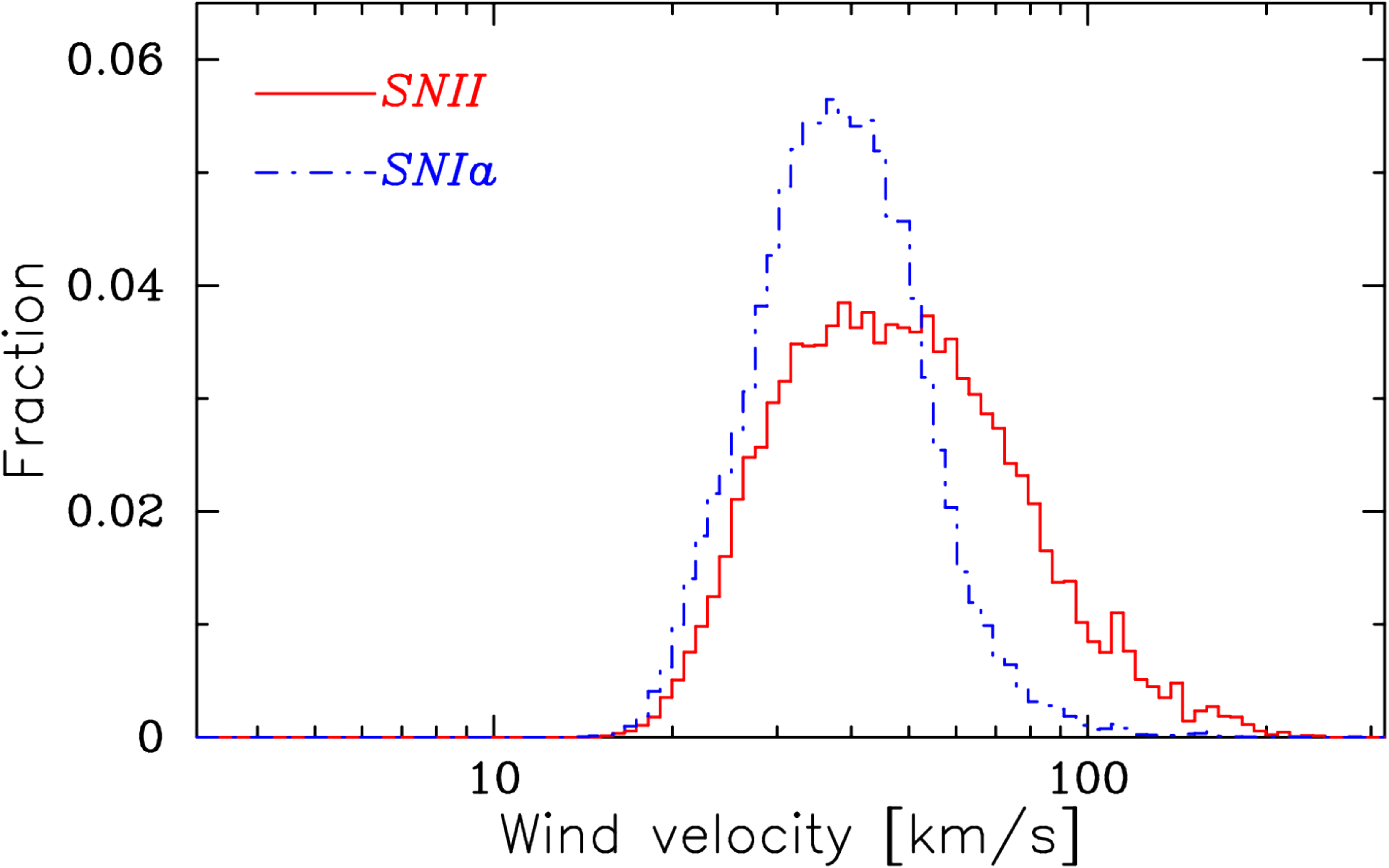}
\caption{Velocity distribution of the wind particles due to SN-Ia and SN-II feedback in the fiducial run, as shown in the legend.} 
\label{fig:wvdm12}
\end{figure}

Figure \ref{fig:wvdm12} represents the PDF of wind velocity by the SN-II feedback in the fiducial run. 
For the comparison, we also plot the PDF of the SN-Ia. 
The peak wind velocity is about $40-50$\,km\,s$^{-1}$, which is lower than the virial velocity of host halo, which is about 200\,km\,s$^{-1}$. 
Therefore almost all wind particles stay in the host galaxy, and only those in the high velocity tail can escape from the host galaxy, which goes up to about 200\,km\,s$^{-1}$. 
These wind velocities are physically plausible, as our simulation is modelling a disk galaxy with a lower stellar mass than the Milky Way galaxy and relatively quiet star formation. 
Wind velocities of a few hundreds to 500\,km\,s$^{-1}$ are observed from interstellar absorption lines of high-redshift star-forming galaxies such as Lyman break galaxies, but these galaxies typically have much higher SFR of $\sim~10-100~{\rm \Msun yr^{-1}}$  than what we simulate here. 
Interestingly, the peak velocity for SN-Ia is smaller than SN-II. 
This is because that SN-Ia explosions can occur in less-density regions than the SN-II explosions due to its time-delay, during which the spawned star particles may drift away from dense star-forming regions. 
This means that the shock radius becomes larger than SN-II explosion and the number of particles in shock radius increases. 
As a result, available energy for each gas particle is smaller and wind velocity also becomes small. 
We can see such time evolution in our simulations. 

\subsection{Dependency on the Feedback Event Number}
\begin{figure*}
\includegraphics[width = 170mm]{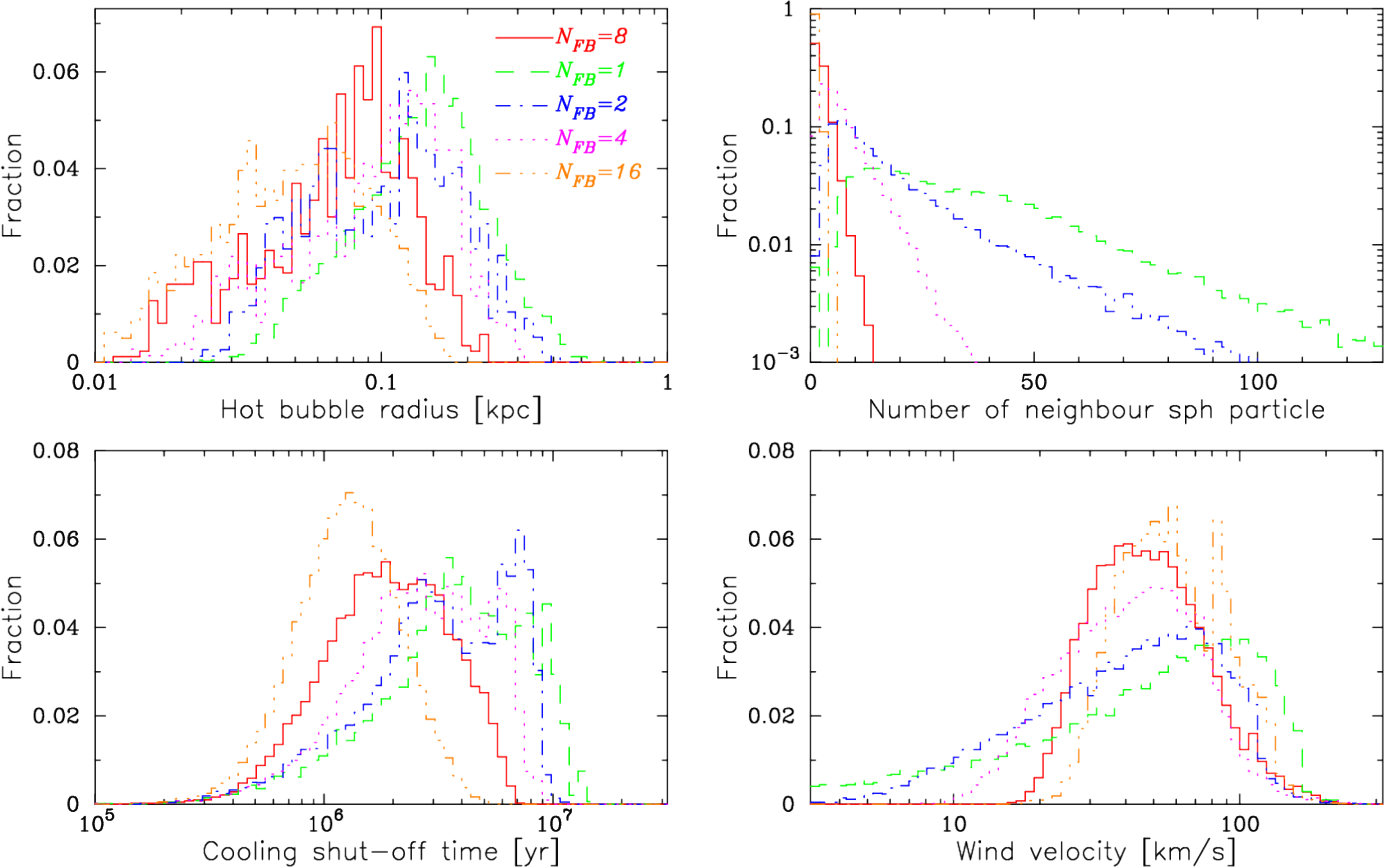}
\caption{Physical quantities of dependency on the $n_{\rm fb}$ at $t = 1$ Gyr for T30K70 (fiducial) run, as indicated in the legend of top-right panel. 
{\it Top left panel}: Hot bubble radius distribution. 
{\it Top right}: Number of gas particles in hot bubble. 
{\it Bottom left}: Cooling shot-off time distribution. 
{\it Bottom right}: Wind velocity distribution. }
\label{fig:nfb}
\end{figure*}

In this subsection, we discuss the choice of the SN feedback event number $n_{\rm fb}$ (see Eqn. \ref{eq:snfbdt}). 
This is an important control parameter of the Osaka feedback model, and our fiducial value is $n_{\rm fb} = 8$. 
In order to explore $n_{\rm fb}$ dependency, we perform additional runs with $n_{\rm fb} = 1, 2, 4$ and $16$. 
This selection means that the available energy range for a single SN event can have a different order of magnitude. 
Note that in order to take advantage of the {\small CELib} library such a metal abundance evolution, it is preferable that we adopt larger $n_{\rm fb}$ values than smaller ones. 

As we show below, there is an interesting balance between the value of $n_{\rm fb}$, bubble radius and wind velocity. 
If we choose a smaller $n_{\rm fb}$ value (e.g., $n_{\rm fb} = 1$), then the input energy becomes larger for a single injection compared to the fiducial case. 
However, at the same time, the bubble radius becomes larger as it depends on the amount of deposited energy. 
When the bubble radius becomes larger, the number of particles inside the radius also increases, and each particle receives less energy. 
These effects cancel each other, and the resulting wind velocity does not depend on the value of $n_{\rm fb}$ very strongly. 

In top left panel of Figure \ref{fig:nfb}, we show the distribution of bubble radius affected by SN-II for different values of $n_{\rm fb}$. 
The peak of the bubble radius distribution shifts toward smaller values gradually with increasing $n_{\rm fb}$, because the bubble radius is weakly dependent on the input energy in Eqn.~\ref{rbub}. 
For our fiducial value, the bubble radius is distributed between 10\,pc and a few hundred parsec, with a peak around 100\,pc. 
As $n_{\rm fb}$ is increased, the energy that each particle receives reduces, and the tail towards smaller bubble radii become more visible. 

Next, we explore the number of gas particles in the bubble radius for different $n_{\rm fb}$. 
In top right panel of Fig. \ref{fig:nfb}, we find that this value is strongly affected by $n_{\rm fb}$, especially the long tail toward the right. 
For $n_{\rm fb}=1$ \& 2, the peak of the distribution is at $5-10$ neighbour particles, and the long tail extends to about 100 particles. 
For $n_{\rm fb}=4, 8,$ \& 16, the long tail shrinks, and the number of neighbour particle quickly decreases. 
For the fiducial value of $n_{\rm fb}=8$, the peak is at a few particles, and the tail extends out to about 20 \& 30 particles for SN-II feedback. 
This suggests that the available energy for each wind particle from SN feedback strongly depend on $n_{\rm fb}$, because the deposited energy is shared by the gas particles in the bubble radius. 
For $n_{\rm fb}= 16$ case, the peak values approach the one. 
This suggests that in this resolution, the maximum value of $n_{\rm fb}$ should be around 16 because if we adopt larger value than 16, the number of gas particles in the bubble radius become one or zero. 
This breaks the condition for the $n_{\rm fb}$ that the number of particles in the hot bubble should be more than two. 
In that case, the wind velocity strongly correlates with the deposited energy. 
Moreover, output SN energy from a star particle becomes below $10^{51}~{\rm erg}$. 
This is unphysical and unacceptable due to breaking another condition for $n_{\rm fb}$. 

Bottom left panel of Fig. \ref{fig:nfb} shows the distribution of cooling shut-off time for different $n_{\rm fb}$. 
The peak of the shut-off time distribution shifts toward smaller values gradually with increasing $n_{\rm fb}$ as well as the case of bubble radius distribution, because the cooling shut-off time is also weakly dependent on the input energy in Eqn.~\ref{timehot}. 
The run with larger $n_{\rm fb}$ values have short shut-off duration. 
It is expected that the suppression of star formation by the feedback becomes weak. 
On the other hand, the time interval between SN feedback of one star cluster become shorter in such runs. 
Consequently, the feedback effectively works even for larger $n_{\rm fb}$ values as well as smaller $n_{\rm fb}$ case by these effects. 

Bottom right panel of Figure \ref{fig:nfb} shows the distribution of wind velocity for different $n_{\rm fb}$. 
As we discussed above, the peak wind velocity does not depend on $n_{\rm fb}$ significantly, shifting only mildly within factor two. 
This is because the available energy per wind particle is self-regulated by the balance between the bubble radius and the deposited SN energy. 
For smaller values of $n_{\rm fb}=1$ and 2, the bubble radius becomes somewhat larger, but the deposited energy is shared by a larger number of particles, extending the low-velocity tail at $V_{\rm wind} < 20$\,km\,s$^{-1}$. 
For $n_{\rm fb}=(8, 16)$, the bubble radius becomes smaller and the energy is shared with a smaller number of particles, therefore the low-velocity tail disappears. 
Interestingly, in the case of $n_{\rm fb} > 2$, the wind velocity distribution of each run is very similar. 
This implies that as long as we adopt $n_{\rm fb} > 2$, the effects of $n_{\rm fb}$ on the result might be very weak. 

\begin{figure}
\includegraphics[width = 83mm]{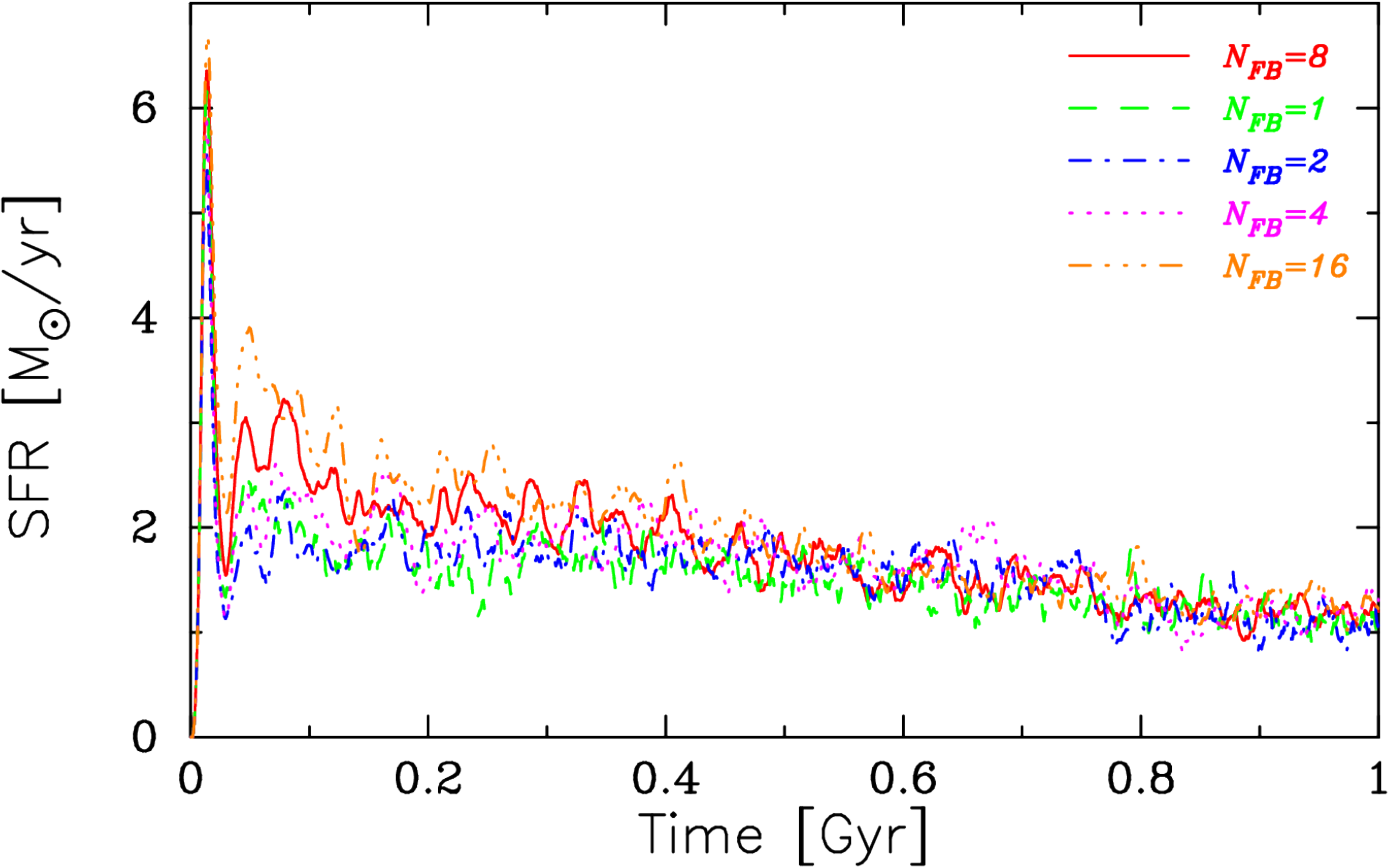}
\caption{Star formation history for various $n_{\rm fb}$ runs. 
Corresponding lines are shown in the figure. }
\label{fig:sfhnfb}
\end{figure}

Figure\,\ref{fig:sfhnfb} represents the star formation history for various runs with $n_{\rm fb}= 1 - 16$. 
Actually, except for the very early phase, we cannot recognize any apparent differences. 

We also explore the radial profiles for some physical properties such as the gas, stellar component, star formation and metallicity. 
Interestingly, we find that the $n_{\rm fb}$ dependency is weak in well-resolved regions at $r< 10~{\rm kpc}$. 

Finally, we should touch on the relation between $n_{\rm spawn}$ and $n_{\rm fb}$. 
In our fiducial model, we use $n_{\rm spawn} = 2$ which is the number of star particles that can be spawned from a gas particle. 
This parameter also changes the released SN energy even if we adopt the same $n_{\rm fb}$ values because the released energy depends on the mass of star. 
This means that $n_{\rm spawn}$ and $n_{\rm fb}$ are degenerate about SN feedback effect. 
Thus, $n_{\rm spawn} \times n_{\rm fb}$ value is an important value for our Osaka model.  
In conclusion, an acceptable range of this value ($n_{\rm spawn} \times n_{\rm fb}$) is less than 32 and $n_{\rm fb} > 2$. 
We note that this value depends on the numerical resolution. 
Thus, we need to adopt adequate value depending on the resolution. 

\begin{figure*}
\includegraphics[width = 170mm]{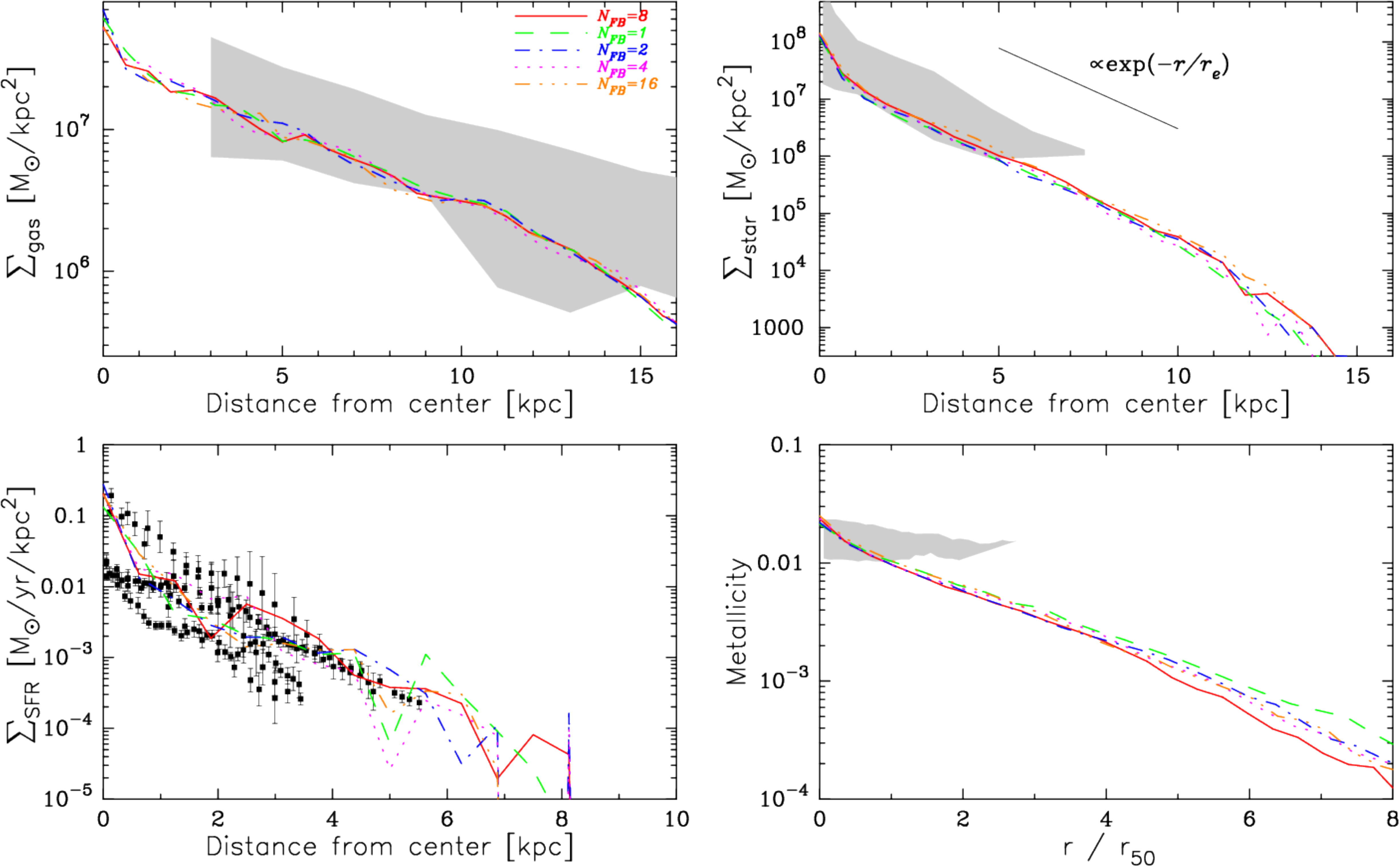}
\caption{Same as Fig.~\ref{fig:rpm12} but for various $n_{\rm fb}$ runs of our fiducial model. }
\label{fig:rpnfb}
\end{figure*}

\subsection{Resolution Dependence}
\begin{figure}
\includegraphics[width = 80mm]{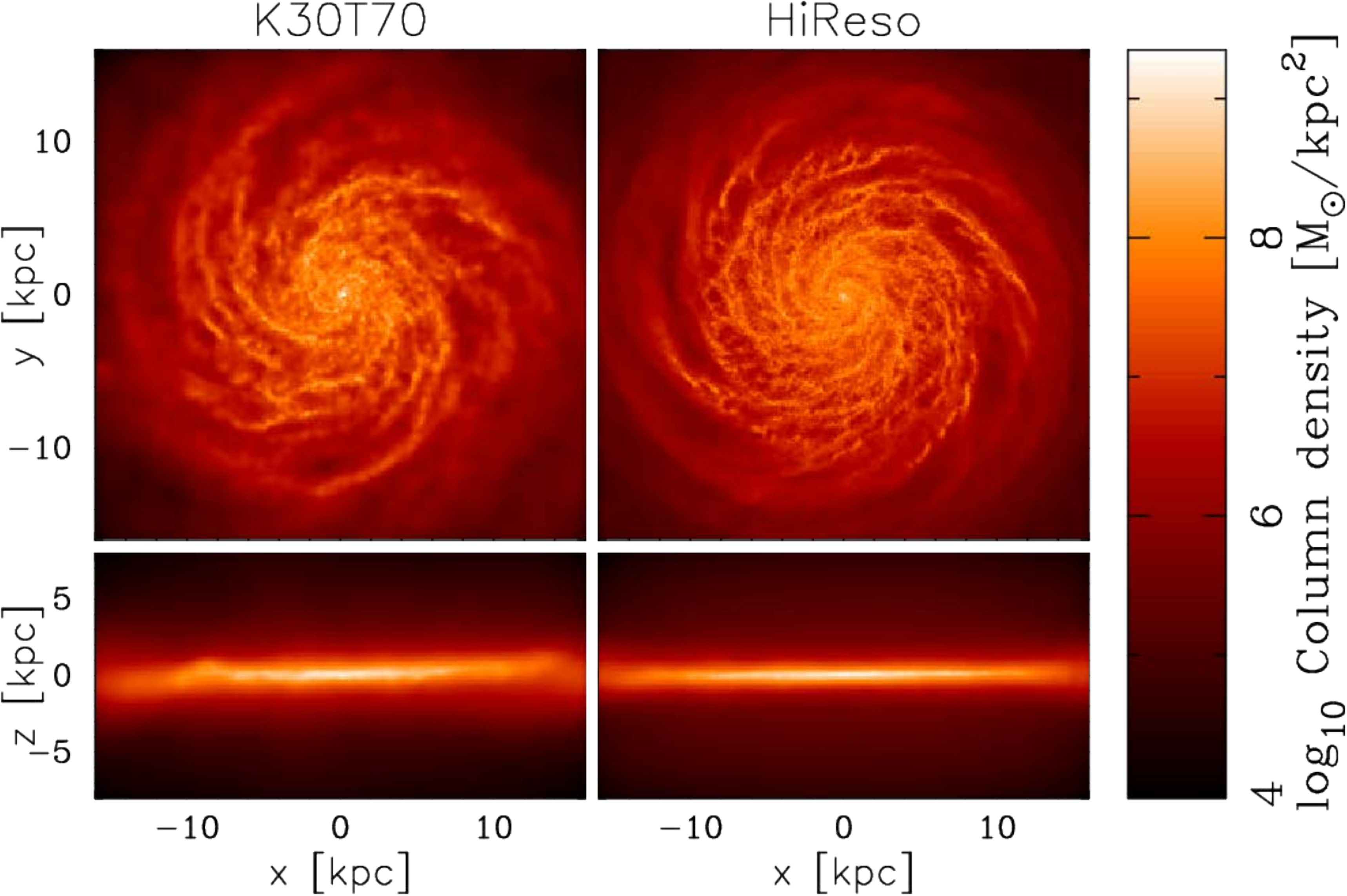}
\caption{Same as Fig.~\ref{fig:gpmm12}, but for the fiducial (left panels) and HiReso (right panels) runs. }
\label{fig:gpmm12hi}
\end{figure}
\begin{figure}
\includegraphics[width = 80mm]{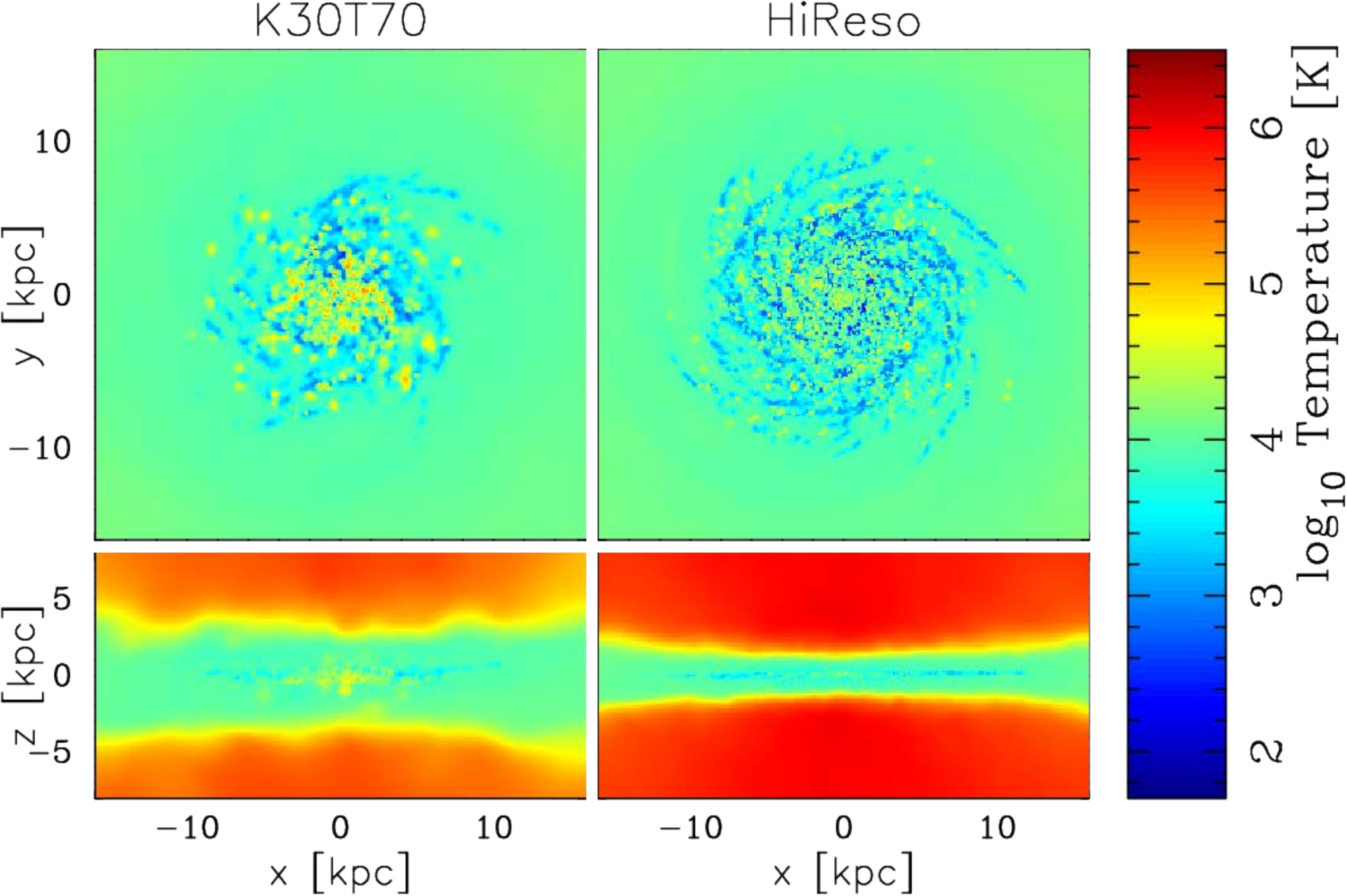}
\caption{Same as Fig. \ref{fig:gpmm12hi}, but for temperature, weighted by density-squared. }
\label{fig:tpmm12hi}
\end{figure}
\begin{figure}
\includegraphics[width = 80mm]{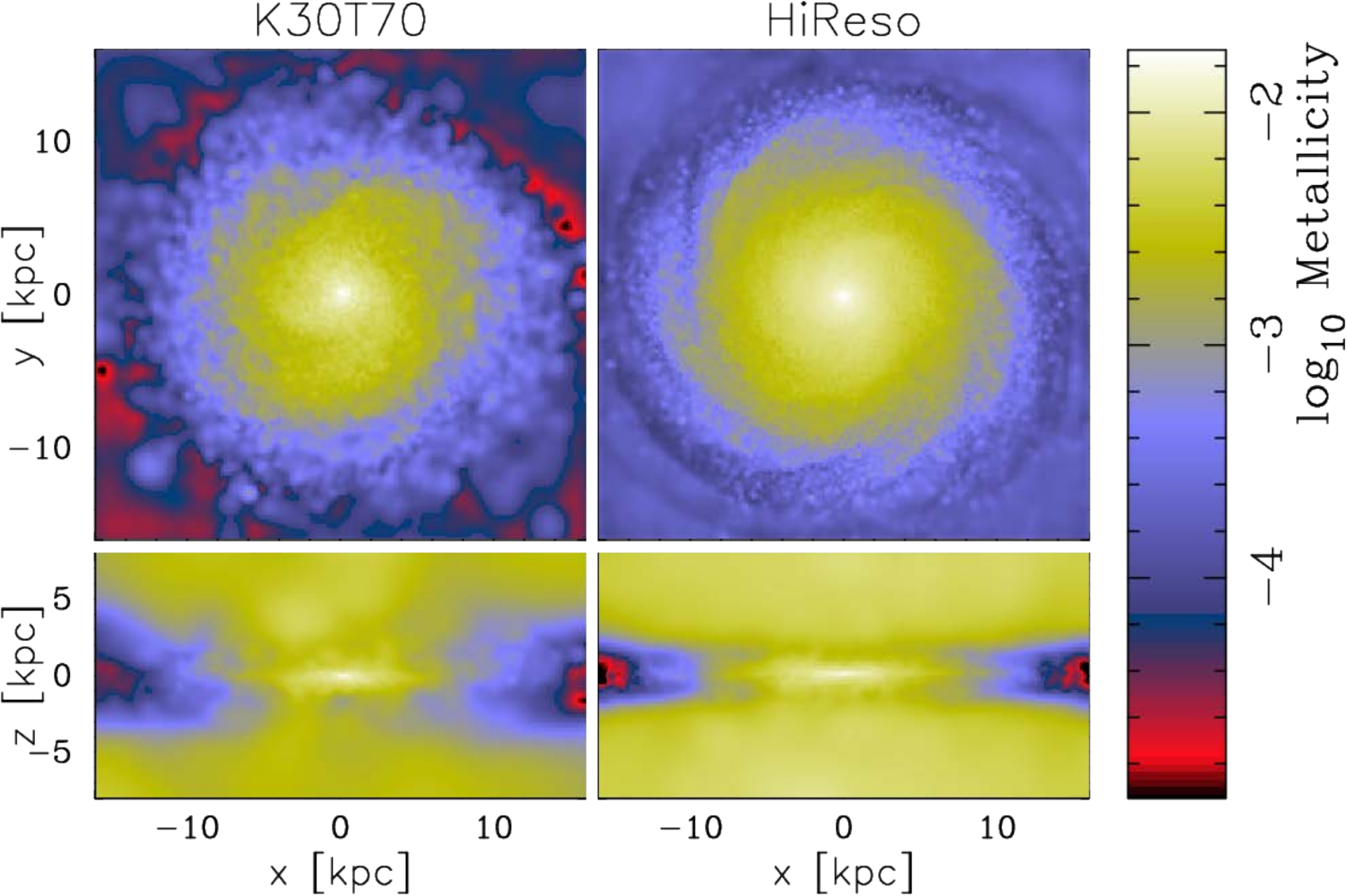}
\caption{Same as Fig.~\ref{fig:gpmm12hi}, but for metallicity. }
\label{fig:mpmm12hi}
\end{figure}
\begin{figure}
\includegraphics[width = 80mm]{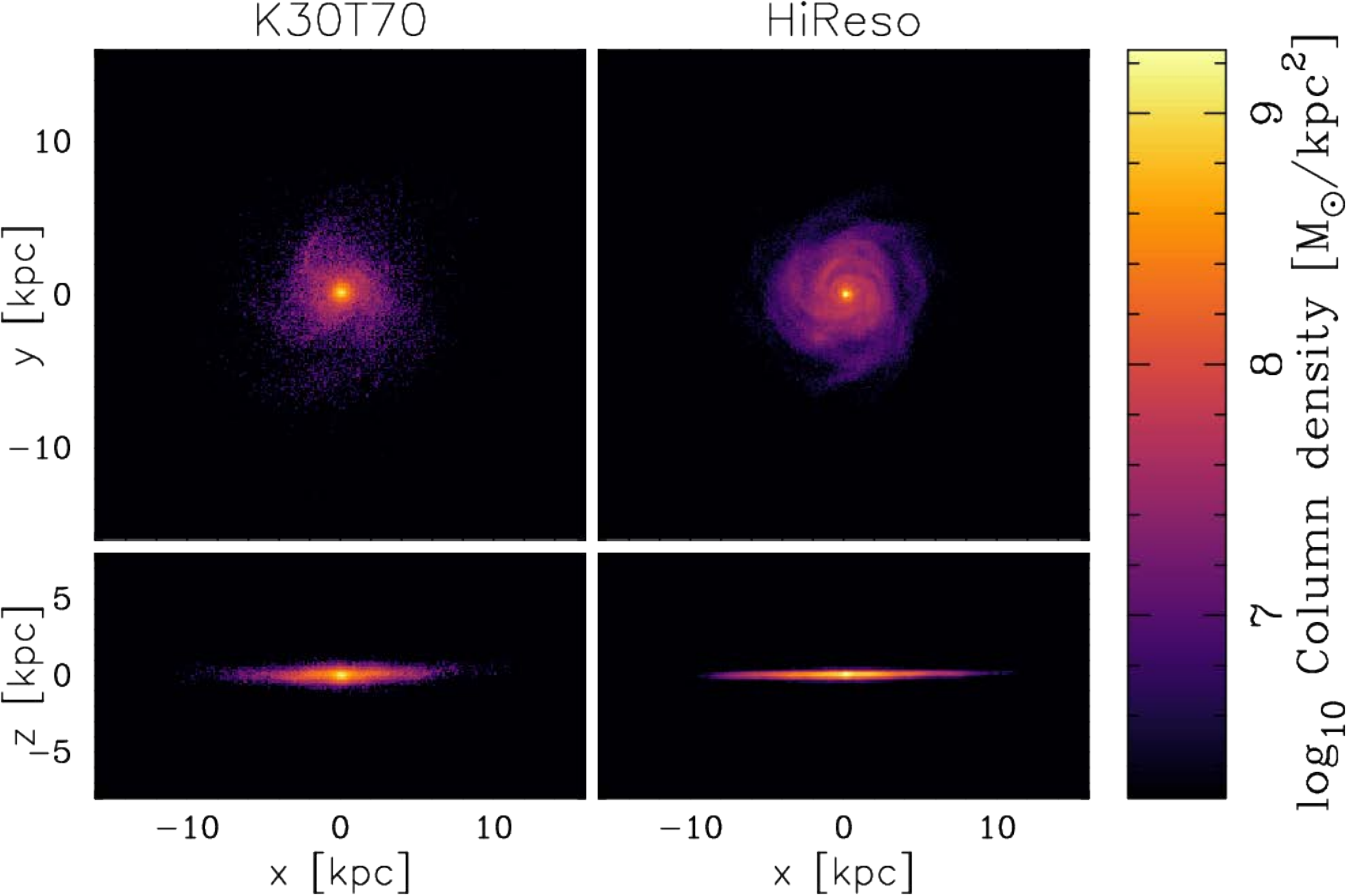}
\caption{Same as Fig. \ref{fig:gpmm12hi}, but for projected stellar mass density. }
\label{fig:spmm12hi}
\end{figure}

In our Osaka model, the feedback efficiency is controlled by the physical condition around star particles such as density and pressure. 
Deposited SN energy which is related to the stellar mass is also an important parameter for our model. 
Therefore, our model can depend on the resolution as well as $n_{\rm fb}$, and it is worth discussing the resolution dependency. 

For this purpose, we perform an additional simulation using higher resolution initial condition from the AGORA comparison project, which we call ``HiReso" run. 
In this additional run, we adopt the exactly same parameters as our fiducial run, except that the gravitational softening length is $40~{\rm pc}$ which is one half of the one in the fiducial run, and the mass resolution is ten times better. 
In general, we need to modify the parameters on star formation (threshold density) and SN feedback to improve the numerical convergence when we change the numerical resolution. 
However, in this study, we do not change the values of any parameters to explore whether our model can self-regulate the star formation by the feedback without any fine tuning. 

In Figures~\ref{fig:gpmm12hi}, \ref{fig:tpmm12hi}, \ref{fig:mpmm12hi} and \ref{fig:spmm12hi}, we compare the fiducial and HiReso runs at $t=1~{\rm Gyr}$, showing the projected gas density, projected temperature weighted by density-squared, projected metallicity, and projected stellar mass density. 
In the HiReso run, sharper spiral structures are seen for both gas and stellar distribution with finer details. 
Moreover, in the edge-on view, the galactic disk in the HiReso run is thinner than that of the fiducial run.  
In the projected temperature map, we see that the SN bubble sizes of HiReso run is smaller than those in the fiducial run, but the number of bubbles is greater in the HiReso run than that of the fiducial run. 
This is because the HiReso run can resolve more high-density regions and the deposited SN energy also becomes small due to finer gas mass elements. 
We also recognize that the heating of CGM is more efficient in the HiReso run with higher and smoother metallicities above the galactic plane. 
This might suggest that high-resolution run can resolve the outer region (low density region) of the galaxy where the gas can escape to the CGM and IGM.  Thus, in this study, we use the inner region of galaxy for the comparison. 

\begin{figure}
\includegraphics[width = 80mm]{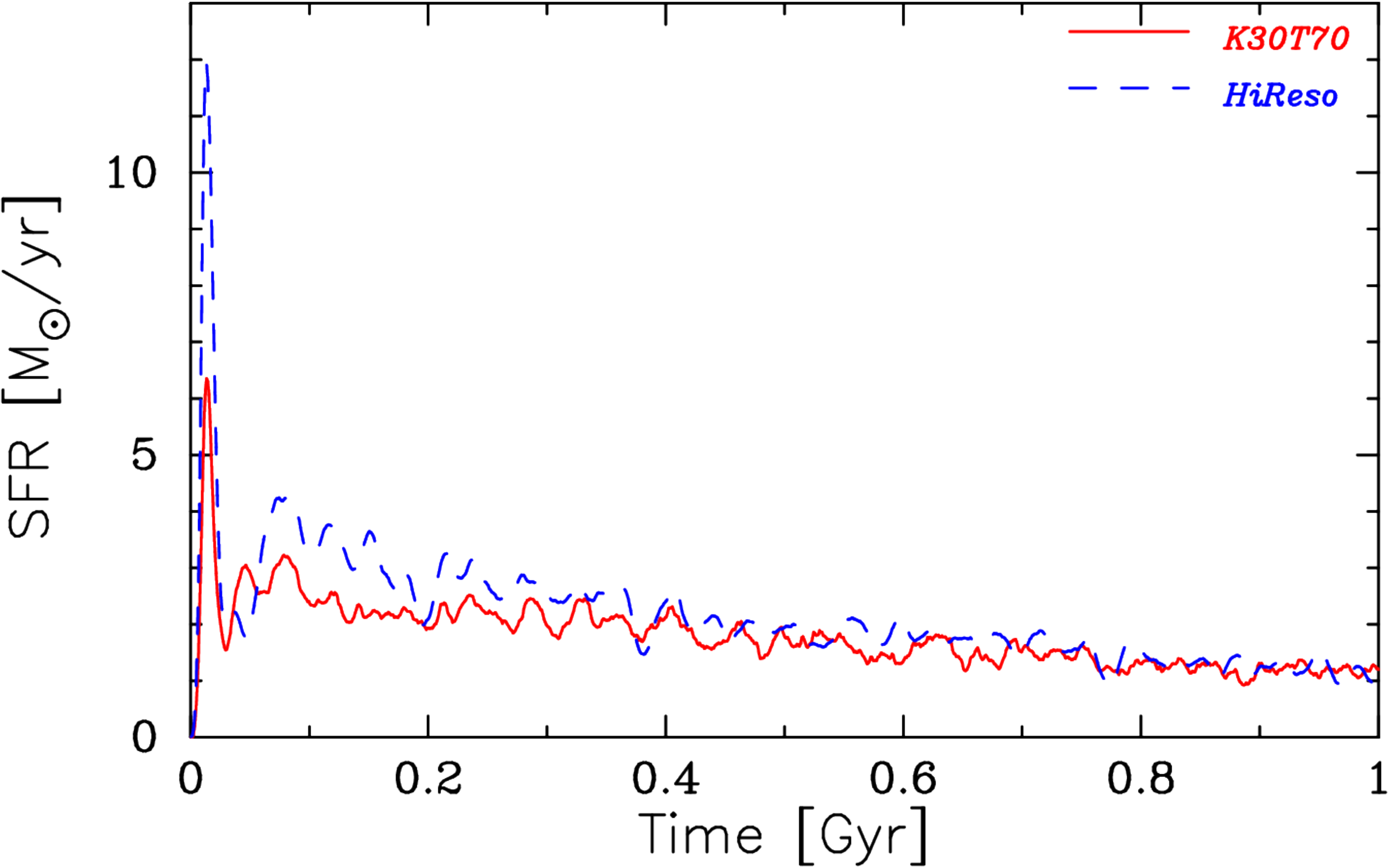}
\caption{Same as Fig.\,\ref{fig:sfhm12}, but for but for the fiducial (solid line) and HiReso (dashed line) runs. }
\label{fig:sfhm12hi}
\end{figure}

Figure~\ref{fig:sfhm12hi} represents the star formation history for our fiducial and HiReso runs. 
Larger difference between them is seen in the very early burst phase but the difference becomes small at later phase. 
This difference is caused by the simulation resolution itself rather than the feedback effect. 

\begin{figure*}
\includegraphics[width = 160mm]{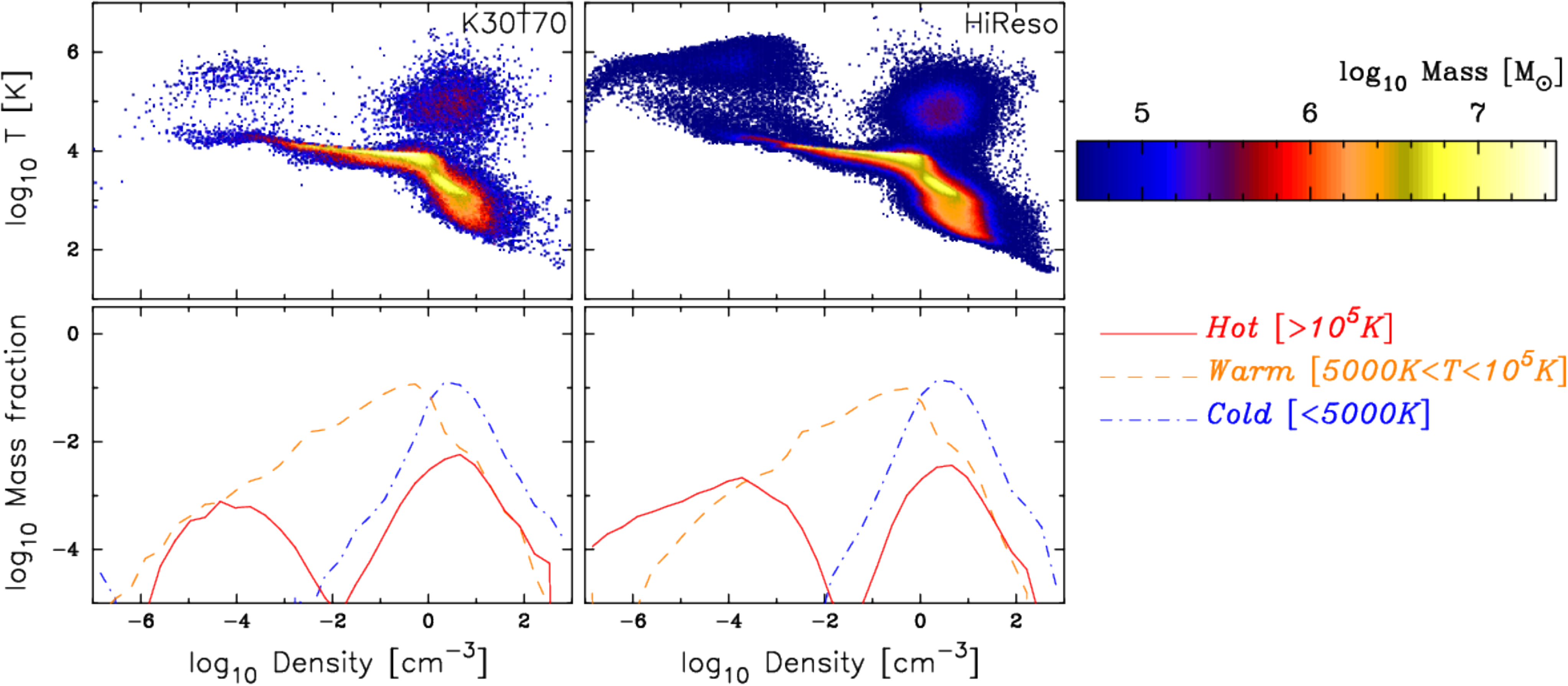}
\caption{Same as Fig.\,\ref{fig:pdm12}, but for the fiducial (left panel) and HiReso (right panel) runs. }
\label{fig:pdm12hi}
\end{figure*}

Next, we investigate the density--temperature phase diagram to see the difference of SN feedback effects. 
Figure~\ref{fig:pdm12hi} represents the phase diagram for our fiducial and HiReso runs. 
HiReso run can produce a lot of low-density and high-temperature gas in visual impression. 
However, the PDFs for each component (cold, warm and hot gas) between both runs are very similar to each other. 
This means that the resolution dependency of the Osaka model is weak. 
We recognize the difference only at very low-density regions. 
Outflow gas from the outer region contributes to such components because these gas easily escape from the galaxy.

\begin{figure*}
\includegraphics[width = 160mm]{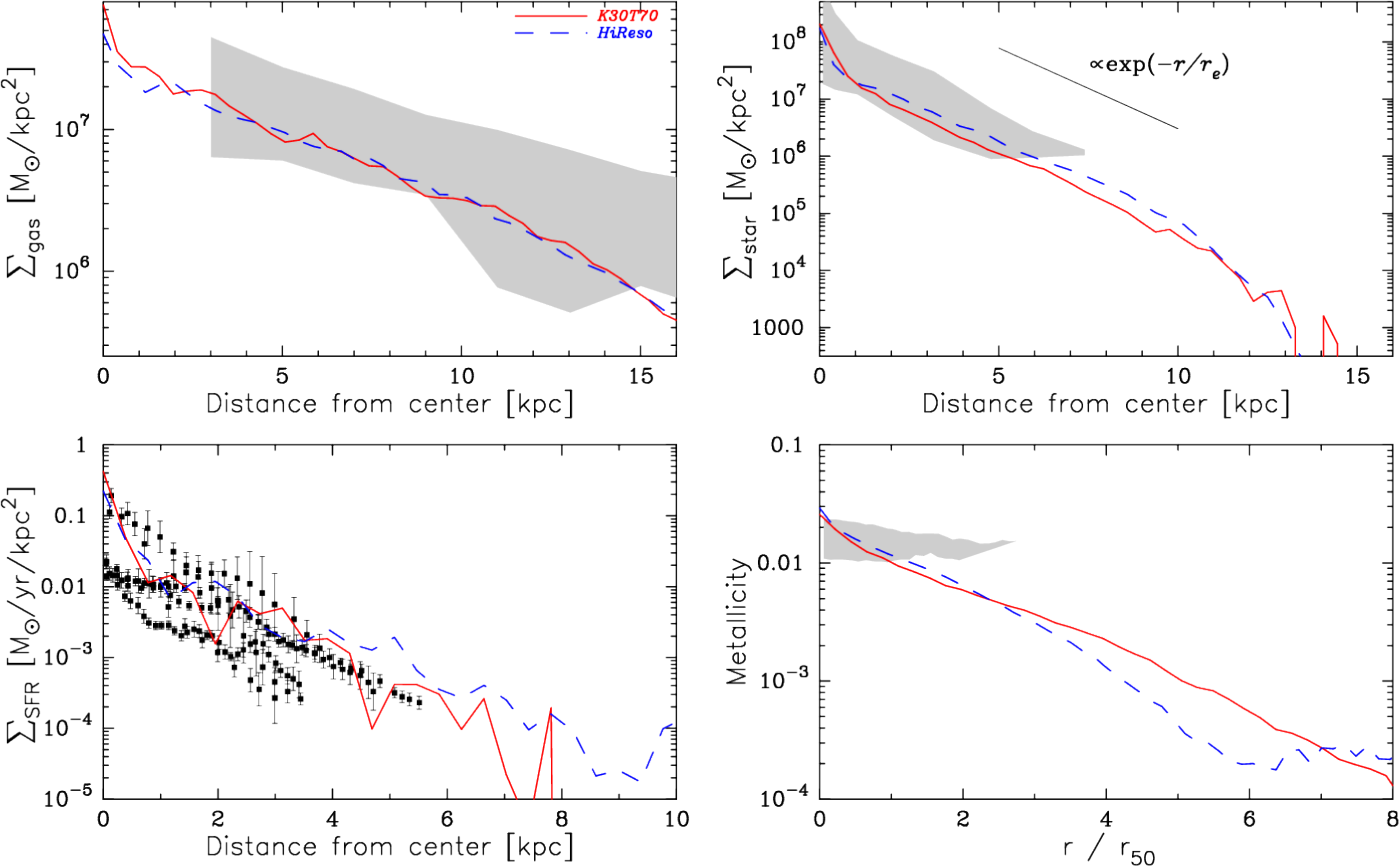}
\caption{Same as Fig.\,\ref{fig:rpm12}, but for the fiducial (solid line) and HiReso (dashed panel) runs. }
\label{fig:rpm12hi}
\end{figure*}

We show the radial profiles of some physical quantities for HiReso run in Figure~\ref{fig:rpm12hi}. 
For comparison, we also plot the profiles of our fiducial run. 
We find that the profiles of HiReso run show similar shapes to our fiducial run in the inner region.
The difference is much smaller than the observational error. 
We find that the profiles for the SFR and metallicity at large radii show a different behaviour from our fiducial run. 
For the SFR profile, HiReso run has larger value than the fiducial run in outer region because HiReso run can resolve the region and hence the star formation as well unlike our fiducial run because we do not change the conditions for the star formation threshold density. 
Interestingly, the metallicity at $2 < r/r_{50} < 7$ for the HiReso run shows smaller value than the fiducial run. 
This suggests that the enriched gas is transferred to the CGM or IGM. 
Actually, the metallicity for the HiReso run at $r/r_{50} > 7$ reverses the situation. 
This implies that in order to know the metal enrichment process, it might be important whether we can resolve the star formation activity in the outer regions of the galaxy. 
We note that the numerical convergence improves if the density threshold is raised similarly to \citet{Parry2012} and \citet{Okamoto2014}. 

\begin{figure}
\includegraphics[width = 80mm]{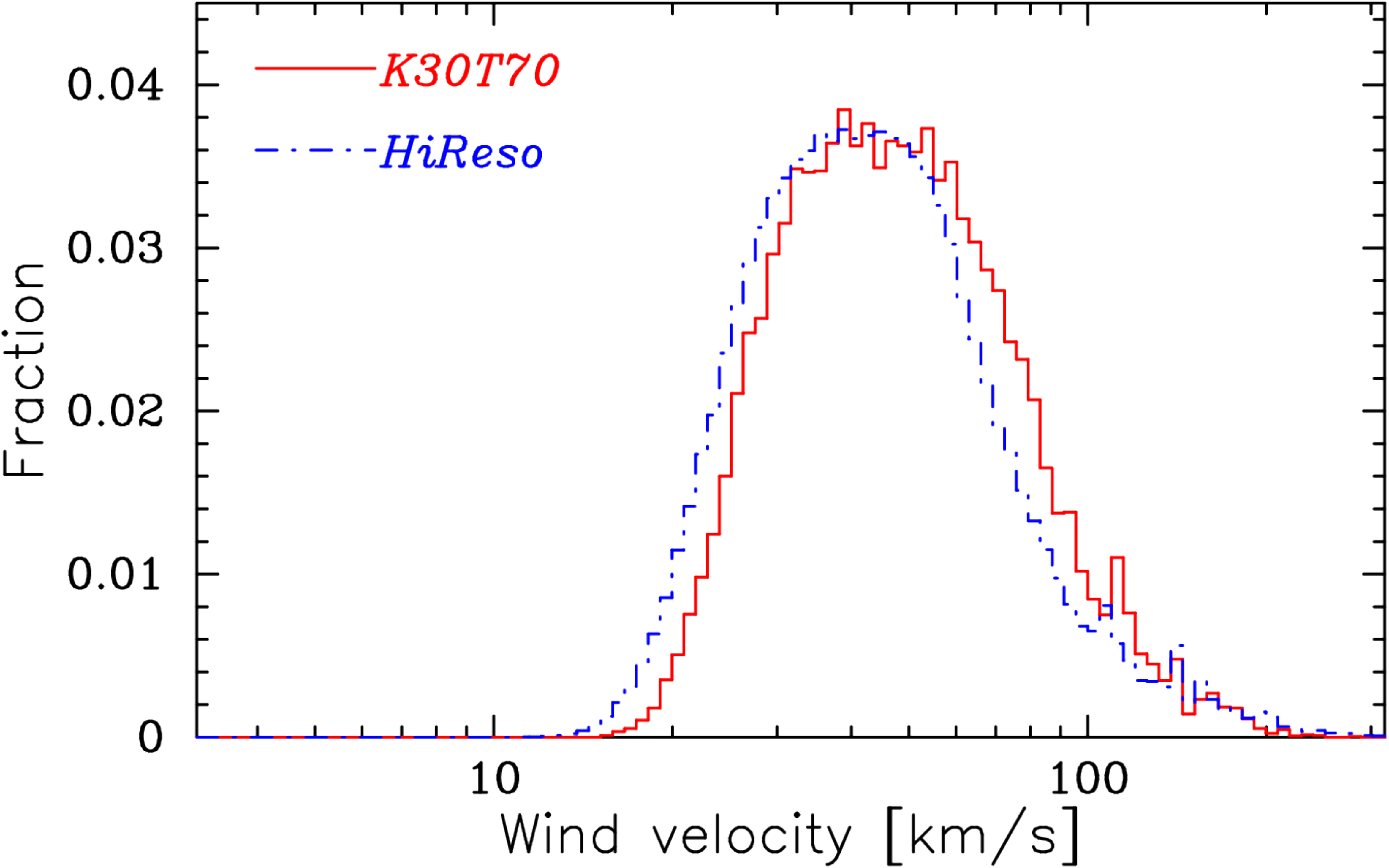}
\caption{Same as Fig.\,\ref{fig:wvdm12}, but only SN-II feedback for the fiducial (solid line) and HiReso (dashed panel) runs. }
\label{fig:wvdm12hi}
\end{figure}

Finally, we explore the PDF of the wind velocity. 
Figure~\ref{fig:wvdm12hi} represents the PDFs of the wind velocity driven by the SN-II feedback in the fiducial and HiReso runs. 
We find that resolution dependency of the wind velocity is not so strong, and our model does relatively a good job in regulating wind velocity distribution, which is one of the advantages of the Osaka model. 


\subsection{Effects of Early Stellar Feedback and SN-Ia Feedback}
We touch on the effects of the early stellar feedback (ESFB) and SN-Ia feedback in this subsection. 
In this study, we use a modified version of \citet{Stinson2013} for the ESFB. 
Our new feature of the ESFB is temperature capping to prevent gas heating up past 20000 K. 
Thus, the thermal pressure of our model might be smaller than that of original Stinson's work. 
As a result, the ESFB of our model may have little effect on the physical state of the gas and the star formation activity. 
Since SN-Ia explosions have time delays from the moment of star formation as shown in Fig. \ref{fig:celib}, the SN-Ia feedback does not work in the early phase of the galaxy formation. 
The SN-Ia feedback may be dominant only the late phase. 

\begin{figure}
\includegraphics[width = 80mm]{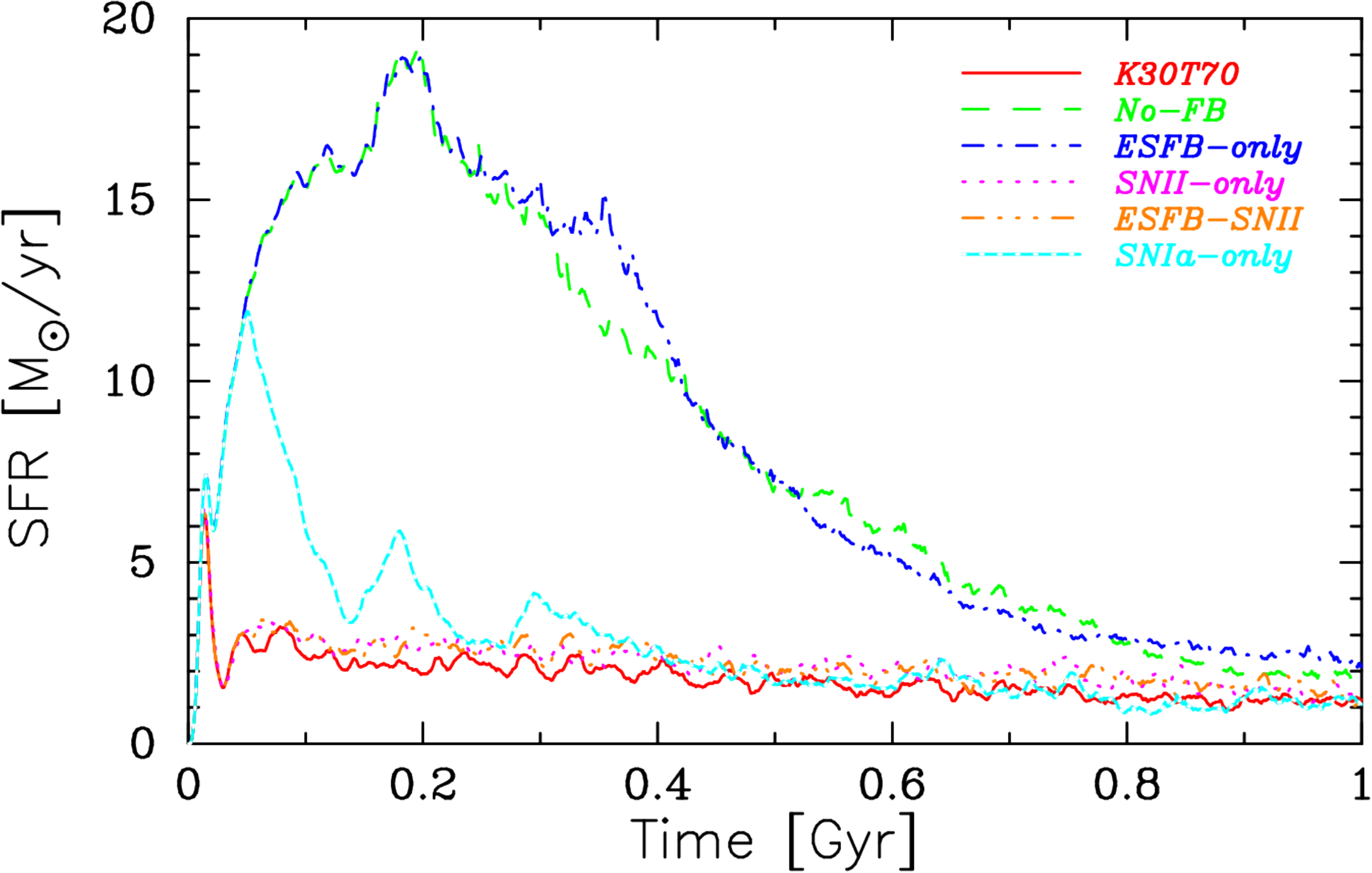}
\caption{Star formation history for another model runs of Fig. \ref{fig:sfhm12}. 
The line styles and their corresponding models are noted in the panel. }
\label{fig:sfhm12other}
\end{figure}

\begin{figure}
\includegraphics[width = 80mm]{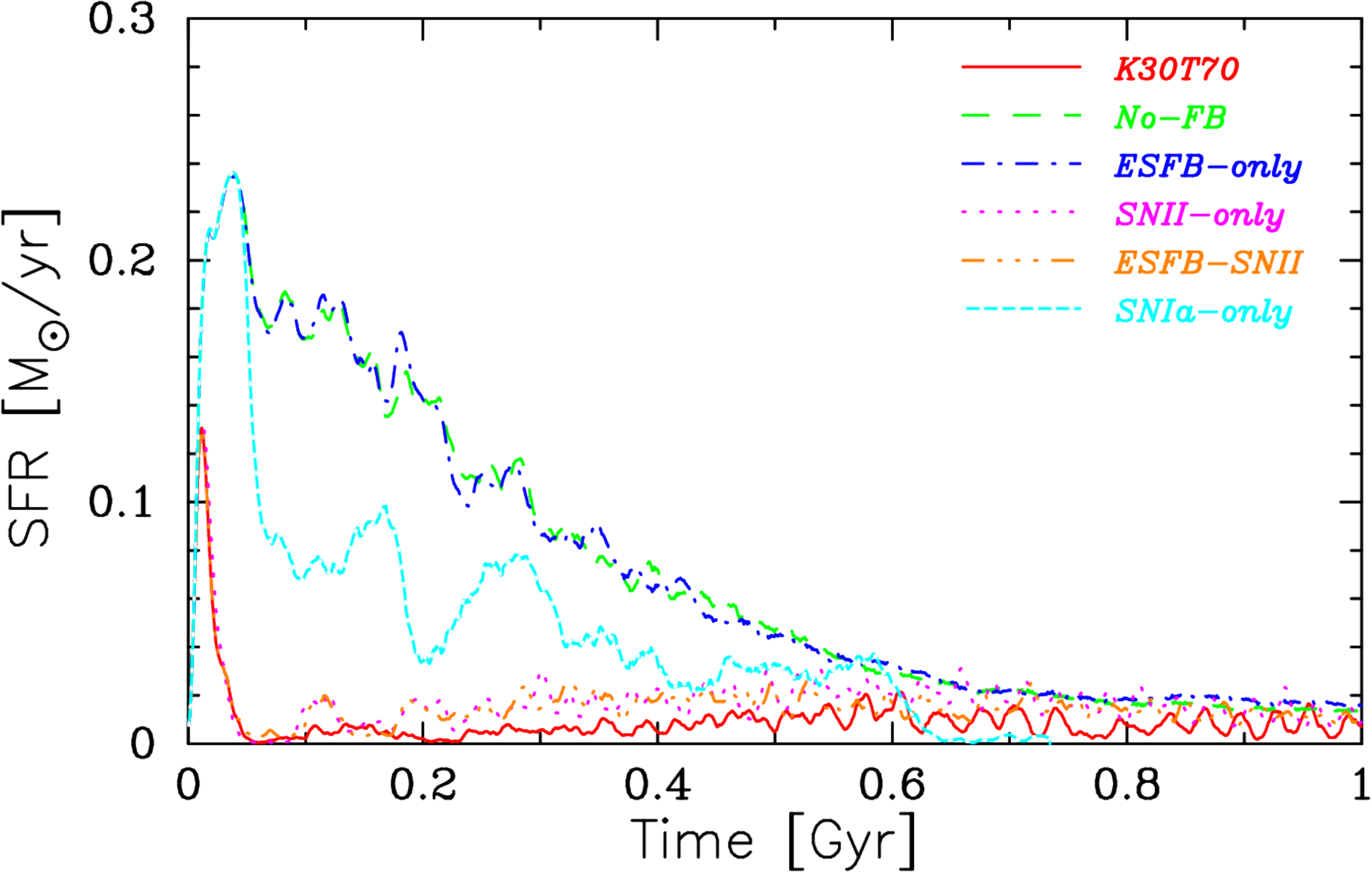}
\caption{Same as Fig. \ref{fig:sfhm12other}, but for M10 galaxy. }
\label{fig:sfhm10other}
\end{figure}

In Figures \ref{fig:sfhm12other} and \ref{fig:sfhm10other}, we show another model runs of above subsection for studying the effects of the ESFB and SN-Ia feedback. 
From comparison between the No-FB and ESFB-only or SNII-only and ESFB-SNII, we find that the ESFB of our model has little impacts on gas dynamics as well as the star formation activity. 
This is inconsistent with the \citet{Stinson2013} in which the effect is noticeable to suppress the star formation activity at the early phase. 
The reason is we newly introduce the temperature capping. 
Moreover, in our ESFB treatment, we distribute the ESFB energy to neighbour particles and the number is 128. 
This means that the energy that each gas particle receives becomes smaller than \citet{Stinson2013}. 
From this results, we conclude that ESFB feedback of our treatment does not increase subsequent SN feedback effect. 

As we expected, the time-delay effect of SN-Ia feedback clearly can be seen in SNIa-only runs. 
Compared to ESFB-SNII or SNII-only run, the star formation for the K30T70 is more suppressed due to the SN-Ia feedback. 
This suggests that the delayed effect of SN-Ia feedback is not negligible even though the SN-II feedback is the dominant contributor for the star formation suppression. 


\subsection{Evolution of Metal Abundance}
\label{sec:abundance}
As we described in Sec. \ref{sec:sim_set}, one of the new features of the present work is the implementation of {\small CELib} package. 
A noticeable feature of {\small CELib} package is that the yield pattern of each element for SN-II, SN-Ia, and AGB stars depends on the stellar age and metallicity. 
In our implementation, we trace 13 elements (H, He, C, N, O, Ne, Mg, Si, S, Ca, Fe, Ni and Eu) using this package. 

In this subsection, we present the evolution of some elements in our fiducial run, taking into account the SF history and time delays between SN-II and SN-Ia, as well as AGB stars. 
Moreover, we compare our results with the observational data.  

\begin{figure*}
\centering
\includegraphics[width = 170mm]{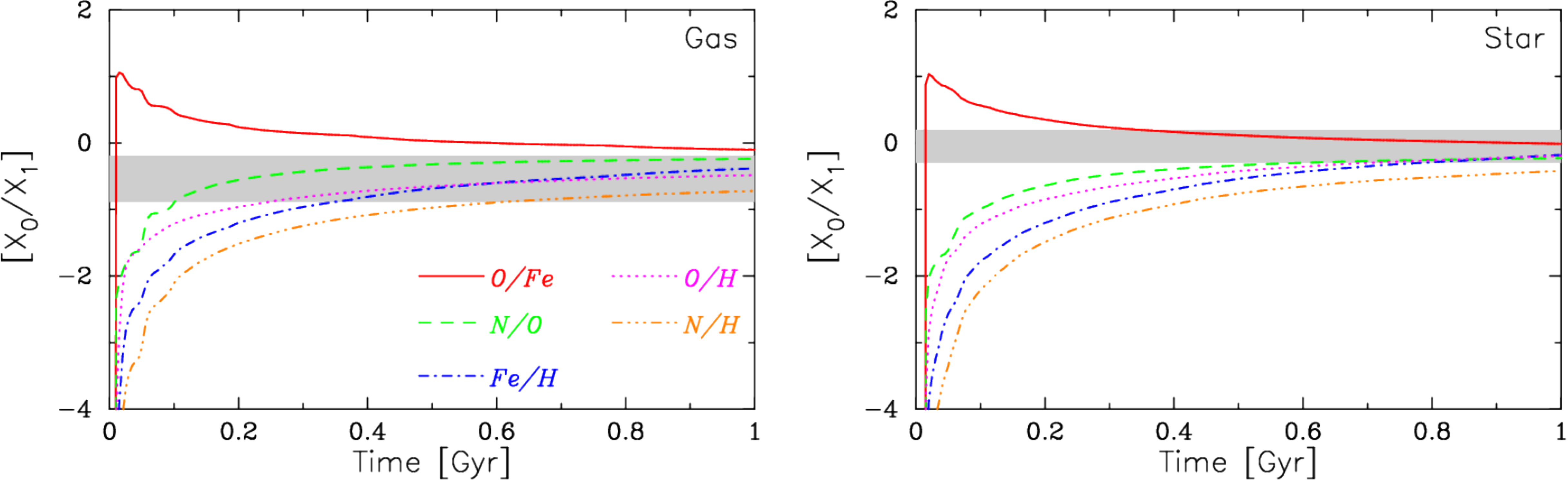}
\caption{Time evolution of metal abundance ratios for the fiducial run. 
Corresponding line styles are shown in the legend. 
Shaded area in the left panel is  [N/O] data from the SDSS galaxies \citep{Andrews2013} for the galaxy stellar mass range $10^{7.5} < M_{*} < 10^{8.5}~{\rm M_{\odot}}$, which covers the same range as our fiducial run. 
Shaded area in the right panel is [Fe/H] data from SDSS APOGEE galaxies \citep{Ness2018, Gutcke2019}. 
}
\label{ear}
\end{figure*}

In Figure~\ref{ear}, we show the time evolution of metal abundance ratios, together with the observed range in shaded regions (left panel: [N/O], right panel: [Fe/H]). 
We find that the abundance ratio of each element (N, O, and Fe) to hydrogen increases with time, and our fiducial run is consistent with the observations at late times.  

It is well known that the abundance ratio [$\alpha$/Fe] is constant during the early evolutionary phase of galaxies (i.e., SN-II dominant phase), and it starts to decrease with increasing SN-Ia contribution at later times. 
For example, [O/Fe] in our fiducial run decreases with time as expected. 
However, in our current result, the SN-Ia contribution might not be the dominant cause for this trend, because in {\small CELib}, the SN-Ia contribution only slowly increases after a few 100\,Myr. 
According to the SN-II yield data used in {\small CELib}, [O/Fe] decreases with increasing stellar metallicity, which  could be the main driver for [O/Fe] to decrease in our fiducial run at $t<1$\,Gyr. 
Our result suggests that it is important to consider the details of metal abundance pattern evolution in order to study the abundance ratio, and not just the number ratio between SN-II and SN-Ia events. 
To further compare the simulation results to the abundant observational data on metals including those from high-$z$ galaxies, we need to perform more realistic cosmological zoom-in hydrodynamic simulations which takes galaxy mergers and gas inflow effects into account properly over cosmic time. 
Studying the impact of $n_{\rm fb}$ on the abundance of each metal element in simulated galaxies is also important for understanding the evolution. 
We will discuss these issues in another paper. 

\section{Summary}
In this paper, we compared different implementations of stellar and supernova feedback, and presented the results of Osaka feedback model, which largely followed the work of  \citet{Stinson2006,Stinson2013} with some changes. 
Our model is based on the numerical and analytic solutions of the evolutionary phases of SNR. 
The Osaka model considers not only the thermal feedback, but also the kinetic mode of feedback, unlike \citet{Stinson2006} where they considered only thermal form feedback.  
In order to explore the impact of various implementations of feedback on  galaxy evolution, we used an idealized Milky Way-type disc galaxy with a halo mass of $10^{12}~{\rm M_{\odot}}$ from the AGORA code comparison project \citep{Kim14, Kim2016} 
and a dwarf galaxy with a mass of $10^{10}\,{\rm M_\odot}$ following \citet{Vecchia2008,Vecchia2012}. 
Another important update was the adoption of the {\small CELib} chemistry package, which allows us to treat SN-Ia and SN-II feedback separately. 
Moreover, we can trace detailed metal enrichment including various metal elements such iron, nitrogen and oxygen that are routinely used to estimate the metallicity in observed galaxies. 
In order to take advantage of this chemistry library, we implemented a method to treat SN feedback in successive time bins rather than a single integrated event. 

The Osaka feedback model proves to alleviate the over-cooling problem without artificial shut-off of hydrodynamic interaction. 
We find that the kinetic feedback is able to suppress the star formation activity and transport gas and metals into the circum-galactic and inter-galactic environment. 
In addition, the thermal feedback helps to drive the outflows (Figs.\,\ref{fig:pdm12} and \ref{fig:pdm10}). 
The fiducial Osaka model produces a nice, smooth disc galaxy with few gas clumps, as well as the radial profiles of gas surface density, stellar mass surface density, SFR, and metallicity within the observed range (Fig.\,\ref{fig:rpm12}). 

We also test the model with the ESFB effect using isolated galaxy simulations. 
The ESFB model is similar to that of \citet{Stinson2013} but we newly incorporate the temperature capping to avoid over-heating ($> 10^6~{\rm K}$). 
We find that our ESFB model does not suppress the star formation nor enhance the SN feedback effect.  
This conclusion is somewhat at odds with other works which stressed strong impact of early momentum feedback from radiation pressure and stellar winds of massive stars \citep[e.g.,][]{Agertz15}. 
Note that, on the other hand, \citet{Vecchia2012} and \citet{Keller2014} reproduced the observations without ESFB. 
Our results are consistent with their results.  

We stress that it is still under the debate whether the ESFB strongly affects the star formation activity and galactic wind or not.  
Given that we only considered the ESFB in the form of a simple thermal feedback and temperature capping in the present work, it is possible that our treatment of ESFB is inadequate, and we plan to investigate this issue further as we improve our numerical resolution. 
For example, we would consider the additional radiation pressure applied onto dust grains whose formation and destruction can be computed self-consistently within our code using the formulation described by \citet{Aoyama2017}. 

One important parameter in the Osaka feedback model is the number of SN feedback event $n_{\rm fb}$, over which we deposit the feedback energy. 
This parameter determines the amount of SN energy deposited each time, as well as the bubble radius over which the energy is distributed. 
We find a nice balance between the value of $n_{\rm fb}$, number of particles that receive the feedback energy, and bubble radius, whose effects cancels with each other and  results in nearly constant peak velocity of wind particle distribution. 

In this paper, we did not take full advantage of detailed yields that the {\small CELib} package offers, except for the general abundance evolution in the fiducial run (see  Section~\ref{sec:abundance}). 
It is encouraging that we obtain consistent abundances on [N/O] and [Fe/H] with observations as shown in Fig.~\ref{ear}. 
In the future, we will perform more in-depth analysis on the chemical enrichment of CGM and IGM by different elements, and compare against various observational data from absorption line studies. 


\section*{Acknowledgments}

This work is supported in part by the JSPS KAKENHI Grant Number JP26247022 and JP17H01111.
We are grateful to Jun-Hwan Choi and Robert Thompson for their help in the early phase of this work. 
We also thank Shohei Aoyama for discussions on the Osaka feedback model, and Takayuki Saitoh on the usage of the {\small CELib} library. 
This work is partly based on the Master's thesis of Keita Todoroki at University of Nevada, Las Vegas. 
Numerical simulations were performed on Cray XC30 and XC50 at CfCA, National Astronomical Observatory of Japan. 
We also utilised the {\small OCTOPUS} at the Cybermedia Centre, Osaka University, as part of the HPCI system Research Project (hp180063).
KN acknowledges the travel support from the Kavli IPMU, World Premier Research Center Initiative (WPI), where part of this work was conducted.




\bibliographystyle{mnras}
\bibliography{mn} 




\appendix



\bsp	
\label{lastpage}
\end{document}